\PassOptionsToPackage{table}{xcolor}
\documentclass[final,5p,times,twocolumn,nopreprintline]{elsarticle}

\usepackage{amssymb}
\usepackage{amsmath}
\usepackage{amsfonts}
\usepackage{algorithmic}
\usepackage{algorithm}
\usepackage{array}
\usepackage{textcomp}
\usepackage{stfloats}
\usepackage{xurl}
\usepackage{verbatim}
\usepackage{graphicx}
\usepackage[T1]{fontenc}
\usepackage{multirow}
\usepackage{xspace}
\usepackage{nicefrac}
\usepackage{hyperref}
\usepackage[capitalise,noabbrev]{cleveref}
\usepackage[most]{tcolorbox}
\usepackage{booktabs}
\usepackage[labelformat=simple]{subfig}
\usepackage{orcidlink}
\usepackage{balance}

\definecolor{darkblue}{HTML}{09357A}
\definecolor{forestgreen}{rgb}{0.13, 0.55, 0.13}

\newcommand{\altparagraph}[1]{%
  \smallskip
  \textbf{#1}
  \noindent
}

\newtcolorbox{insightbox}{
    enhanced,                           %
    colback=gray!10,                    %
    colframe=gray!10,                   %
    left=5mm,                           %
    right=5mm,                          %
    borderline west={1mm}{0mm}{gray},   %
}
\newcounter{insight}
\newenvironment{insight}{
  \stepcounter{insight}
  \begin{insightbox}
    \textbf{Insight \theinsight.}
}{
  \end{insightbox}
}

\definecolor{barblue}{RGB}{31, 119, 180}
\newlength{\barunit}
\newcommand{\db}[1]{\textcolor{barblue}{\rule{#1\barunit}{1.8ex}}}

\def\AzureFirst{\textsc{Azure19}\xspace}
\def\AzureSecond{\textsc{Azure21}\xspace}
\def\AzureBlob{\textsc{Azure20}\xspace}
\def\Alibaba{\textsc{Alibaba21}\xspace}
\def\HuaweiFirst{\textsc{Huawei23}\xspace}
\def\HuaweiSecond{\textsc{Huawei25}\xspace}
\def\Globus{\textsc{Globus24}\xspace}
\def\IBM{\textsc{IBM26}\xspace}

\begin{document}

\begin{frontmatter}

\title{Characterizing Serverless Workloads:\\A Large-Scale Analysis of Function-as-a-Service Platforms}

\author[1]{Saman Akbari~\orcidlink{0009-0009-8446-4200}\corref{cor1}}
\ead{akbari@tu-berlin.de}

\author[1]{Yassine Lazreg~\orcidlink{0009-0003-1533-0459}}
\ead{y.lazreg@tu-berlin.de}

\author[1]{Mohammadreza Malekabbasi~\orcidlink{0009-0002-0627-3600}}
\ead{malekabbasi@tu-berlin.de}

\author[1,2]{Manfred Hauswirth~\orcidlink{0000-0002-1839-0372}}
\ead{manfred.hauswirth@tu-berlin.de}

\cortext[cor1]{Corresponding author}

\affiliation[1]{
    organization={Technische Universität Berlin},
    city={Berlin},
    country={Germany}
}

\affiliation[2]{
    organization={Fraunhofer Institute for Open Communication Systems (FOKUS)},
    city={Berlin},
    country={Germany}
}

\begin{abstract}
Serverless computing is a popular approach to application deployment without having to manage the underlying infrastructure. Its most common implementation is Function-as-a-Service (FaaS), in which applications are structured as fine-grained functions. FaaS providers face non-trivial tasks in the management of their platforms, including cold starts, auto-scaling, and scheduling. The design of optimized platforms requires an accurate understanding of realistic serverless workloads. However, existing research and platforms rely on insights from single, often outdated datasets.
In this work, we conduct a large-scale, cross-platform comparative analysis of eight publicly available FaaS workload traces from major cloud providers (Alibaba, Azure, Huawei, IBM) and a scientific computing platform (Globus). We provide a comprehensive and up-to-date view of the serverless ecosystem by characterizing key dimensions such as invocation patterns, latency, cold start behavior, and resource utilization.
Our analysis shows that although invocation patterns remain universal, execution times have shortened, shifting the performance bottleneck toward platform scheduling and dependency management. We also observe an architectural shift toward ``Serverless 2.0'' models, where concurrent, long-lived containers are increasingly utilized to bypass cold start penalties. Furthermore, we found resource over-provisioning and extreme skew in user request volumes. These findings provide actionable insights for designing more efficient FaaS platforms, informing improved approaches to targeted cold start mitigation, predictive resource management, and dependency-aware scheduling.
\end{abstract}


\end{frontmatter}

\section{Introduction}
\label{sec:introduction}

Function-as-a-Service (FaaS) has fundamentally altered application development by eliminating infrastructure management. Serverless platforms like AWS Lambda, Azure Functions, and Google Cloud Functions allow developers to deploy and execute code on demand, with the provider managing scaling, provisioning, scheduling, and other infrastructure tasks~\cite{shahrad2020serverless}. This application and execution model has seen rapid adoption, driven by the paradigm's promise of reduced operational overhead and a pay-per-use pricing model, where users are billed only for the computation time their functions consume~\cite{awslambda}.

However, this user-centric convenience creates significant challenges for FaaS providers. To provide the illusion of infinite, instantaneous resources, providers must navigate a complex trade-off between performance and cost-efficiency. A central challenge is the ``cold start,'' the latency incurred when a new function instance must be initialized from scratch, which can dramatically increase response times~\cite{shahrad2020serverless, joosen2025serverless}. Providers mitigate this by keeping recently used instances ``warm,'' but this consumes additional resources. Without a deep understanding of workload patterns, providers risk either over-provisioning resources, leading to high operational costs, or under-provisioning, resulting in poor user experience.

Effective optimization strategies, such as predictive pre-warming, intelligent keep-alive policies, or optimized scheduling, depend on an accurate characterization of production FaaS workloads.
Beyond cold start mitigation, providers face numerous other operational challenges, such as dynamic auto-scaling, efficient memory and CPU bin-packing, data locality and caching. The design of efficient management algorithms requires precise knowledge of serverless workloads.
However, much of the existing research relies on a limited number of public datasets, with most studies referencing the Azure trace from 2019~\cite{shahrad2020serverless}. This is problematic for two reasons: First, we cannot assume that the workload of a single platform is representative of the entire FaaS ecosystem. Second, the FaaS landscape is rapidly evolving and workload characteristics from several years ago may no longer reflect current usage patterns.

This paper addresses this gap by conducting a large-scale, cross-platform comparative analysis of eight publicly available FaaS workload traces from major cloud providers (Alibaba, Azure, Huawei, IBM) and a scientific computing platform (Globus). By comparing these diverse datasets, we aim to identify common patterns, platform-specific differences, and temporal trends in serverless workloads. Our analysis focuses on metrics that directly impact system design and performance, such as request arrival patterns, latencies, cold start behavior, resource usage, and application dependencies.

Our contributions are twofold: (1) We provide a comprehensive characterization of modern serverless workloads across multiple platforms to identify universal patterns, as well as distinct nuances across diverse deployments. (2) We translate these findings into concrete recommendations for the design of FaaS systems, informing areas such as predictive resource management, targeted cold start mitigation, and efficient scheduling. By making these datasets and their characteristics systematically comparable, we aim to provide a solid empirical foundation for future research in serverless computing.

Our findings show that the serverless ecosystem has evolved. Although strong diurnal request patterns and extreme skewness in function popularity and user request volumes remain universal, modern FaaS executions have become dramatically shorter, often completing in just a few milliseconds. This shift turns orchestration overhead, particularly scheduling delays and dependency deployment, into the primary performance bottleneck. Furthermore, we expose pervasive resource over-provisioning, a growing architectural shift toward ``Serverless 2.0'' models that favor persistent, concurrent containers, and a distinct divide between commercial ``nano-service'' platforms and heavy-tailed scientific execution environments.

The rest of this paper is organized as follows: \cref{sec:background} provides the necessary context on FaaS concepts and optimization techniques. \cref{sec:methodology} describes the datasets and our analysis methodology. \cref{sec:characterization} presents our findings on workload characteristics. We discuss the implications of our results in \cref{sec:discussion}, review related work in \cref{sec:related_work}, and finish with our conclusions in \cref{sec:conclusion}. To support reproducibility and further research, we have open-sourced the complete analysis: \textcolor{darkblue}{\url{https://github.com/saman-akbari/characterizing-serverless-workloads}}.

\section{Background and Motivation}
\label{sec:background}
This section recaps the FaaS execution model and reviews existing serverless workload traces.

\subsection{Serverless Computing and Function-as-a-Service}
\begin{figure}[!t]
    \centering
        \includegraphics[width=0.8\columnwidth]{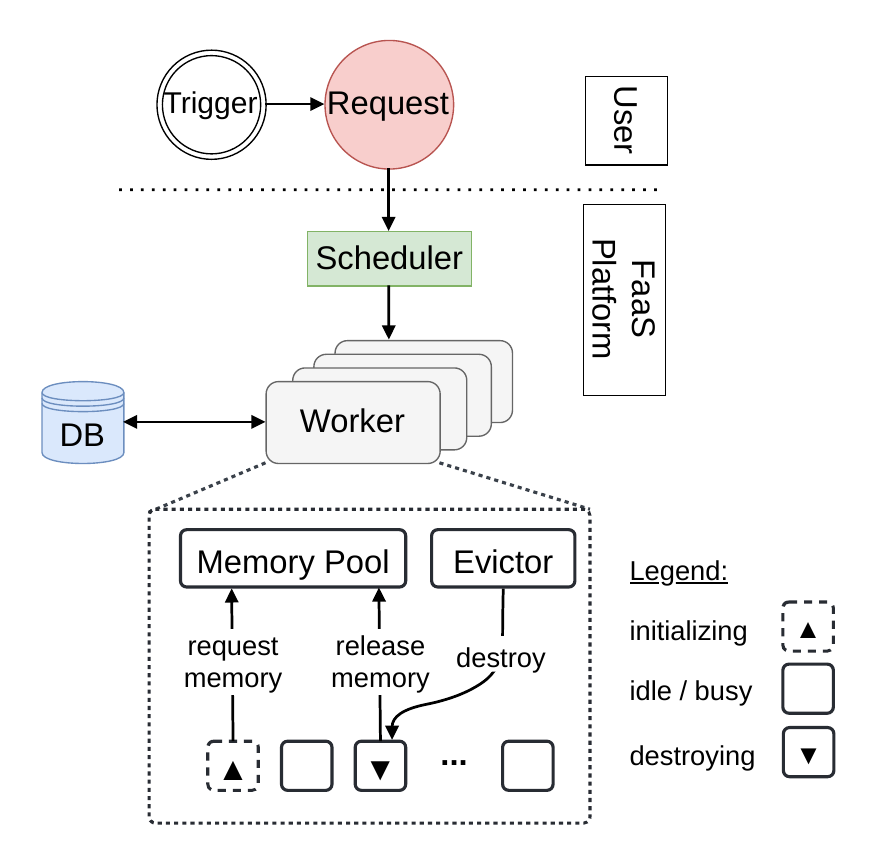}
    \caption{Architectural overview of Function-as-a-Service platforms. Workers manage a memory pool and execute functions in response to events in virtualized execution environments (sandboxes).}
    \label{fig:faas_platform_architecture}
\end{figure}

\Cref{fig:faas_platform_architecture} shows the high-level architecture and execution flow of a typical FaaS platform. Following an event-driven programming model, workers execute stateless functions in response to specific triggers, such as HTTP requests, scheduled events, or messages from other services. A scheduler routes the incoming request to a worker node. Because FaaS functions are inherently stateless, workers often interact with external storage or databases to read or persist data during and between executions. Internally, each worker manages a dynamic set of virtualized execution environments (e.g., containers or microVMs). When a request arrives and no preexisting instance is available, the worker allocates resources from its memory pool. This creates a new instance, incurring a ``cold start'' latency as the environment boots and dependencies are loaded. Once prepared, the instance processes the request and subsequently remains idle in memory for some time to serve potential future requests (``warm starts''). To maintain cluster efficiency and free up resources for new functions, an evictor component actively monitors the worker. It destroys instances that have been idle beyond a specified threshold or during memory pressure, releasing their allocated resources back to the memory pool.

\begin{table*}[!t]
    \centering
    \caption{Available metrics across serverless workload traces. Recent datasets provide significantly richer metadata, including cold start breakdowns and dependency information, compared to earlier traces.}
    \label{tab:metrics_analyzed}
    \begin{tabular}{l c c c c c c c c}
        \hline
         \textbf{Metric} & \AzureFirst & \AzureBlob & \AzureSecond & \Alibaba & \HuaweiFirst & \HuaweiSecond & \Globus & \IBM \\
         \hline
         Request Arrival Rates & \checkmark & & \checkmark & & \checkmark & \checkmark & \checkmark & \checkmark \\
         \rowcolor{gray!10} Execution Time & \checkmark & & \checkmark & & \checkmark & \checkmark & \checkmark & \checkmark \\
         Response Time & & & & & \checkmark & \checkmark & \checkmark & \checkmark \\
         \rowcolor{gray!10} Cold Start Latency & & & & \checkmark & & \checkmark & \checkmark & \\
         Memory Usage & \checkmark & & & & \checkmark & \checkmark & & \\
         \rowcolor{gray!10} Memory Limit & & & & \checkmark & \checkmark & \checkmark & & \checkmark \\
         CPU Usage & & & & & \checkmark & \checkmark & & \\
         \rowcolor{gray!10} CPU Limit & & & & & \checkmark & \checkmark & & \checkmark \\
         Instance Count & & & & & \checkmark & \checkmark & & \checkmark \\
         \rowcolor{gray!10} Instance Concurrency Limit & & & & & & & & \checkmark \\
         Minimum Instance Scale & & & & & & & & \checkmark \\
         \rowcolor{gray!10} Imported Libraries & & & & & & & \checkmark & \\
         Function Code Metrics & & & & & & & \checkmark & \\
         \rowcolor{gray!10} Runtimes & & & & \checkmark & & \checkmark & & \\
         Triggers & \checkmark & & & & & \checkmark & & \\
         \rowcolor{gray!10} Users & \checkmark & \checkmark & & & & \checkmark & \checkmark & \checkmark \\
         Data Access & & \checkmark & & & & & & \\
         \rowcolor{gray!10} I/O & & & & & & \checkmark & \checkmark & \\
         \hline
    \end{tabular}
\end{table*}

\subsection{Serverless Workload Traces}
\label{sec:serverless_workload_traces}
Our analysis is based on a collection of publicly available serverless workload traces. These datasets originate from several platforms, including public cloud providers and scientific computing infrastructures, providing a broad perspective on FaaS usage.
Previous work has analyzed these datasets individually, and many studies rely on the first Azure trace from 2019~\cite{shahrad2020serverless}. This isolated and outdated approach can lead to conclusions that are not generalizable across the constantly evolving serverless ecosystem. Our work addresses this limitation by conducting a systematic comparison of existing traces to provide a more comprehensive and structured characterization of serverless workloads.

\altparagraph{Microsoft Azure Functions.}
Microsoft has released three datasets characterizing its FaaS platform, each spanning two weeks. The 2019 trace (\AzureFirst)~\cite{shahrad2020serverless} provides request counts per minute, execution duration percentiles, and memory usage metrics for over 70\,000 functions. The 2020 trace (\AzureBlob)~\cite{romero2021faa} focuses on data-centric behavior, logging interactions between functions and the Azure Blob Storage, including blob types, sizes, and access patterns. The 2021 trace (\AzureSecond)~\cite{zhang2021faster} offers finer granularity with per-invocation start and end timestamps, enabling a precise concurrency and latency analysis.

\altparagraph{Alibaba Cloud Function Compute.}
The Alibaba Cloud trace (\Alibaba)~\cite{wang2021faasnet} characterizes 15~days of production workloads, specifically detailing cold start events. Unlike general invocation traces, this dataset records system-level latencies such as image pulling and container initialization. It includes metadata for function names, allocated memory, and runtime environments, making it particularly valuable for analyzing the costs of container provisioning.

\altparagraph{Huawei Cloud.}
Huawei has contributed two extensive datasets. The first (\HuaweiFirst)~\cite{joosen2023does} spans over seven months and consists of two parts: a ``Public'' trace containing per-minute arrival rates for over 5\,000 functions, and a ``Private'' trace capturing internal workloads with high-fidelity per-second metrics for requests, delays, and resource usage. The subsequent trace (\HuaweiSecond)~\cite{joosen2025serverless}, spanning 31~days across five data center regions, provides a unique component-level breakdown of cold start latencies, separating pod allocation, code deployment, dependency installation, and scheduling overheads.

\altparagraph{Globus Compute Functions.}
The 2024 trace from Globus Compute (\Globus)~\cite{bauer2024globus} captures 31~weeks of activity on a federated FaaS platform used primarily for scientific computing. It logs task executions across various endpoints, ranging from local devices to HPC clusters. A key characteristic of this dataset is that it provides function source code metrics (lines of code, complexity, libraries) and detailed lifecycle timestamps to analyze queuing and overhead in federated environments.

\altparagraph{IBM Cloud Code Engine.}
The 2026 trace from IBM (\IBM)~\cite{nasiri2026serverless} characterizes a ``Serverless 2.0'' platform based on Knative. Spanning 66 days ($\approx 9.5$~weeks) with 1.9~billion invocations, it captures the behavior of general-purpose containers rather than just functions. Furthermore, it includes detailed user configurations absent in other traces, such as CPU and memory limits, minimum scale settings (allowing users to opt-out of scale-to-zero), and container concurrency limits.

\begin{table}[!t]
    \centering
    \caption{Dataset statistics. Number of raw rows and the number of rows excluded during data cleaning.}
    \label{tab:processing_stats}
    \begin{tabular}{lrr}
        \toprule
        Dataset & \# Rows Raw & \# Rows Excluded \\
        \midrule
        \AzureFirst   & 1\,488\,281 & 7 \\
        \AzureBlob    & 44\,282\,912 & 50\,063 \\
        \AzureSecond  & 1\,980\,951 & 2\,606 \\
        \Alibaba      & 1\,629\,445 & 2 \\
        \HuaweiFirst  & 13\,806\,720 & 0 \\
        \HuaweiSecond & 17\,373\,334 & 5 \\
        \Globus       & 2\,399\,438 & 0 \\
        \IBM          & 1\,900\,376\,220 & 0 \\
        \bottomrule
    \end{tabular}
\end{table}

\begin{figure}[!t]
    \centering
    \subfloat[\AzureFirst]{
        \includegraphics[width=0.45\linewidth]{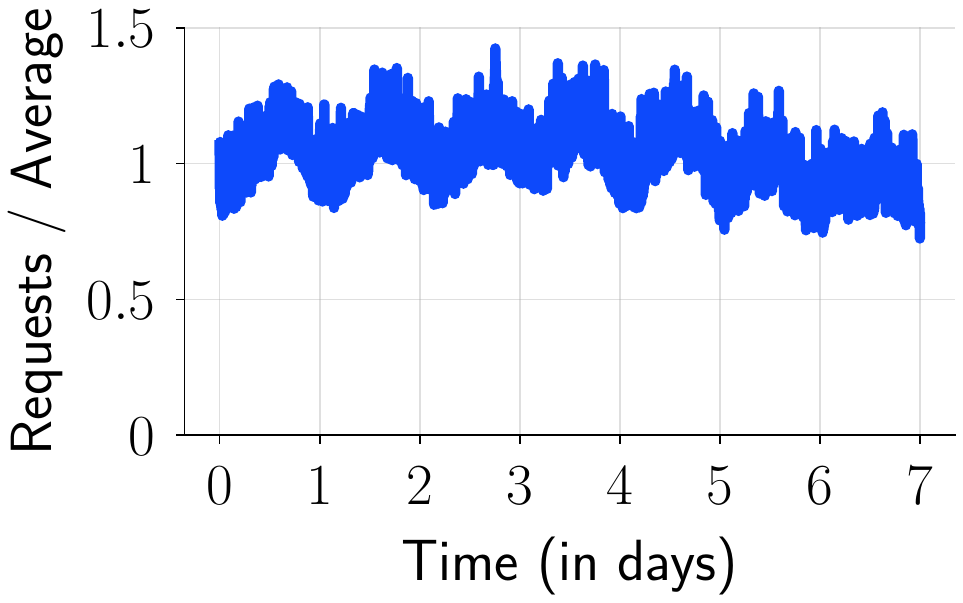}
        \label{fig:azure19_time_series}
    }
    \hfill
    \subfloat[\AzureSecond]{
        \includegraphics[width=0.45\linewidth]{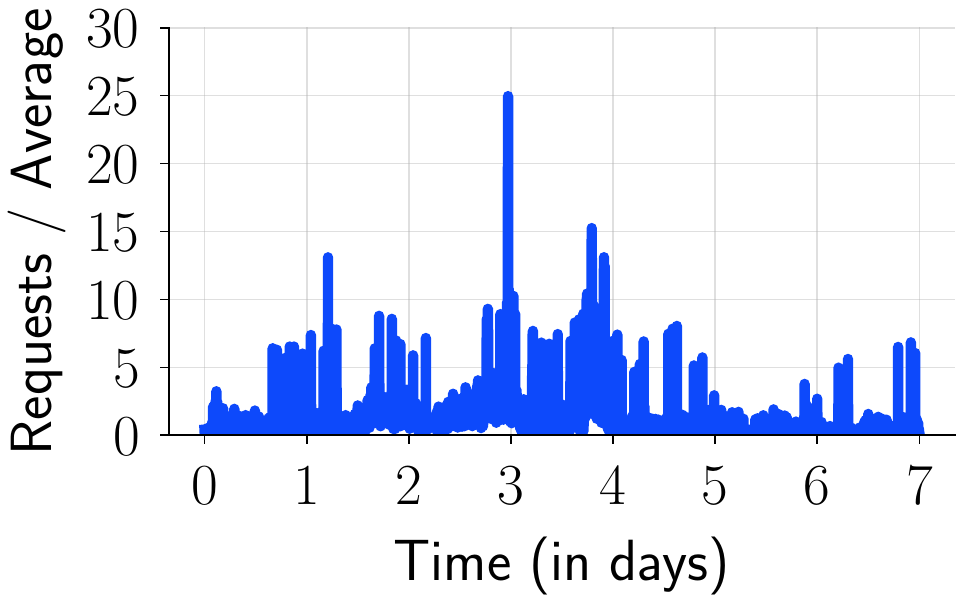}
        \label{fig:azure21_time_series}
    }
    \\
    \subfloat[\HuaweiFirst]{
        \includegraphics[width=0.45\linewidth]{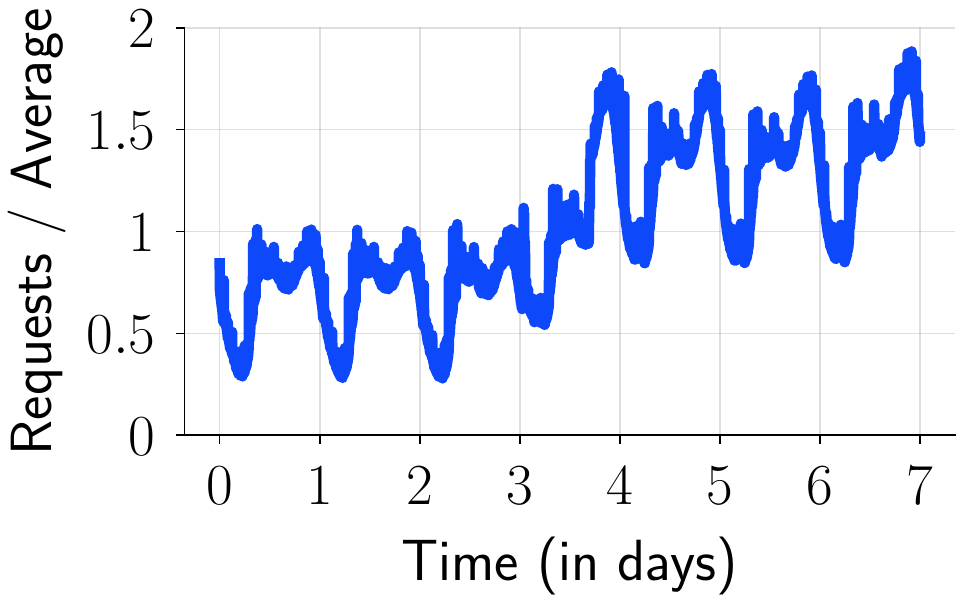}
        \label{fig:huawei23_time_series}
    }
    \hfill
    \subfloat[\HuaweiSecond]{
        \includegraphics[width=0.45\linewidth]{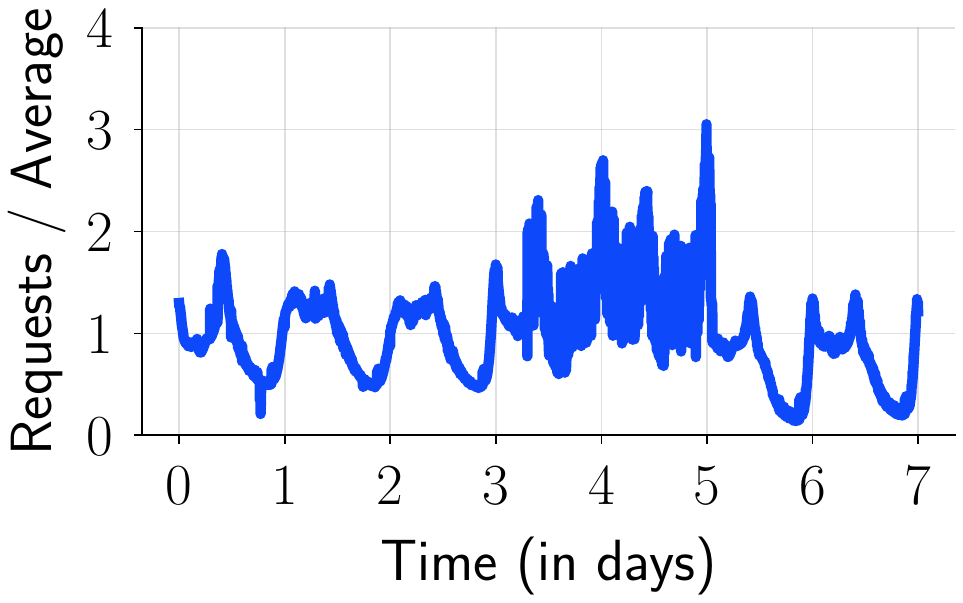}
        \label{fig:huawei25_time_series}
    }
    \\
    \subfloat[\Globus]{
        \includegraphics[width=0.45\linewidth]{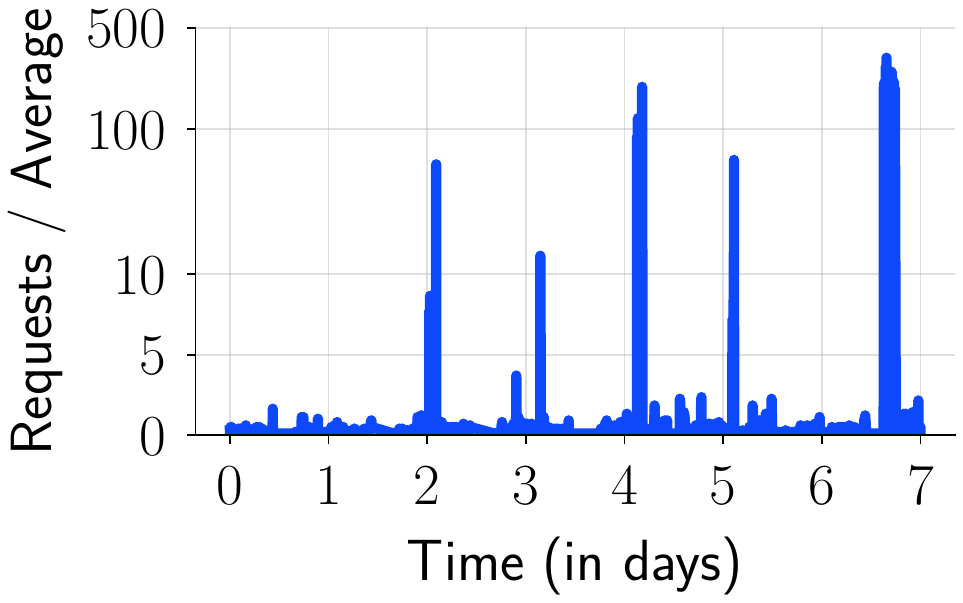}
        \label{fig:globus24_time_series}
    }
    \hfill
    \subfloat[\IBM]{
        \includegraphics[width=0.45\linewidth]{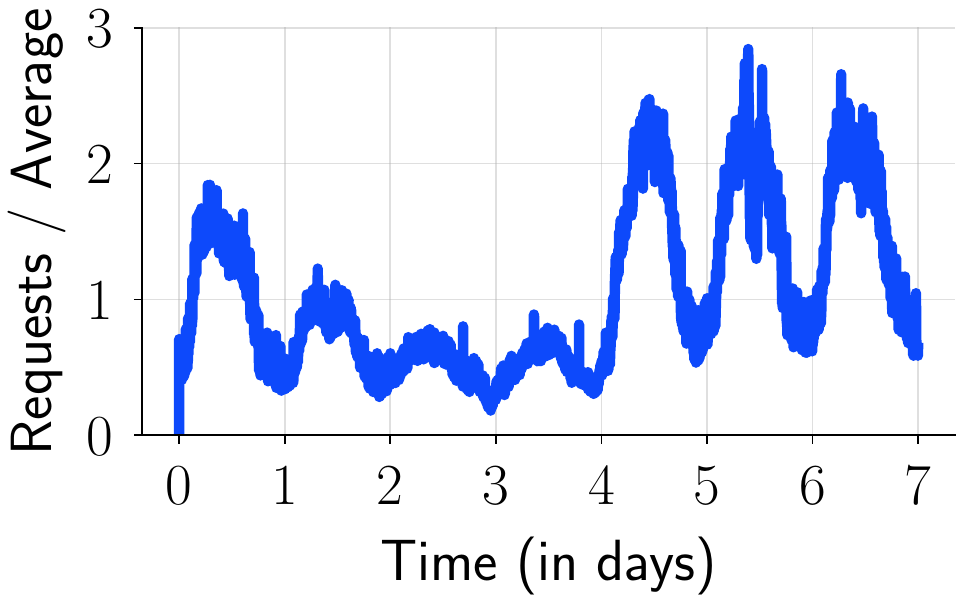}
        \label{fig:ibm26_time_series}
    }
    \caption{Time series of request arrival rates (requests per minute, normalized by the mean) for the first week in all datasets. While platforms like \AzureFirst and \HuaweiFirst maintain tight bands near their average, \AzureSecond and \Globus show extreme burstiness, with sudden spikes reaching $25\times$ and $310\times$ their respective means.}
    \label{fig:arrival_time_series}
\end{figure}

\begin{figure}[!t]
    \centering
    \subfloat[\AzureFirst]{
        \includegraphics[width=0.29\linewidth]{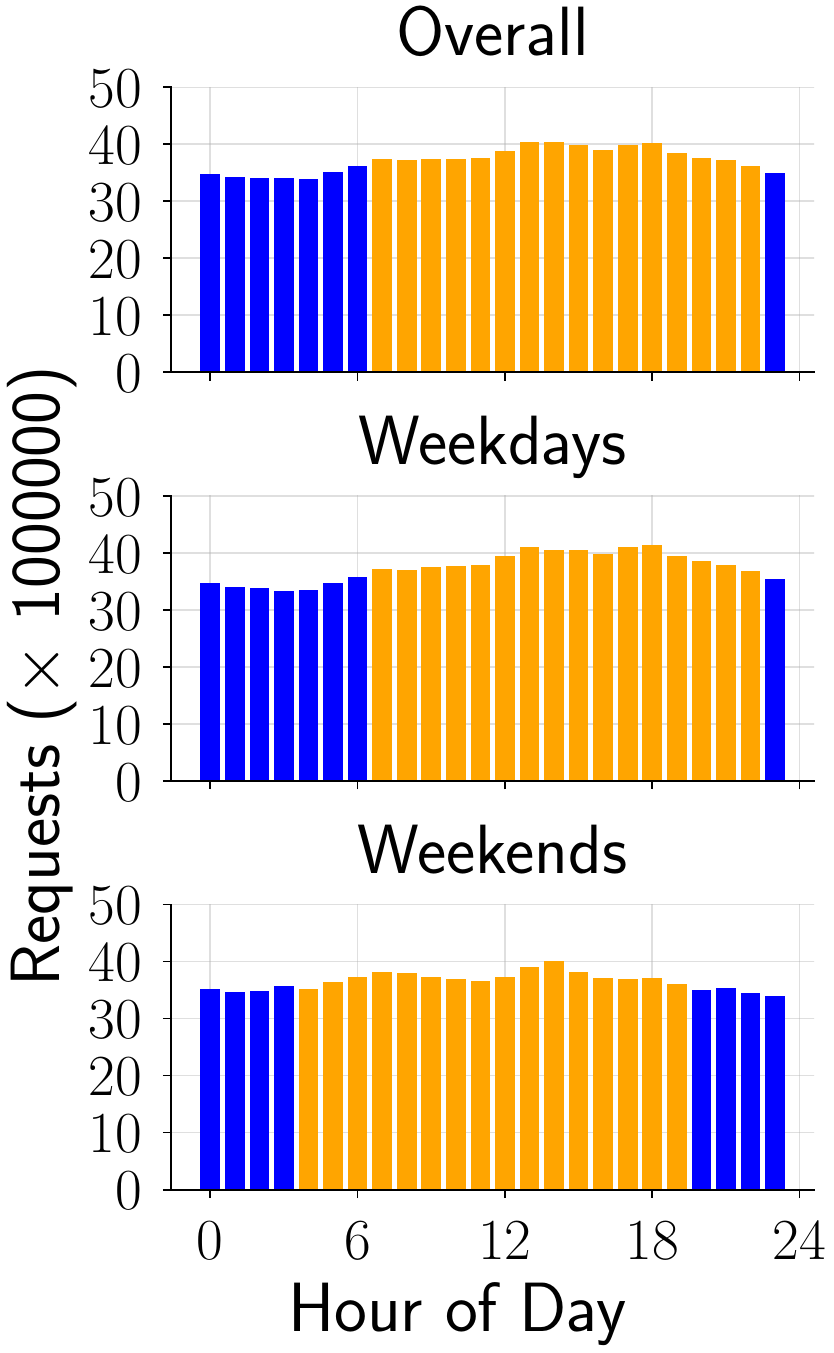}
        \label{fig:azure19_avg_requests}
    }
    \hfill
    \subfloat[\AzureSecond]{
        \includegraphics[width=0.29\linewidth]{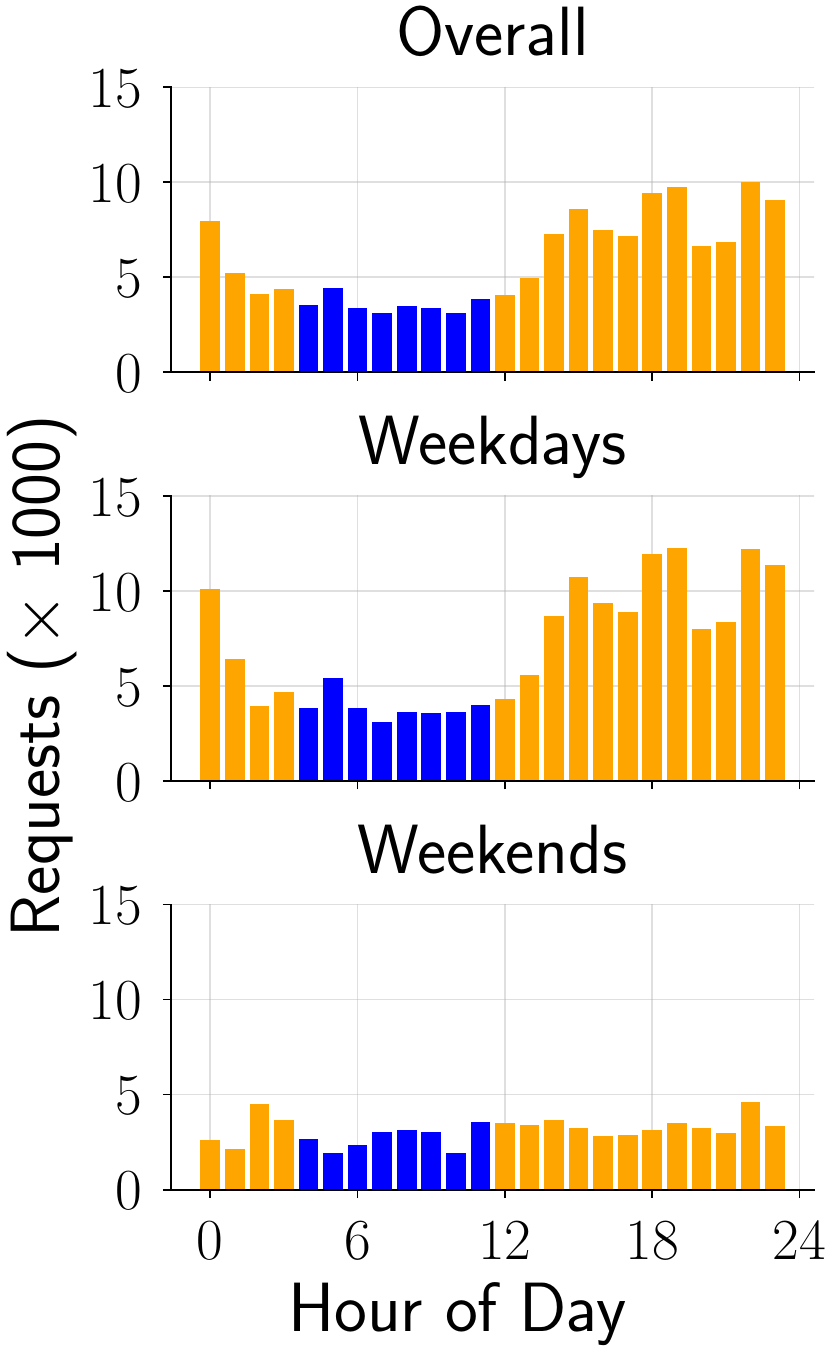}
        \label{fig:azure21_avg_requests}
    }
    \hfill
    \subfloat[\HuaweiFirst]{
        \includegraphics[width=0.29\linewidth]{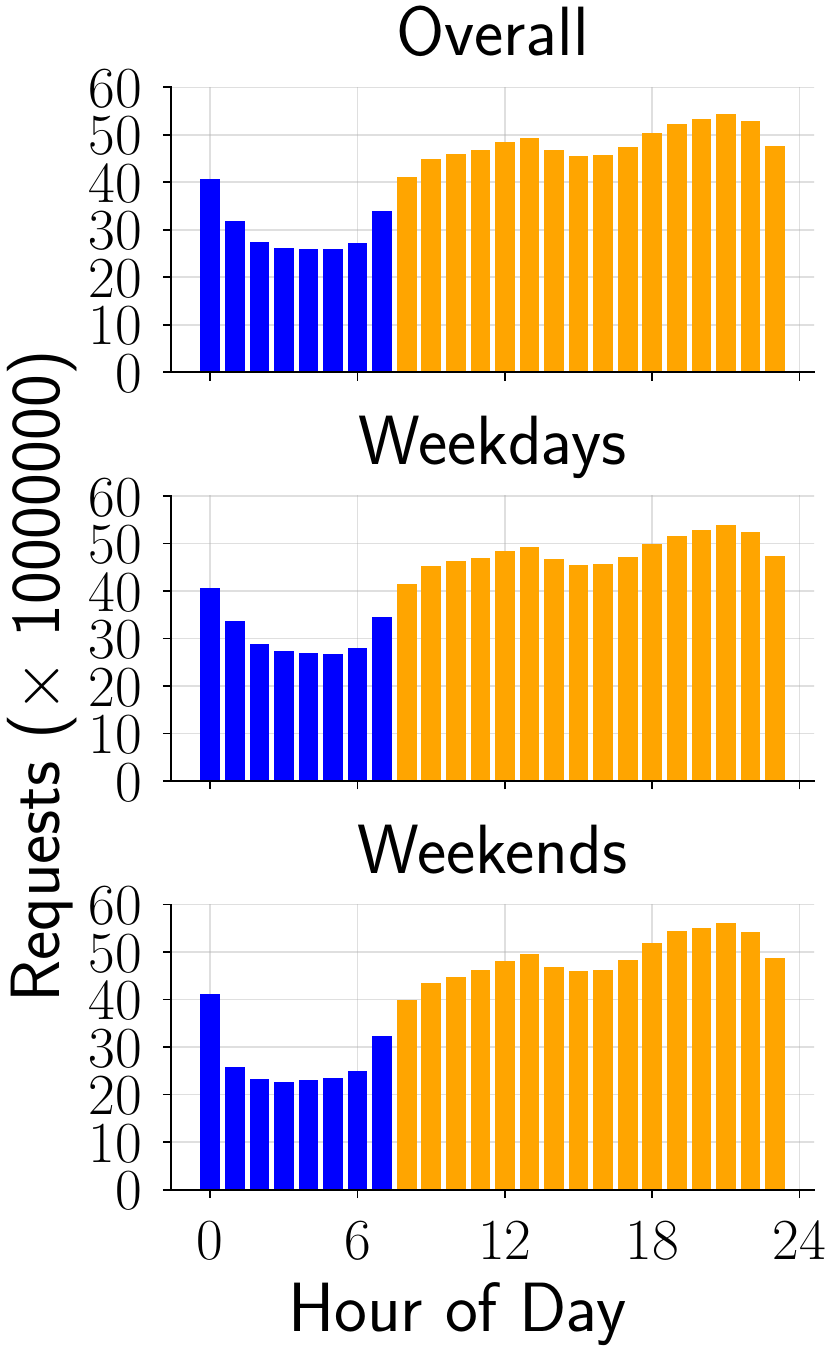}
        \label{fig:huawei23_avg_requests}
    }
    \\
    \subfloat[\HuaweiSecond]{
        \includegraphics[width=0.29\linewidth]{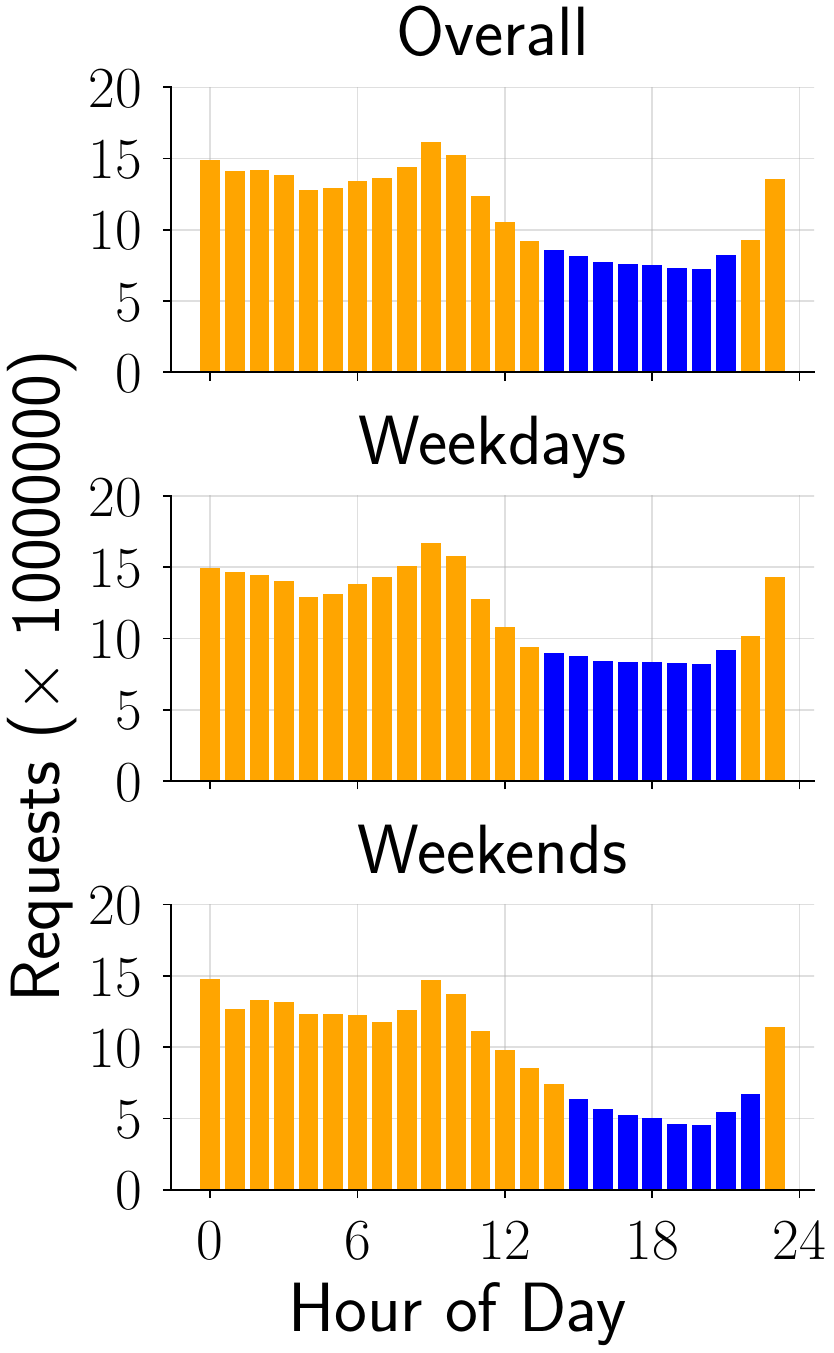}
        \label{fig:huawei25_avg_requests}
    }
    \hfill
    \subfloat[\Globus]{
        \includegraphics[width=0.29\linewidth]{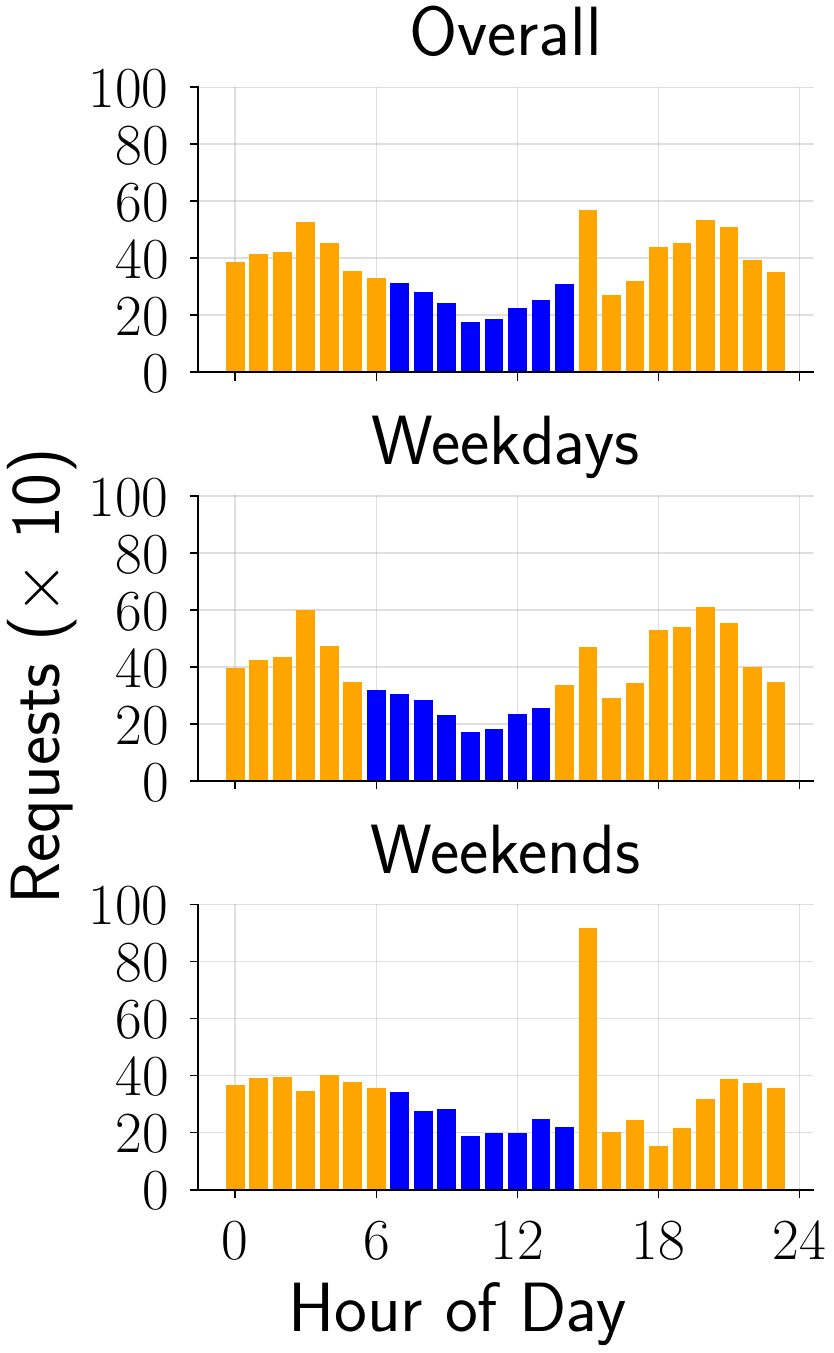}
        \label{fig:globus24_avg_requests}
    }
    \hfill
    \subfloat[\IBM]{
        \includegraphics[width=0.29\linewidth]{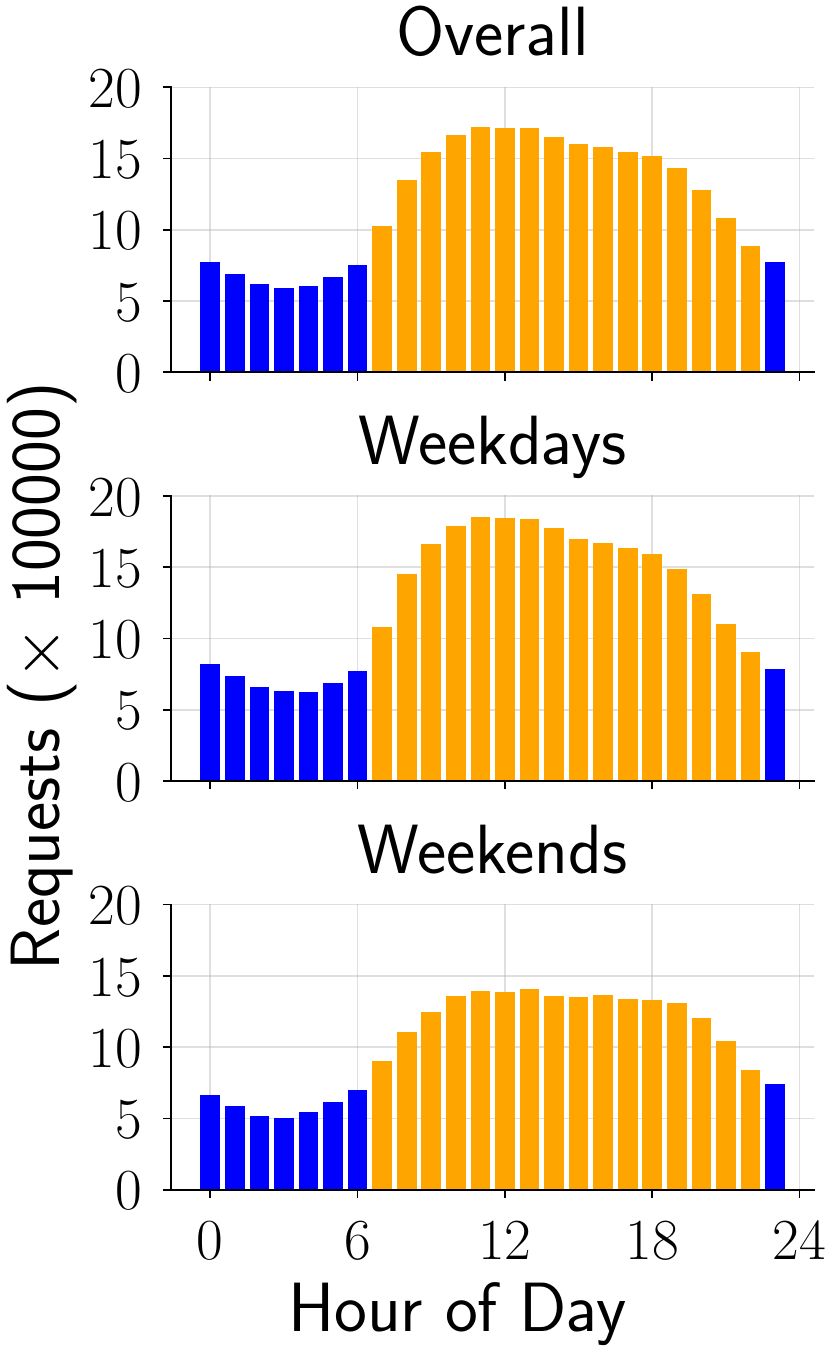}
        \label{fig:ibm26_avg_requests}
    }
    \caption{Average request arrival rates per hour show distinct daily usage patterns. Each column presents the hourly average for all days (top), weekdays (middle), and weekends (bottom). The blue bars highlight the continuous 8-hour window with the lowest overall activity, whereas orange bars represent the more active periods.}
    \label{fig:diurnal_patterns}
\end{figure}

\section{Methodology}
\label{sec:methodology}

Our methodology is centered around a comparative analysis of the publicly available FaaS workload traces. We aim to identify both commonalities and platform-specific differences to build a comprehensive understanding of serverless workloads.

\subsection{Datasets}
We selected all eight publicly available serverless workload traces from Azure~\cite{shahrad2020serverless, romero2021faa, zhang2021faster}, Alibaba~\cite{wang2021faasnet}, Huawei~\cite{joosen2023does, joosen2025serverless}, Globus~\cite{bauer2024globus}, and IBM~\cite{nasiri2026serverless}. For \HuaweiFirst, although the original release includes both public and private workloads, we exclusively analyzed the private trace because the public trace only provides a single metric (requests per minute), whereas the private trace offers higher-fidelity per-second data, such as latencies and resource metadata. In the following, we refer to the datasets using the name of the dataset and its publication year as shorthand, for example, \Alibaba for the dataset published by Alibaba in 2021.

\subsection{Metrics}
\cref{tab:metrics_analyzed} provides an overview of the metrics available in the analyzed datasets. The traces collectively capture a wide variety of metrics, including request arrival rates, execution and response times, cold start latencies, and resource metrics such as CPU and memory usage versus their configured limits. Additionally, they provide insight into instance dynamics (e.g., active instance counts and scaling bounds), and application metadata, such as runtimes, triggers, I/O payloads, and code-level characteristics.
The availability and granularity of these metrics reflect a clear evolution in serverless tracing. Earlier datasets, such as \AzureFirst, offered only few metrics, such as timestamps and execution times, which were often aggregated per minute. In contrast, recent traces provide significantly higher granularity. For example, \HuaweiSecond provides per-invocation records along with detailed latency breakdowns at the component-level of cold starts. Furthermore, modern traces expose configuration and structural details that were previously unavailable, such as instance concurrency settings in \IBM. This increasing number of metrics enables a deeper characterization of modern serverless platforms.

\subsection{Data Processing and Normalization}

To perform our computations, we used Dask~\cite{dask}, an out-of-core parallel computing framework that can process datasets larger than memory by operating on data in chunks. We dropped a small number of erroneous records, such as negative latencies resulting from clock synchronization issues. We also normalized the timestamp formats across all datasets to a relative time from the start of the trace, facilitating time-series analysis. We implemented a taxonomy for categorical metadata, mapping proprietary trigger names (e.g., Huawei’s ``APIG'' or ``OBS'') to general equivalents (``HTTP'' and ``Storage'') and generalizing runtime environments (e.g., mapping \texttt{python-3.9} to \texttt{python}). We provide statistics for the subsets of the eight traces that we used for our characterization in \cref{tab:processing_stats}. The \IBM dataset has a distinct structure in which the raw data consists of weekly records that each contain a list of invocation events (i.e., a list of lists). Therefore, we count each individual invocation event as one row in our statistics for this trace.

\section{Characterization}
\label{sec:characterization}

This section presents the findings of our cross-platform workload analysis. We structure our results by key characteristics.

\subsection{Request Patterns}
We analyze request arrival patterns in \AzureFirst, \AzureSecond, \HuaweiFirst, \HuaweiSecond, \Globus, and \IBM, because these traces provide the necessary timestamped invocation data or arrival rates to study burstiness and diurnal cycles.

\altparagraph{Bursty Workloads.}
A time-series view of request arrivals shows that the scale of short-term variability, or burstiness, differs between FaaS platforms. \Cref{fig:arrival_time_series} plots the number of requests per minute over several days, normalized by the overall mean of the platform. Platforms like \AzureFirst and \HuaweiFirst show relatively smooth aggregate behavior, with tightly bounded maximum loads that do not exceed $2\times$ their average. \HuaweiSecond and \IBM show moderate volatility, with peaks reaching roughly $3\times$ the mean. In contrast, \AzureSecond and \Globus experience extreme and sudden spikes in demand. \AzureSecond shows bursts that are up to $25\times$ its average, whereas the scientific computing workloads in \Globus reach peaks of up to $310\times$ their baseline, due to large-scale, sudden batch submissions. This disparity implies that autoscaling and resource provisioning mechanisms cannot rely on a universal assumption of burstiness. They should be tailored to the specific volatility profile of the hosted workload, ranging from conservative buffering to highly aggressive, instantaneous scale-out.

\begin{figure*}[!t]
    \centering
    \newcommand{\colwidth}{0.18\linewidth}

    \subfloat[\AzureSecond]{
        \begin{minipage}{\colwidth}
            \includegraphics[width=\linewidth]{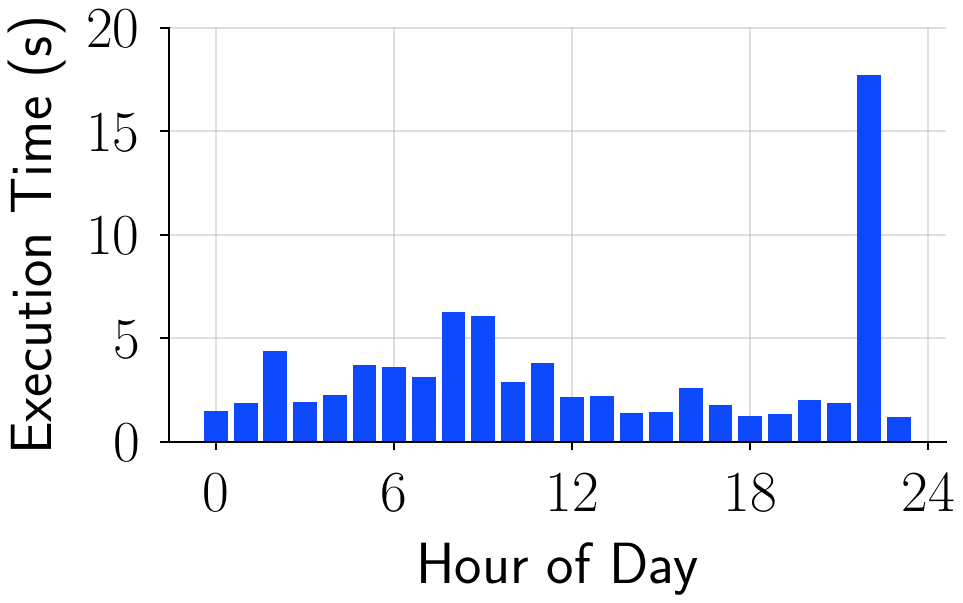}
            \label{fig:azure21_avg_latency}
        \end{minipage}
    }
    \hfill
    \subfloat[\Globus]{
        \begin{minipage}{\colwidth}
            \includegraphics[width=\linewidth]{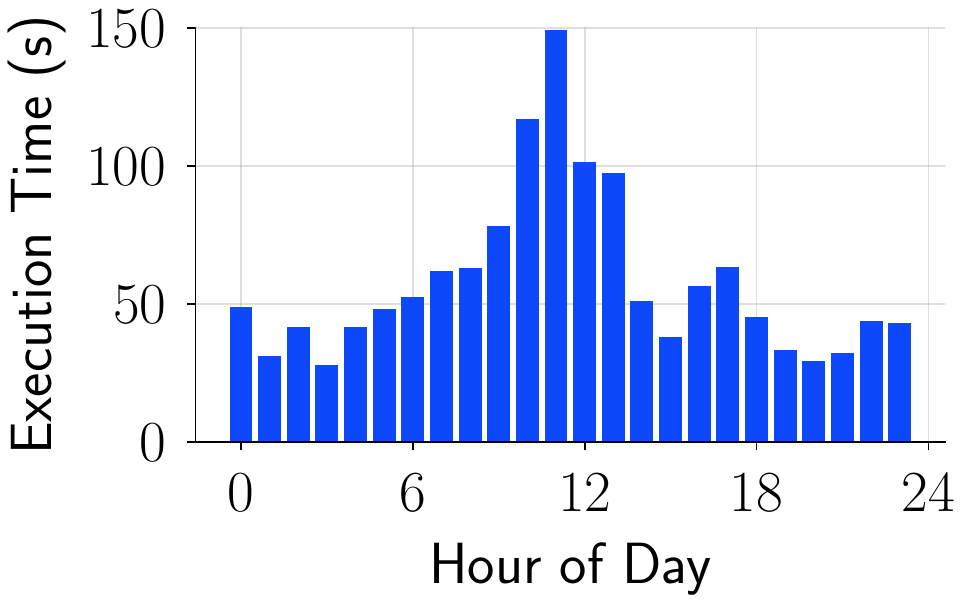}
            \label{fig:globus24_avg_latency}
        \end{minipage}
    }
    \hfill
    \subfloat[\HuaweiFirst]{
        \begin{minipage}{\colwidth}
            \includegraphics[width=\linewidth]{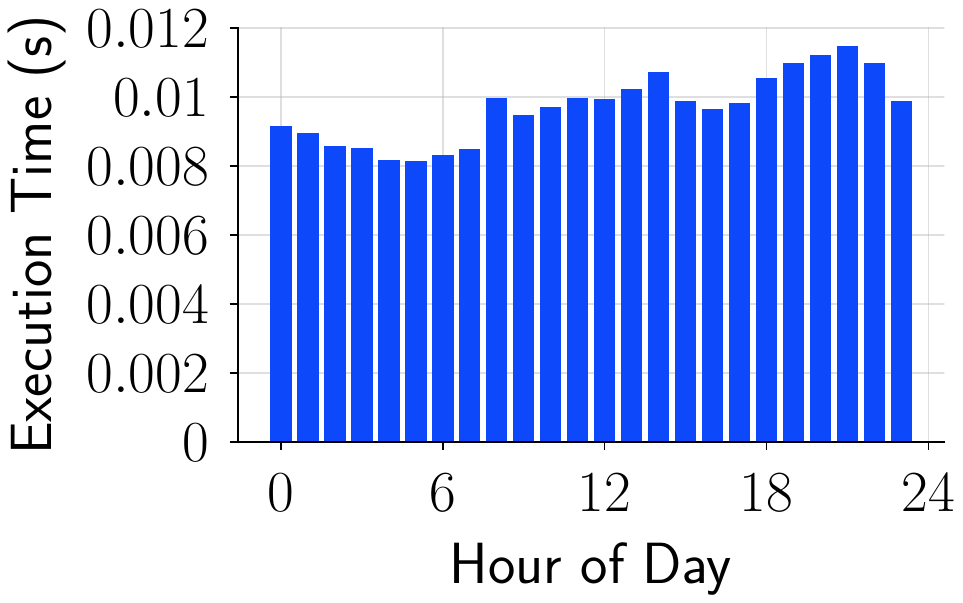}
            \label{fig:huawei23_avg_latency}
        \end{minipage}
    }
    \hfill
    \subfloat[\HuaweiSecond]{
        \begin{minipage}{\colwidth}
            \includegraphics[width=\linewidth]{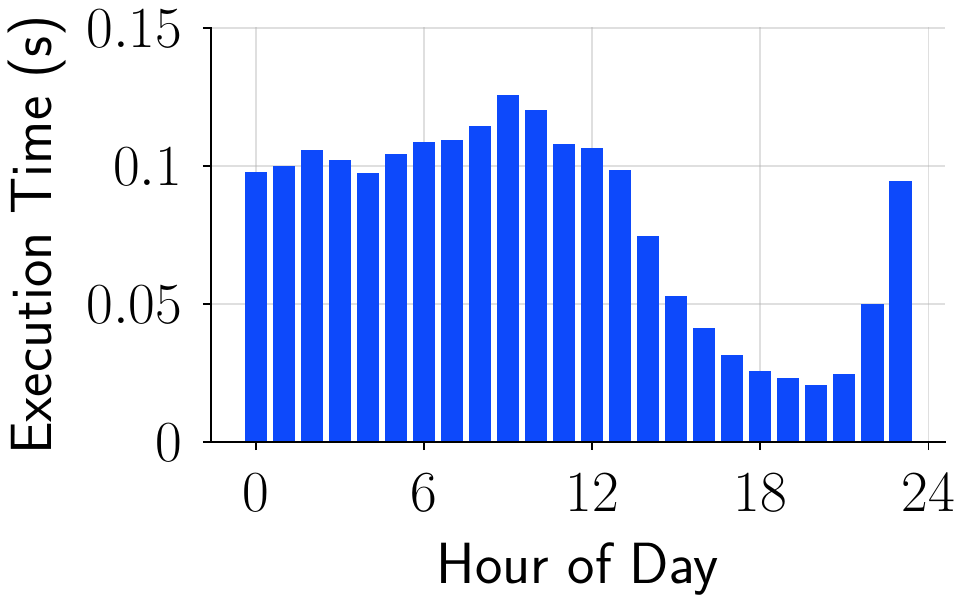}
            \label{fig:huawei25_avg_latency}
        \end{minipage}
    }
    \hfill
    \subfloat[\IBM]{
        \begin{minipage}{\colwidth}
            \includegraphics[width=\linewidth]{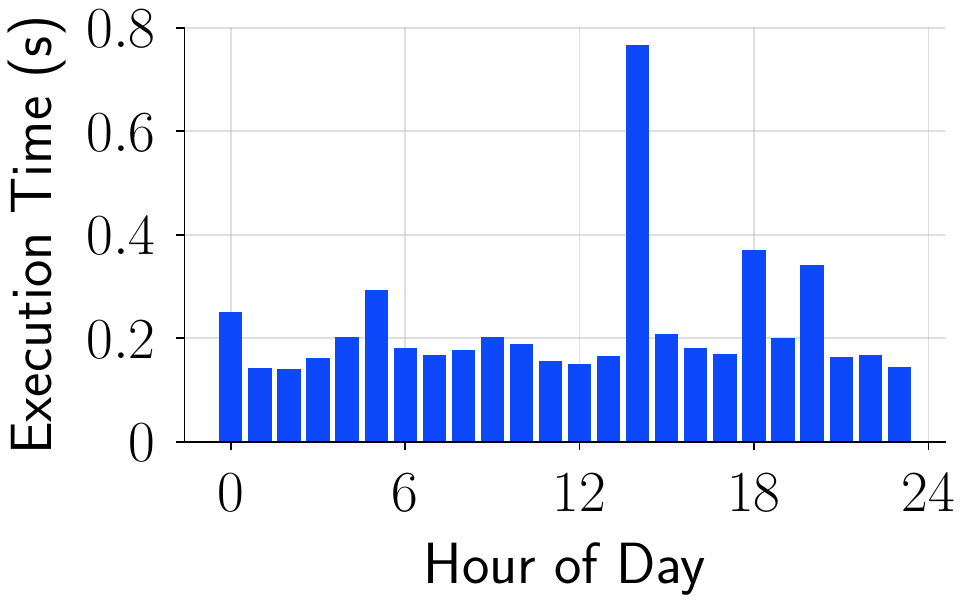}
            \label{fig:ibm26_avg_latency}
        \end{minipage}
    }
    \caption{Hourly variations in average execution time across five FaaS platforms over a 24-hour cycle. \AzureSecond and \Globus show significant fluctuations with latency spikes, whereas \HuaweiFirst, \HuaweiSecond, and \IBM maintain relatively stable execution times throughout the day.}
    \label{fig:combined_latency_load_patterns}
\end{figure*}

\altparagraph{Diurnal Cycles.}
When aggregated hourly, the workloads show distinct diurnal (24-hour) patterns with varying intensities (\cref{fig:diurnal_patterns}). Although commercial platforms generally peak during typical working hours, they show different off-peak behaviors. For instance, \IBM's volume decreases by 68.8\% from its daily peak to its overnight minimum, whereas \HuaweiSecond maintains a higher overnight baseline, with a smaller decrease of 55.1\%. Weekend traffic reductions are similarly platform-dependent. Across the full trace, \IBM averages a 58.0\% weekend volume reduction, but this decrease is highly concentrated, exceeding 65\% around midday compared to only 32\% at 5:00 AM. \HuaweiSecond experiences a much smaller weekend reduction of 17.6\%. In contrast, \AzureFirst shows a significantly flatter profile, with a 16.1\% diurnal reduction and a 2.3\% weekend reduction, suggesting that globally distributed or background workloads smooth out daily peaks. The \Globus trace exhibits an irregular pattern dominated by batch submission windows rather than human-driven demand patterns. Despite these variations, the least busy period consistently falls within a continuous eight-hour window (blue bars), offering a predictable daily timeframe for system maintenance or resource reclamation.

In contrast to the predictable diurnal cycles of request arrivals, average function execution times can be highly volatile throughout the day, as shown in \cref{fig:combined_latency_load_patterns}. Although platforms like \HuaweiFirst maintain a stable average execution latency across all hours, other traces show significant swings during a day. For example, \AzureSecond ranges from $\approx 1.1$~s during quiet periods to more than 17~s during peak spikes. End-to-end response times follow a similar pattern of diurnal volatility, but are further exacerbated by platform congestion. Commercial platforms keep this scheduling overhead minimal. However, systems like \Globus experience extreme queueing delays during peak submission windows. The average response time in \Globus reaches nearly 1\,400~s, which substantially exceeds its execution time fluctuations. Interestingly, comparing \cref{fig:globus24_avg_requests} and \cref{fig:globus24_avg_latency} shows that the execution time peak in \Globus (around 11:00) strongly negatively correlates ($-0.84$) with the request volume. Therefore, service bottlenecks in batch-oriented systems are not necessarily caused by a high volume of requests, but rather by resource-heavy tasks.

\begin{figure}[!t]
    \centering
    \subfloat[Skewed Function Popularity]{
        \includegraphics[width=0.45\columnwidth]{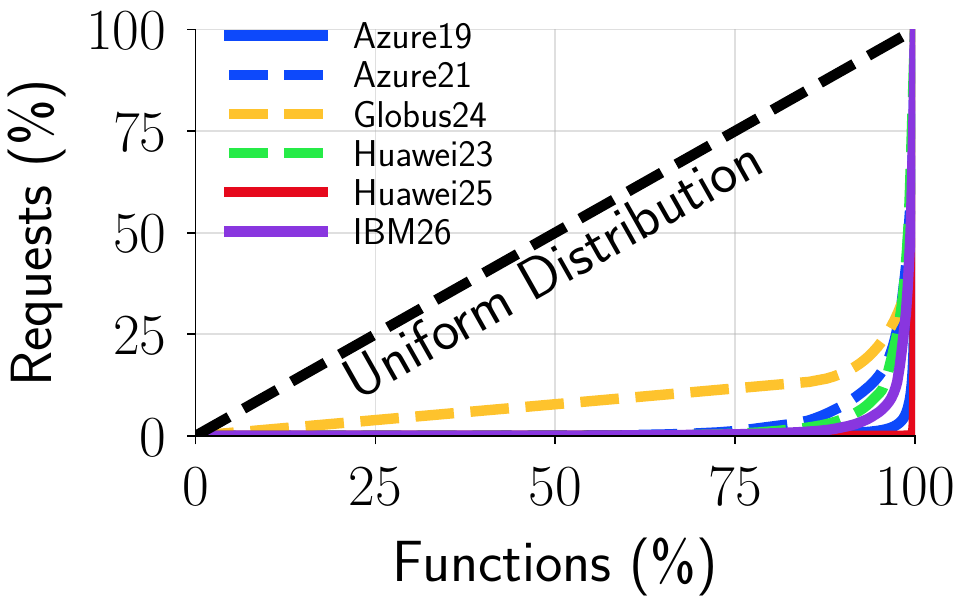}
        \label{fig:function_popularity_cdf}
    }
    \hfill
    \subfloat[Function Idle Times]{
        \includegraphics[width=0.45\columnwidth]{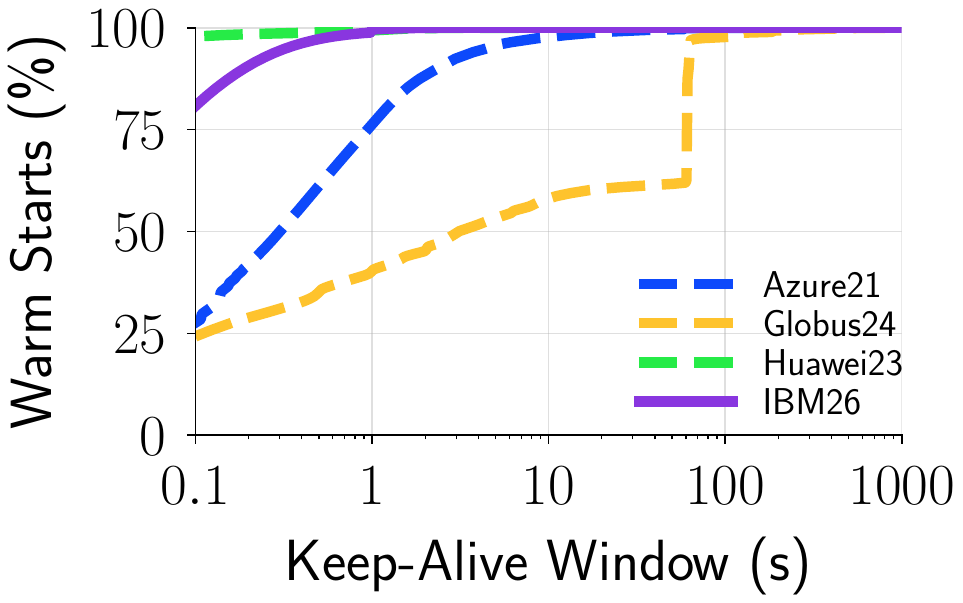}
        \label{fig:function_idle_times}
    }
    \caption{Function popularity and idle times. (a) A small fraction of functions consistently accounts for the majority of invocations across all platforms. (b) The probability of a warm start as a function of the keep-alive window duration shows diverse reuse patterns, from rapid reuse (\AzureSecond) to periodic jobs (\Globus).}
    \label{fig:popularity_and_idle_times}
\end{figure}

\altparagraph{Skewed Function Popularity.}
The distribution of requests across individual functions is skewed across all platforms. As shown by the Lorenz curve in \cref{fig:function_popularity_cdf}, a small fraction of functions consistently accounts for the majority of invocations.  For commercial platforms, the magnitude varies slightly: In \IBM, \AzureFirst, and \HuaweiSecond, the skew is extreme; the top 1--5\% of functions process more than 90\% of requests. Specifically, in \HuaweiSecond, the top 1\% of functions account for over 99\% of all traffic. In \AzureSecond and \HuaweiFirst, the distribution is slightly broader but still highly concentrated, with the top 10\% of functions accounting for over 90\% of requests. The \Globus trace shows a slightly less extreme skew, rising earlier than the others. This aligns with its scientific workload, where a wider array of specialized batch tasks are executed. However, it still starkly deviates from a uniform distribution. The ``hot'' functions are likely to remain warm due to frequent reuse, naturally avoiding cold starts, whereas the long tail of rarely invoked functions is more likely to incur cold start penalties.

\altparagraph{Function Idle Times.}
To assess the potential effectiveness of keep-alive strategies, we examine the inter-arrival time (IAT) distributions, which represent function idle times between consecutive requests. \Cref{fig:function_idle_times} plots the theoretical probability of warm-start versus the length of the keep-alive window (equivalent to the IAT CDF). The platforms show distinctly different temporal behaviors. \AzureSecond demonstrates high temporal locality, smoothly reaching a warm-start rate of 76.4\% with a 1~s window. \HuaweiFirst exhibits an almost continuous invocation, achieving near 98\% warm starts even with a 0.1~s window. Because the \HuaweiFirst dataset aggregates multiple request counts into 1~s buckets, we modeled concurrent invocations within the same second as a Poisson process using exponential inter-arrival times to approximate these sub-second dynamics. This trace captures 200 highly popular internal functions, rather than a general user workload, which explains this extreme frequency. In contrast, \Globus shows a pronounced step-function pattern with a large vertical jump at 60~s from 62.5\% to 87.4\%. This strongly indicates periodic workloads driven by minute-level cron jobs or batch polling intervals. Like \HuaweiFirst, the \IBM dataset truncates arrival timestamps to a 1~s integer resolution. After applying the same Poisson approximation to recover intra-second arrivals, \IBM reveals a rapid progression, achieving an 80\% warm start rate for a keep-alive window of just 100~ms, and crossing 90\% at $\approx 188$~ms. By 1~s, the warm start rate exceeds 99.5\%. These contrasting behaviors show that a single static keep-alive window mismatches with the diverse temporal profiles of serverless workloads.

\begin{insight}
The combination of high popularity skew, bursty arrivals, and diverse idle time profiles indicates that uniform resource management policies are inefficient. FaaS platforms should adopt asymmetric, per-function strategies: High-volume functions can adopt extended keep-alive windows and dedicated capacity to handle sudden spikes, whereas the long-tail of sporadic or periodic functions can be scaled to zero coupled with predictive pre-warming.
\end{insight}

\begin{figure}[!t]
\centering
    \includegraphics[width=0.6\columnwidth]{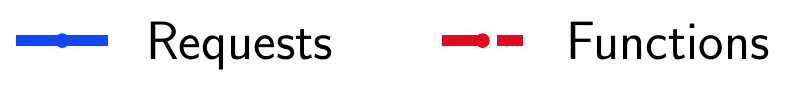}
    \vspace{-0.2cm}

    \subfloat[\HuaweiSecond]{
        \includegraphics[width=0.29\columnwidth]{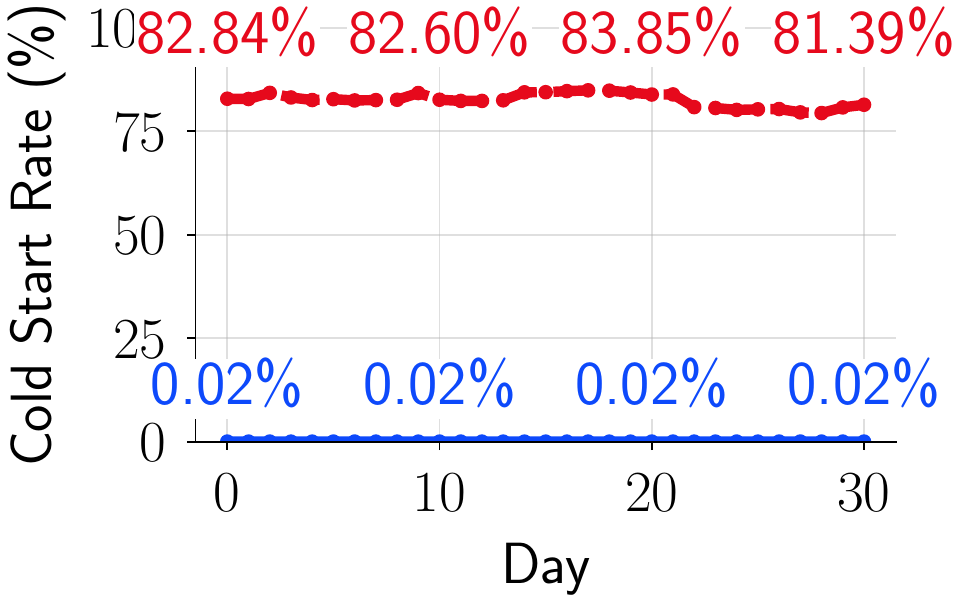}
        \label{fig:cold_start_rate_huawei25}
    }
    \hfill
    \subfloat[\Globus]{
        \includegraphics[width=0.29\columnwidth]{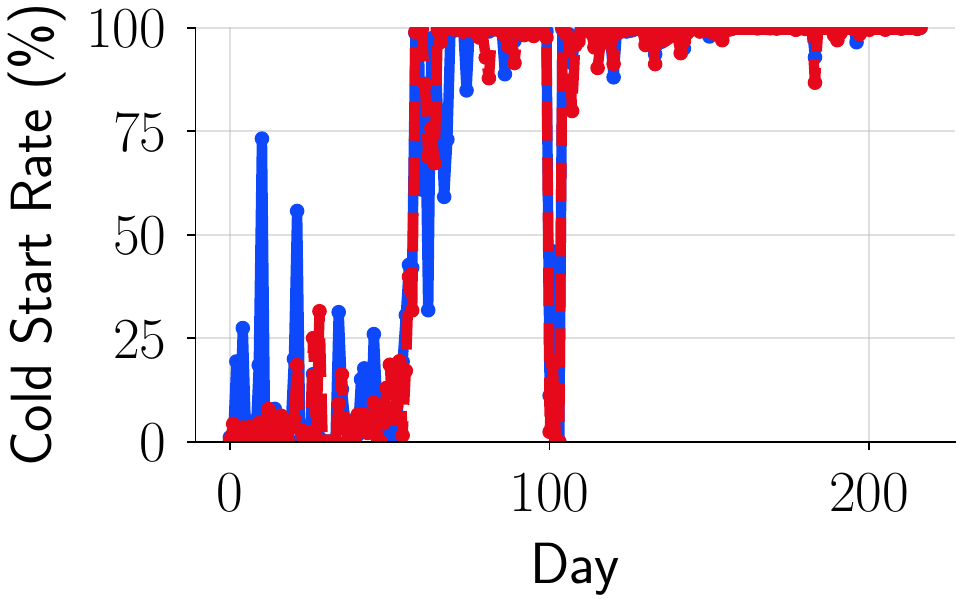}
        \label{fig:cold_start_rate_globus24}
    }
    \hfill
    \subfloat[\IBM]{
        \includegraphics[width=0.29\columnwidth]{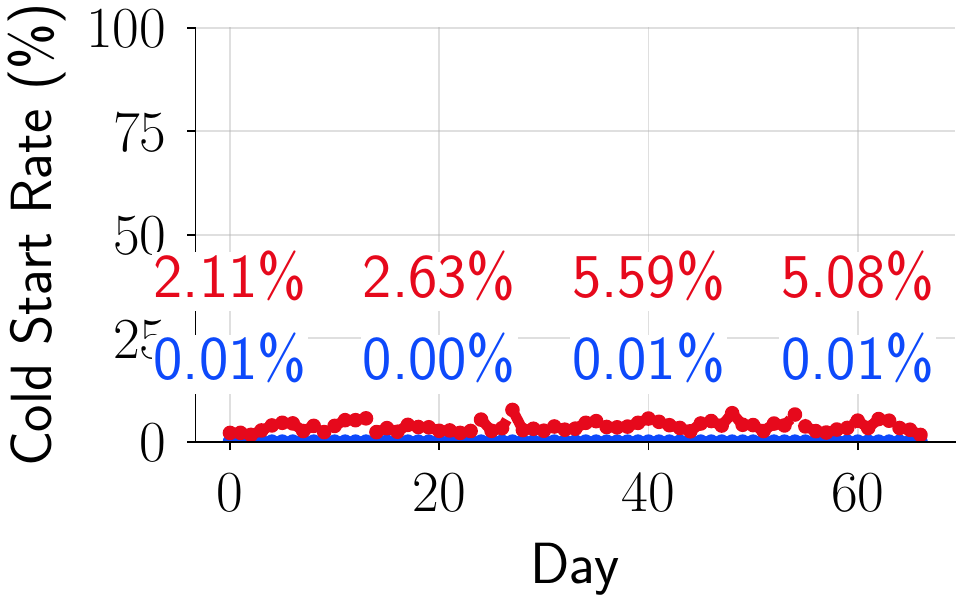}
        \label{fig:cold_start_rate_ibm26}
    }
    \caption{Cold start rate over time. A clear contrast exists between the per-request and per-function cold start rates in \HuaweiSecond due to extreme popularity skew. \IBM maintains near-zero cold starts across both metrics due to minimum instance scale and high concurrency limits, whereas \Globus shows high volatility typical of batch-oriented provisioning of scientific workloads.}
\label{fig:cold_start_rate}
\end{figure}

\begin{table}[!t]
    \centering
    \caption{Cold start latency statistics (s). Commercial platforms show sub-second medians with $>30$~s tails, whereas \Globus shows extreme tail latencies driven by batch scheduling queues.}
    \label{tab:cold_start_latency_stats}
    \resizebox{\columnwidth}{!}{%
    \begin{tabular}{l r r r r r}
        \toprule
        \textbf{Dataset} & \textbf{Median} & \textbf{Mean} & \textbf{P95} & \textbf{P99} & \textbf{Max} \\
        \midrule
        \Alibaba      & 0.86 & 3.85 & 19.2 & 36 & 193 \\
        \Globus       & 0.01 & 298.89 & 371.2 & 8\,645.8 & 130\,818 \\
        \HuaweiSecond & 1.29 & 3.25 & 11 & 36.9 & 287 \\
        \bottomrule
    \end{tabular}%
    }
\end{table}

\begin{figure}[!t]
    \centering
    \includegraphics[width=0.9\columnwidth]{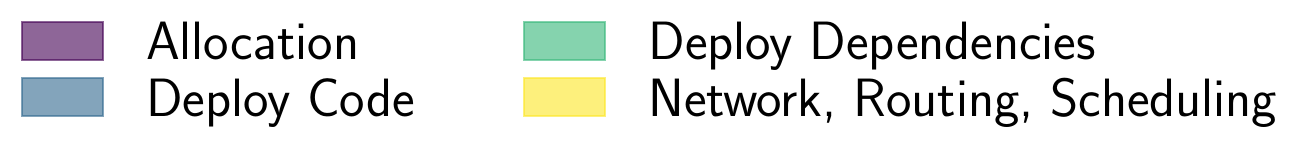}
    \vspace{-0.2cm}

    \subfloat[Cold Start Latency CDF]{
        \includegraphics[width=0.45\columnwidth]{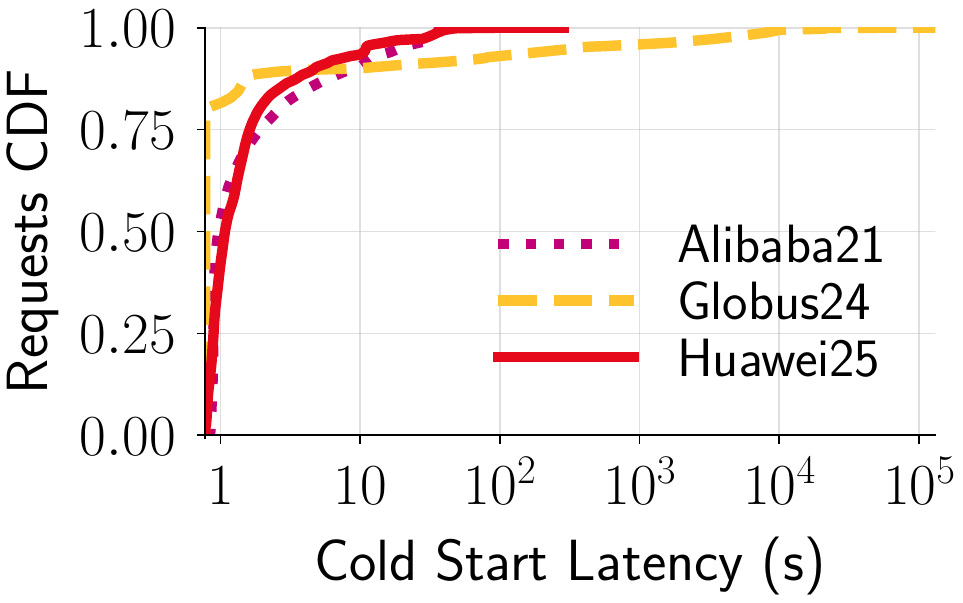}
        \label{fig:cold_start_latency_cdf}
    }
    \hfill
    \subfloat[Latency Breakdown (\HuaweiSecond)]{
        \includegraphics[width=0.45\columnwidth]{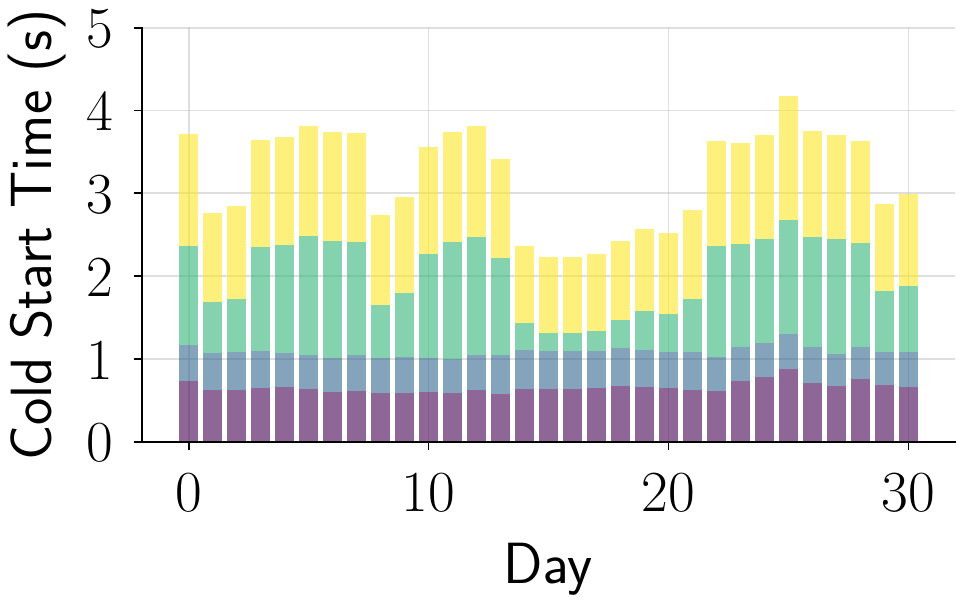}
        \label{fig:cold_starts_breakdown}
    }
    \caption{Cold start latencies and components. (a) Cumulative distribution of cold start latencies shows commercial platforms with long tails up to 30~s, whereas scientific workloads (\Globus) show queuing delays. (b) Breakdown of latency components shows that network, routing, scheduling, and dependency deployment dominate the total initialization time. The chart also shows significant day-to-day variance in total latency.}
    \label{fig:cold_starts_combined}
\end{figure}

\subsection{Cold Starts}
\label{sec:cold_starts}
We examine cold starts in \Alibaba, \HuaweiSecond, \Globus, and \IBM. These datasets are selected because they explicitly log cold start events, detailed latency breakdowns, or provide pod allocation timestamps that allow inferring initialization penalties.

\altparagraph{Cold Start Rate.}
We examine the frequency of cold starts on FaaS platforms in \cref{fig:cold_start_rate}. We distinguish between the cold start rate calculated across the total request volume versus the unweighted average rate across individual functions. For commercial platforms, this distinction shows a substantial disparity driven by workload skew. In \HuaweiSecond, the cold start rate per-request is negligible at $\approx 0.02$\%, which means that the majority of user traffic is served by warm instances. However, the cold start rate per-function consistently exceeds 80\%. This indicates that although heavily invoked ``hot'' functions remain persistently warm, the long tail of rarely invoked functions almost always suffers a cold start upon execution. As the \IBM dataset does not explicitly include a cold start attribute, we infer these events by identifying the initialization of new pod instances that exceed the application's configured minimum scale. Under this assumption, \IBM shows a particularly low cold start rate for both requests ($\approx 0.002$--$0.01\%$) and functions ($\approx 2$--$6\%$). This sustained warmth across the function population is attributable to the platform's ``Serverless 2.0'' configuration patterns: Users actively use minimum scale settings ($>0$) and high concurrency limits, with a default of $100$ requests per instance on \IBM, which eliminate cold start latency for most workloads. We discuss such instance metrics in more detail in \cref{sec:instances}. In contrast, \Globus shows extreme volatility. During peak periods, the cold start rate for requests and functions increases to nearly 100\%. This behavior confirms that its execution model is primarily on-demand, where resources are provisioned specifically for incoming task batches rather than being kept warm continuously.

\altparagraph{Cold Start Latency.}
Cold start latency remains a primary performance bottleneck, although its magnitude varies by platform type. \Cref{fig:cold_start_latency_cdf} compares the cumulative distribution function (CDF) of cold start latencies. For commercial public clouds, such as \Alibaba and \HuaweiSecond, median cold start times are comparable at $\approx 0.86$~s and 1.29~s, respectively (\cref{tab:cold_start_latency_stats}). However, these distributions are heavily skewed. The 99th percentile (P99) reaches $\approx 36$~s for both platforms. This implies that although the median user experience is acceptable, a significant fraction of requests suffers delays that are orders of magnitude longer than typical execution times. In contrast, the scientific workloads of \Globus show a distinct, bimodal profile. Although the median cold start latency is negligible ($\approx 0.01$~s), likely reflecting rapid task dispatch to already-active endpoints, the distribution has a long tail. The mean reaches 299~s ($\approx 5$~minutes) and the P99 exceeds 2.4~hours. This reflects the batch-scheduling nature of scientific computing, where a ``cold start'' often implies waiting for the underlying batch scheduler to allocate HPC nodes to the federated endpoint.

\altparagraph{Breakdown of Cold Start Latency.}
To understand the sources of these delays, we analyze the daily breakdown of the cold start components in the \HuaweiSecond dataset (\cref{fig:cold_starts_breakdown}). Our statistical analysis shows that network, routing, and scheduling overhead is the dominant factor, with a mean of $\approx 1.17$~s. Dependency deployment closely follows at 0.95~s, and pod allocation averages 0.66~s. The code deployment is the fastest component, averaging only 0.43~s. This suggests that platform orchestration overhead and the management of heavy runtime environments contribute more to latency than the network transfer of the function logic itself. Furthermore, heavier resource requests (larger CPU/memory limits) increase these delays, because the scheduler requires more time to find suitable nodes for bin-packing. We also observe a cyclic pattern over the 31-day period. The total cold start times fluctuate noticeably between 2.2~s and 4.1~s depending on the day. This variance indicates that initialization performance is significantly affected by overall cluster congestion and diurnal load patterns.

\altparagraph{Impact of Runtimes.}
The choice of runtime environment affects initialization overhead. As visualized in \cref{fig:cold_start_by_runtime}, custom containers consistently show higher median latencies ($\approx 30$~s in \Alibaba and $\approx 10$~s in \HuaweiSecond) along with a high variance. This confirms the performance penalty associated with unmanaged environments that cannot leverage the platform's standard pre-warmed pools. In contrast, lightweight runtimes like Node.js, Python, and Go maintain low median latencies ($\approx 1$~s) across both datasets. In particular, Go, being a natively compiled language, benefits from a small runtime footprint and the absence of language-level VM startup overhead, leading to highly predictable performance with minimal variance and almost no extreme outliers. This is particularly evident in the \Alibaba dataset, where Go largely mitigates the ``long-tail'' problem. However, this advantage is environment-dependent: In \HuaweiSecond, Go exhibits tail latencies similar to .NET (C\#) and PHP, although still lower than most other runtimes. We also observe a significant difference in Java performance. In the \Alibaba dataset, Java functions show high cold start penalties (median $\approx 15$~s), reflecting the JVM startup overhead. In contrast, the \HuaweiSecond platform achieves sub-second median cold starts for Java, putting it on par with lightweight interpreted runtimes like Python and Node.js. This behavior occurs because the platform maintains reserved resource pools of pre-warmed environments for default runtimes like Java.

\begin{figure}[!t]
    \centering
    \subfloat[\Alibaba]{
    \includegraphics[width=0.8\columnwidth]{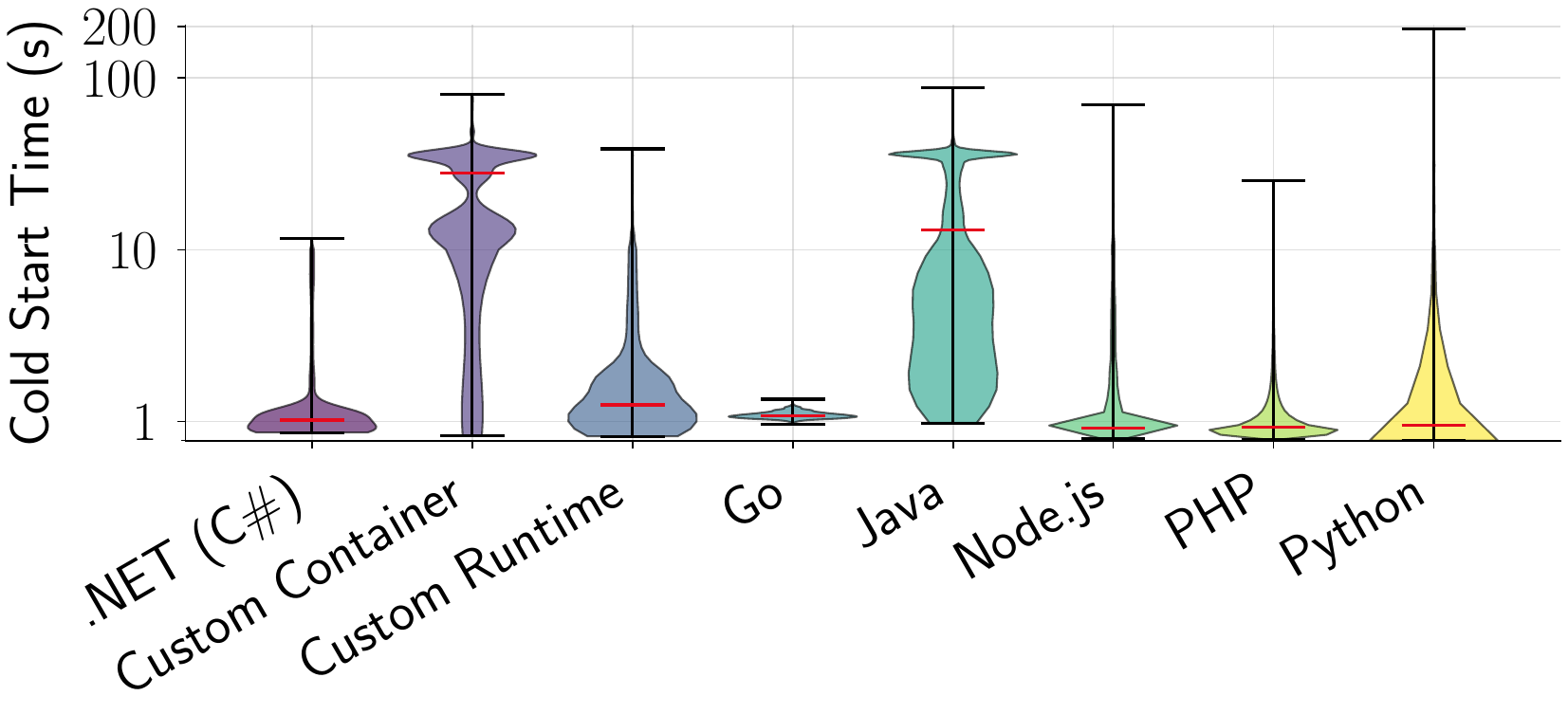}
    \label{fig:cold_start_by_runtime_alibaba21}
    }
    \hfill
    \subfloat[\HuaweiSecond]{
    \includegraphics[width=0.8\columnwidth]{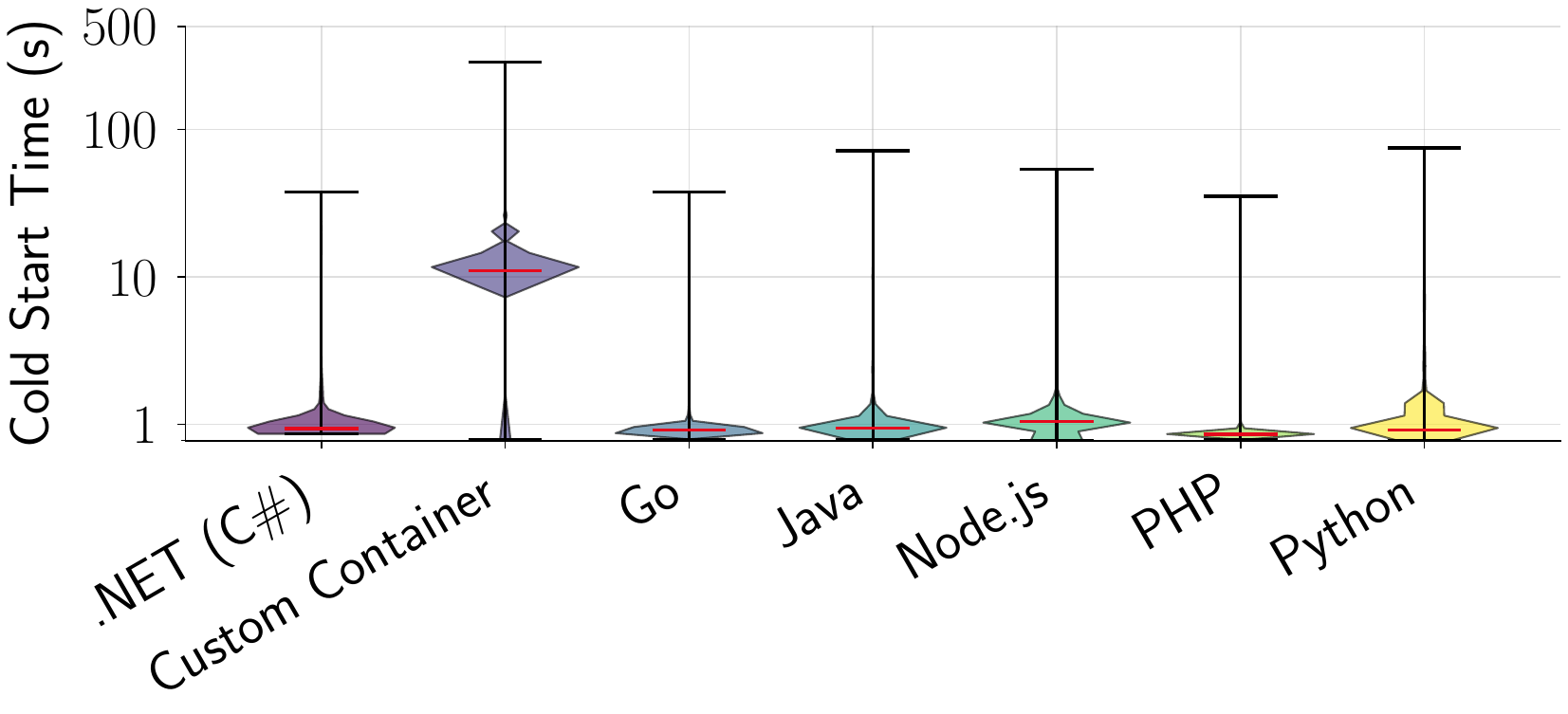}
    \label{fig:cold_start_by_runtime_huawei25}
}
\caption{Cold start latencies by runtime. Custom containers consistently incur the highest initialization penalties. Although Java showed high latency in \Alibaba, it shows sub-second performance in the newer \HuaweiSecond dataset, matching Python and Node.js.}
\label{fig:cold_start_by_runtime}
\end{figure}

\begin{insight}
The dominance of scheduling and dependency deployment in cold start latency indicates that orchestration overhead and environment setup are critical bottlenecks. Optimizations such as dependency-aware scheduling or predictive pre-fetching can yield large performance gains. Additionally, the disparity between per-request and per-function cold start rates confirms that initialization latency is primarily a problem for the ``long tail'' of unpopular functions. Finally, the near-elimination of cold starts in the \IBM trace suggests that exposing control, such as minimum instance scaling and concurrency limits, allows users to effectively bypass these penalties for latency-sensitive applications.
\end{insight}

\subsection{Latency}
To characterize execution and response times, we focus on \AzureFirst, \AzureSecond, \Globus, \HuaweiFirst, \HuaweiSecond, and \IBM, which record latency metrics.

\begin{figure*}[!t]
    \centering
    \begin{minipage}{0.48\linewidth}
        \centering
        \subfloat[CDF]{
            \includegraphics[width=\columnwidth]{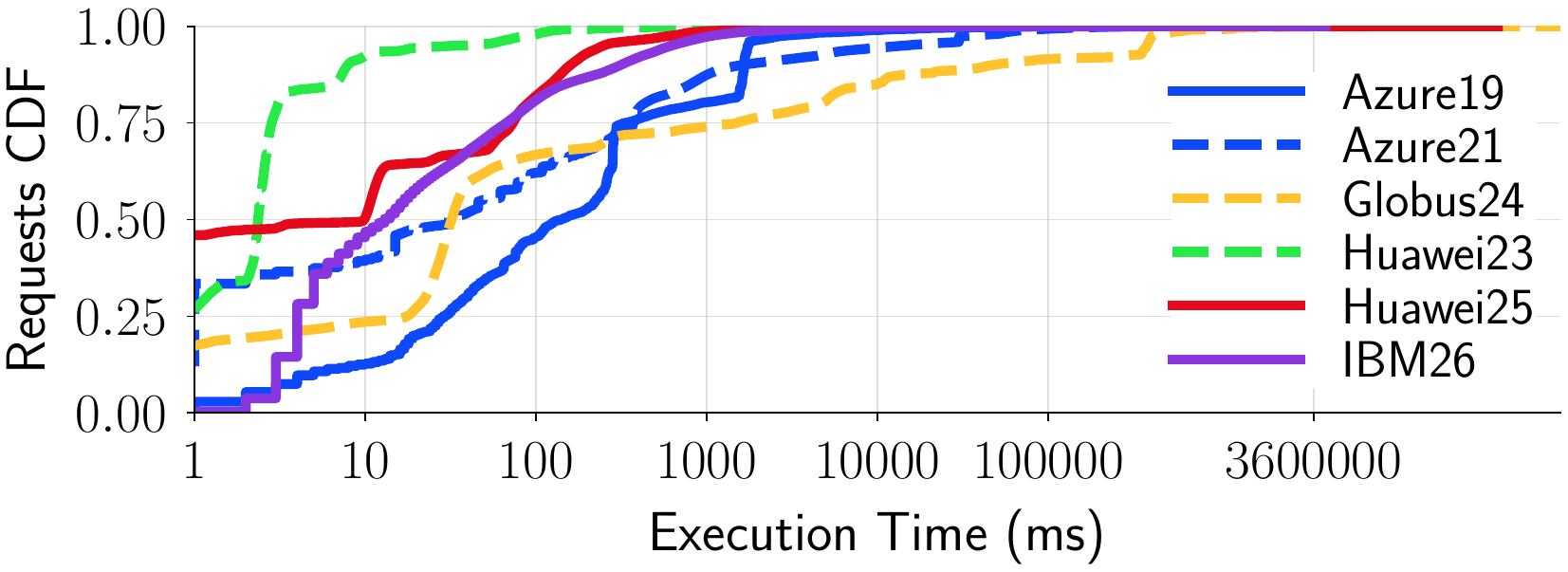}
            \label{fig:execution_time_cdf}
        }
        \hfill
        \subfloat[Distribution]{
            \includegraphics[width=\columnwidth]{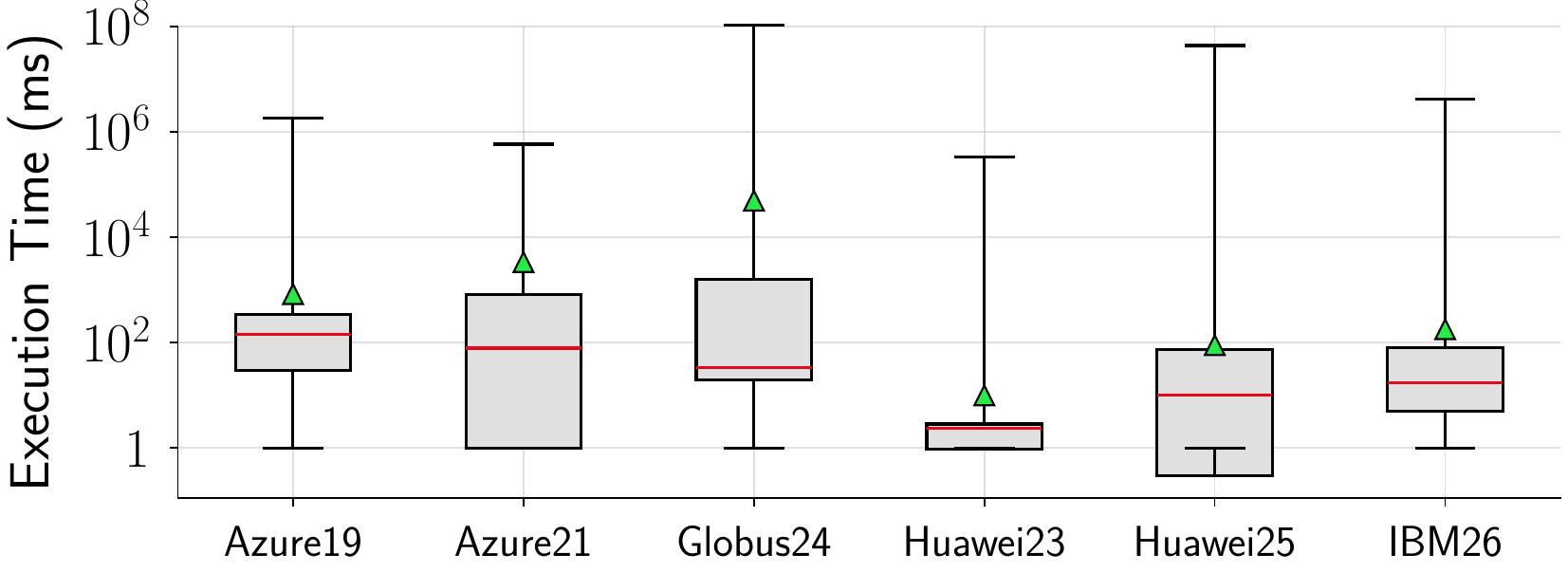}
            \label{fig:execution_time_boxplot}
        }
        \caption{Function execution times. Modern workloads (\AzureSecond, \Globus, \HuaweiFirst, \HuaweiSecond, \IBM) show significantly lower median latencies ($<35$~ms) compared to the older \AzureFirst trace, although all show heavy tails.}
        \label{fig:execution_time}
    \end{minipage}
    \hfill
    \begin{minipage}{0.48\linewidth}
        \centering
        \addtocounter{figure}{1}
        \subfloat[CDF]{
            \includegraphics[width=\columnwidth]{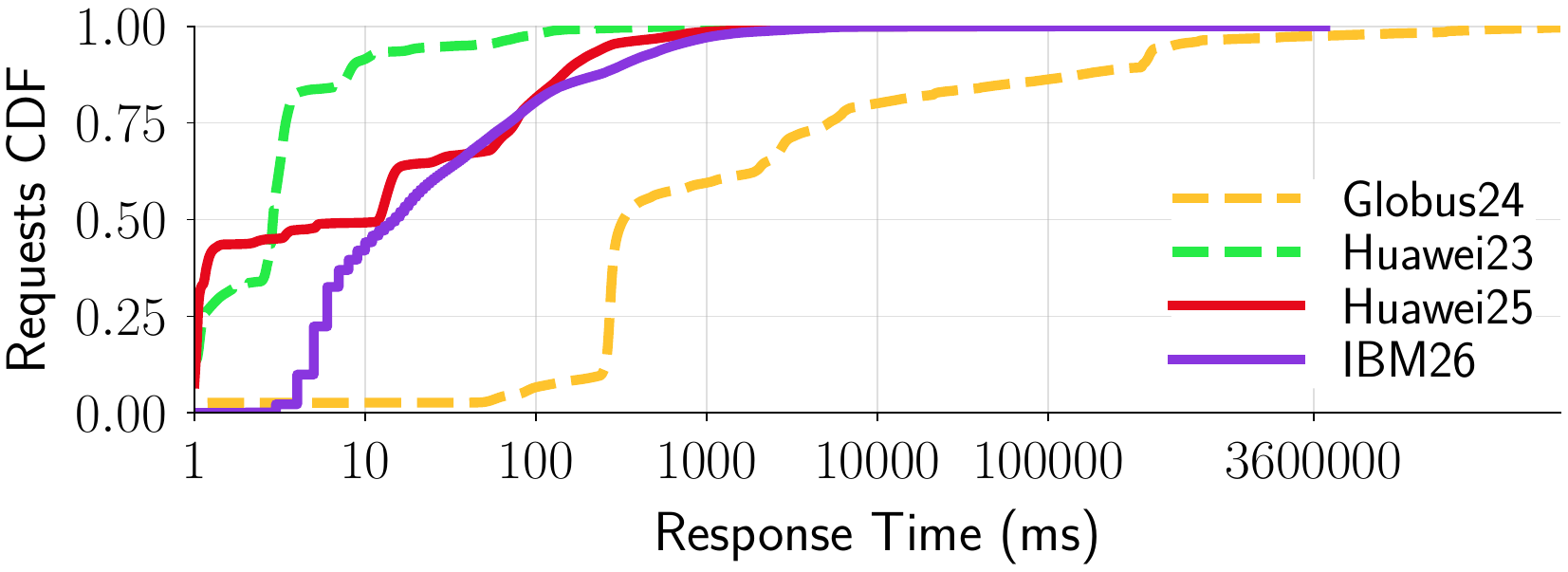}
            \label{fig:response_time_cdf}
        }
        \hfill
        \subfloat[Distribution]{
            \includegraphics[width=\columnwidth]{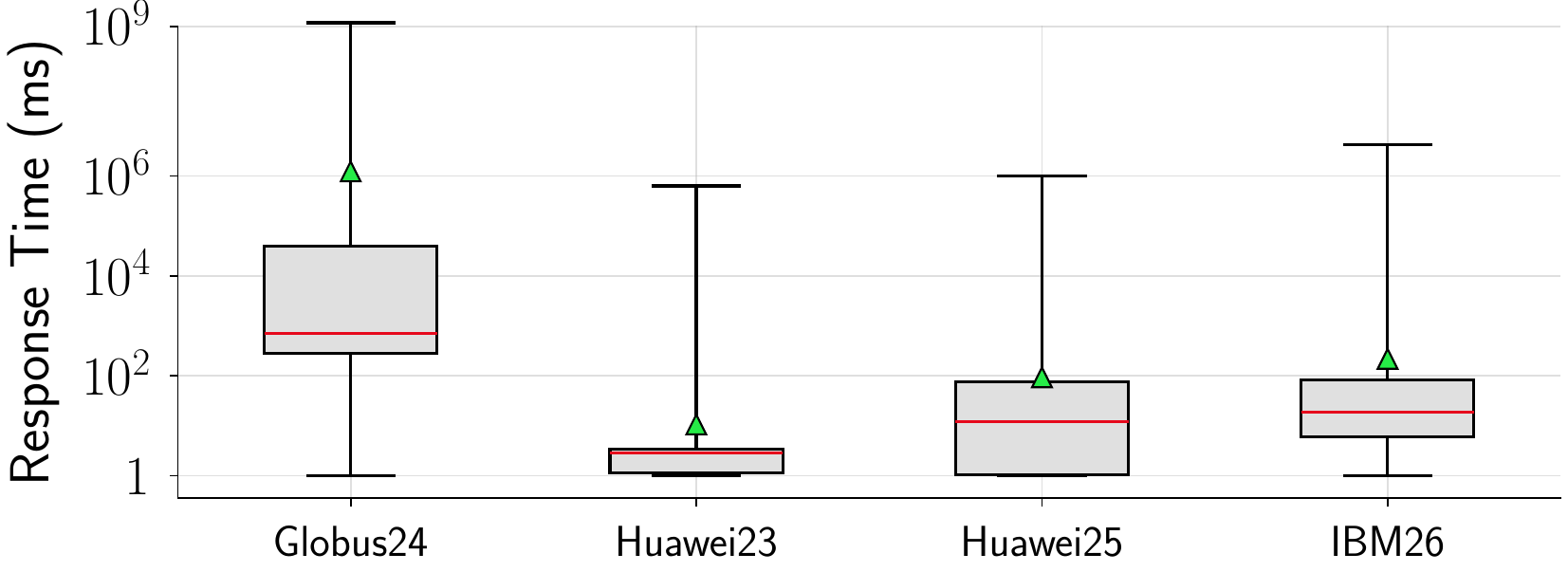}
            \label{fig:response_time_boxplot}
        }
        \caption{End-to-end response times. Commercial platforms (\HuaweiFirst, \HuaweiSecond, \IBM) maintain low median response times ($<15$~ms), whereas \Globus shows higher mean latency and a heavy tail.}
        \label{fig:response_time}
        \addtocounter{figure}{-1}
    \end{minipage}
\end{figure*}

\begin{figure}[!b]
    \centering
    \addtocounter{figure}{-1}
    \includegraphics[width=\columnwidth]{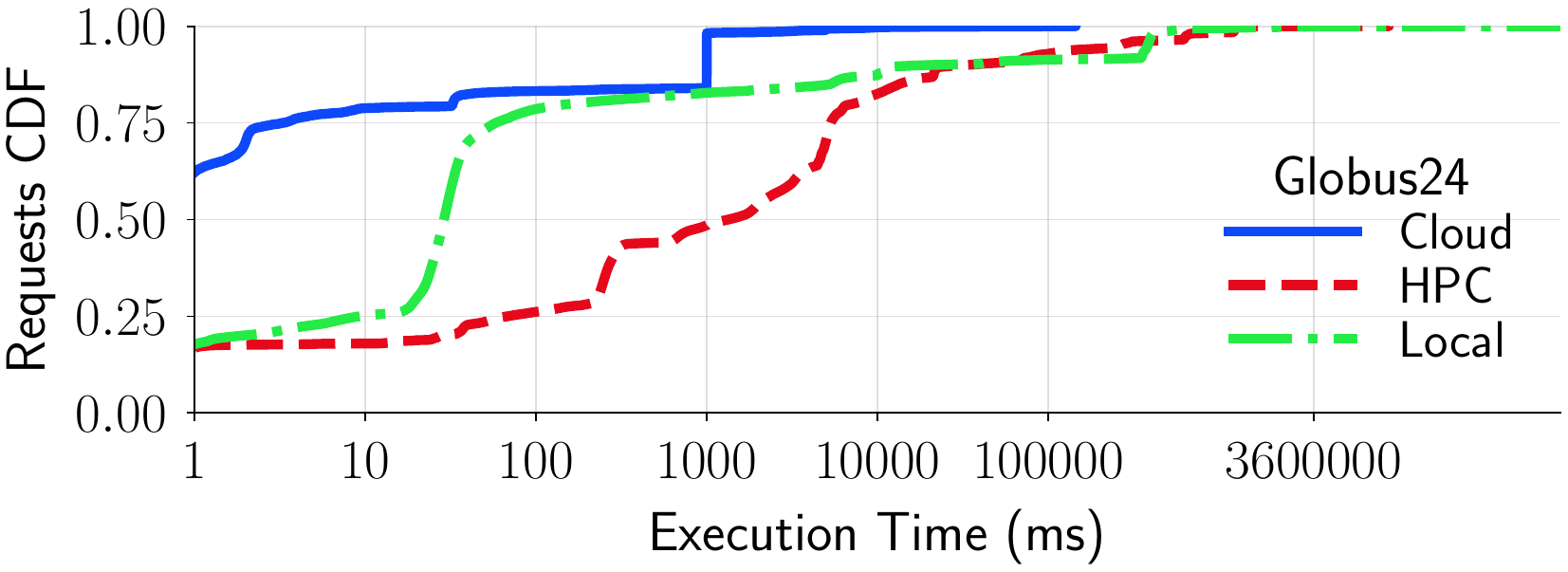}
    \caption{CDF of execution time by infrastructure type in the \Globus trace. Cloud endpoints execute sub-second tasks, whereas HPC endpoints handle heavy-tailed scientific workloads.}
    \label{fig:globus24_infra_cdf}
    \addtocounter{figure}{1}
\end{figure}

\altparagraph{Execution Time.}
The median execution time of functions in modern traces (\Globus, \HuaweiFirst, \HuaweiSecond, \IBM) is low, typically ranging from 2~ms to 35~ms, as shown by the CDF in \cref{fig:execution_time}. This represents a significant performance evolution compared to the older \AzureFirst trace, where the median was $\approx 140$~ms (\cref{tab:latency_combined_stats}). This trend towards ``nanoservice''-style functions makes the relative impact of system overheads, especially cold starts, even more pronounced.

To further contextualize this behavior, we also analyze the execution time aggregated \textit{by function} rather than by individual request. Because the invocation frequency is highly skewed across FaaS platforms, per-request metrics are heavily dominated by a few popular, fast-running functions. When averaging execution time per function, the median increases significantly (e.g., from 31~ms to 677~ms for \AzureSecond, and 13~ms to 131~ms for \IBM). This indicates that although the typical \textit{request} is handled in milliseconds, the typical \textit{function logic} deployed by developers often involves longer-running tasks that are invoked less frequently.

The \Globus dataset offers a unique perspective on how the underlying computing infrastructure influences execution time. \Cref{fig:globus24_infra_cdf} illustrates the execution time CDF for functions running on cloud, local, and high-performance computing (HPC) endpoints. As a federated platform, Globus Compute allows users to deploy endpoints on arbitrary remote resources. Tasks routed to cloud endpoints are short-lived, with a median below 1~ms. ``Local'' endpoints, which represent single-node resources such as edge devices and account for $\approx 75$\% of the platform's endpoints, show a median execution time of $\approx 63$~ms. In contrast, tasks executed on HPC endpoints show a pronounced heavy tail. The median execution time on HPC resources jumps to $\approx 3$~s, with the tail extending up to several hours.

The choice of runtime environment also introduces distinct performance characteristics. \HuaweiSecond breaks down execution times by runtime for the dataset. Java functions show the lowest median latencies ($\approx 0.3$~ms), likely due to pre-warmed internal microservices keeping the JVM hot. Node.js and Python also display low medians (1.7~ms and 6.5~ms), confirming their suitability for high-frequency ``glue'' code. In contrast, custom containers show a higher median latency ($\approx 43$~ms).

\altparagraph{Response Time.}
The end-to-end response time represents the total latency experienced by the user, which includes execution time, cold starts, queueing, and network transit. As shown in \cref{fig:response_time}, although the median response times for commercial platforms are low (2.9~ms for \HuaweiFirst and 15~ms for \IBM), the distributions are heavy-tailed (\cref{tab:latency_combined_stats}). The P99 response times are often two to three orders of magnitude higher than the median. For example, \HuaweiSecond has a median response time of 12~ms but a P99 of 1.78~s. This discrepancy indicates that although the warm start path is highly optimized, a significant minority of requests suffers from severe performance degradation due to cold starts or resource contention.

Furthermore, we observe that the response time is subject to diurnal pressure, even when the execution time is relatively constant. In the \IBM dataset, the average hourly response time fluctuates based on system load, ranging from 161~ms to 309~ms. Since the function logic itself does not change, this latency increase is attributable to platform-level congestion and queueing delays during peak load hours. In contrast, the \Globus platform shows the extreme impact of queueing in federated and scientific systems. Although its median response time is relatively fast at $\approx 339$~ms, its mean response time substantially increases to nearly 1\,400~s ($\approx 23$~minutes), because some tasks wait in queues for HPC nodes to become available.

\begin{table}[!bt]
    \centering
    \caption{Latency characterization (ms). Recent traces show double-digit millisecond median execution times and minimal system overhead ($\approx 1$~ms), whereas \Globus incurs high wait times due to remote scheduling.}
    \label{tab:latency_combined_stats}
    \resizebox{\columnwidth}{!}{%
    \begin{tabular}{l l r r r r}
        \toprule
        \textbf{Dataset} & \textbf{Metric} & \textbf{Median} & \textbf{Mean} & \textbf{P95} & \textbf{P99} \\
        \midrule
        \multirow{1}{*}{\AzureFirst} & Execution & 139.84 & 826.39 & 1.7e3 & 9.4e3 \\
        \midrule
        \multirow{1}{*}{\AzureSecond} & Execution & 31.00 & 3.3e3 & 1.4e4 & 7.2e4 \\
        \midrule
        \multirow{3}{*}{\Globus} & Execution & 32.04 & 4.9e4 & 3.8e5 & 6.5e5 \\
         & Response & 339.28 & 1.4e6 & 5.5e5 & 3.2e7 \\
         & Overhead & 112.07 & 4.3e5 & 2.0e4 & 2.2e6 \\
        \midrule
        \multirow{3}{*}{\HuaweiFirst} & Execution & 2.36 & 9.90 & 35.57 & 132.53 \\
         & Response & 2.89 & 10.59 & 39.02 & 136.72 \\
         & Overhead & 0.50 & 0.69 & 0.97 & 3.26 \\
        \midrule
        \multirow{3}{*}{\HuaweiSecond} & Execution & 9.86 & 88.93 & 249.09 & 1.7e3 \\
         & Response & 12.03 & 92.82 & 270.03 & 1.8e3 \\
         & Overhead & 1.94 & 4.84 & 10.67 & 28.42 \\
        \midrule
        \multirow{3}{*}{\IBM} & Execution & 13.00 & 177.99 & 624.00 & 2.5e3 \\
        & Response & 15.00 & 217.98 & 644.00 & 2.8e3 \\
     & Overhead & 1.00 & 39.99 & 2.00 & 7.00 \\
        \bottomrule
    \end{tabular}%
    }
\end{table}

\altparagraph{System Overhead.}
We isolate the system overhead by calculating the difference between the end-to-end response time and the execution time of the function. This metric represents the ``tax'' paid by the user for the convenience of the serverless abstraction. We clamped negative values to zero to compensate for distributed clock skews. This correction does not affect our tail latency estimates (e.g., P95 and P99), because clock synchronization errors are orders of magnitude smaller than tail latencies. For \Globus, overhead specifically measures the queue delay before execution. Because records with missing timestamps (e.g., timeouts) are filtered independently per metric, mean overheads do not strictly equal the arithmetic difference of the mean response and execution times. As detailed in \cref{tab:latency_combined_stats}, commercial FaaS platforms have optimized this overhead to negligible levels for the median request: \HuaweiFirst, \IBM, and \HuaweiSecond show median overheads of 0.50~ms, 1.00~ms, and 1.94~ms, respectively. This implies that for warm requests, the platform introduces almost no perceptible delay. For \IBM, although the median overhead is 1~ms, the mean is $\approx 40$~ms, and the maximum observed overhead stretches to over 1\,082~s. In serverless workloads, where the execution itself often lasts only a few milliseconds, even a moderate P95 overhead of 2--11~ms (as seen in \IBM and \HuaweiSecond) can represent a performance penalty of 50--100\% relative to the computation. For \Globus, the overhead is naturally high (median 112~ms, mean 426~s), because it includes the time tasks spend traversing the batch scheduler queues of remote endpoints, reflecting an architectural trade-off that favors high-throughput batch computing over low-latency responsiveness.

\begin{insight}
The shift towards ``nanoservices'', where median execution times are often in the double-digit milliseconds, changes the performance bottleneck of FaaS platforms. When a function executes in 2~ms, even a highly optimized system overhead of 1--2~ms imposes a 50--100\% overhead. The divergence between fast executions and a heavy tail of slow, resource-intensive outliers suggests that FaaS architectures should move toward workload-aware isolation, differentiating the execution paths and resource management policies for high-volume, latency-sensitive functions versus long-running, resource-intensive outliers to prevent platform-level congestion.
\end{insight}

\subsection{Instances}
\label{sec:instances}
Our analysis of instance dynamics uses \HuaweiFirst, \HuaweiSecond, and \IBM, because these traces uniquely capture active instance counts, container lifespans, and scaling configurations (e.g., concurrency limits).

\begin{figure}[!t]
    \centering
    \subfloat[Active Instance Lifespan]{
        \includegraphics[width=0.45\columnwidth]{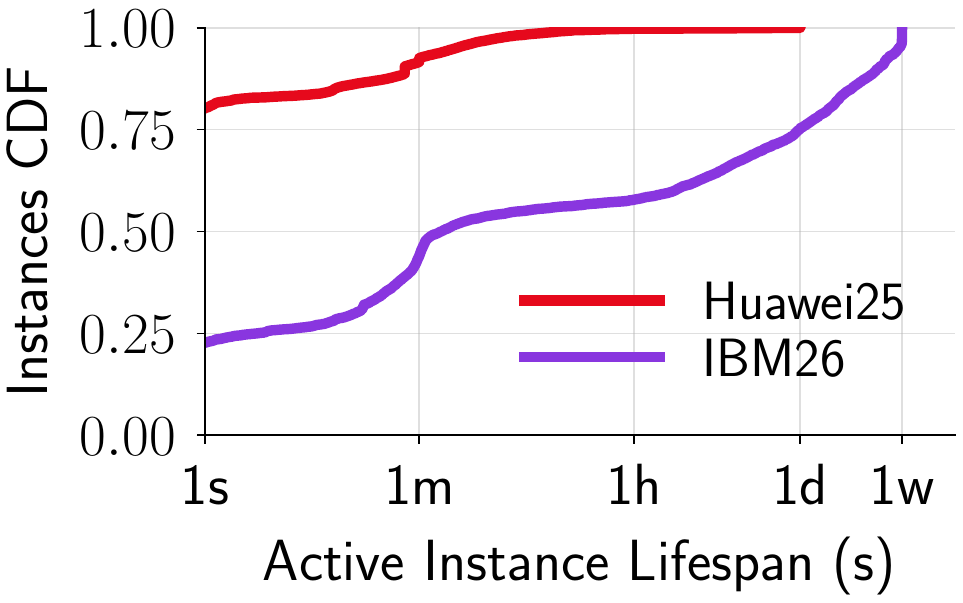}
        \label{fig:lifespan_cdf}
    }
    \hfill
    \subfloat[Peak Concurrent Instances]{
        \includegraphics[width=0.45\columnwidth]{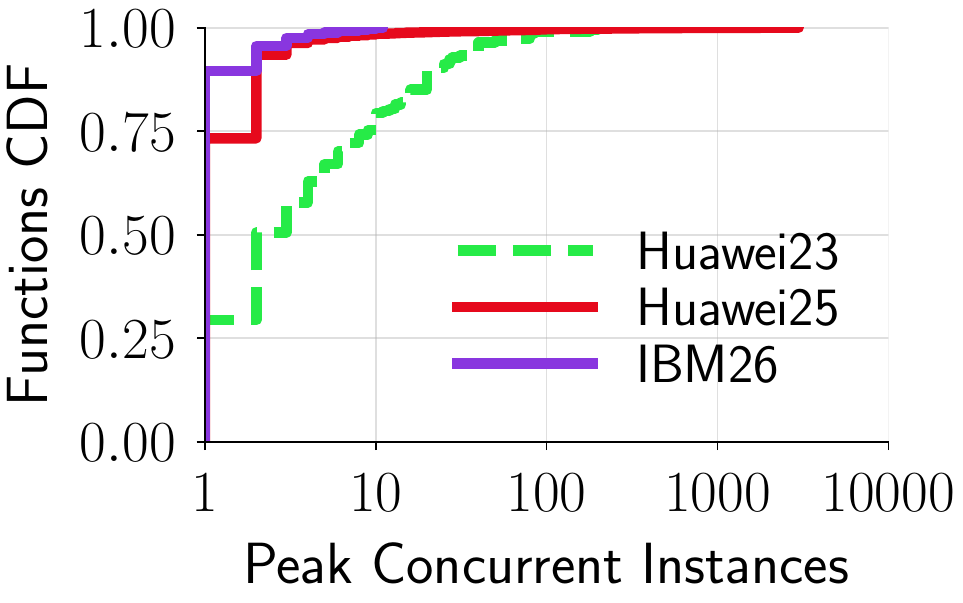}
        \label{fig:max_instances_cdf}
    }
    \caption{Instance dynamics. (a) \IBM containers persist for days due to concurrency mechanics and minimum scale configurations, whereas \HuaweiSecond instances are highly ephemeral. (b) Consequently, \IBM workloads require fewer simultaneous instances (P99 of 6) compared to \HuaweiFirst (P99 of 90) to handle traffic peaks.}
    \label{fig:instance_metrics}
\end{figure}

\altparagraph{Instance Lifespan and Peak Instances.}
The lifecycle of a function instance varies significantly depending on the underlying execution model. Although traditional FaaS platforms typically enforce request isolation (one request per instance), newer platforms allow general-purpose containers to handle multiple concurrent requests. This difference is observable in the lifespan of active instances shown in \cref{fig:lifespan_cdf}. Instances in the \HuaweiSecond trace are highly ephemeral, with a 90th percentile lifespan of only 45~s. This aligns with a request-isolated model where instances are provisioned on-demand and reclaimed shortly after execution. In contrast, the \IBM trace shows significantly greater persistence with a heavy tail: The 90th percentile reaches over 7~days ($6.16 \times 10^5$~s) and some instances persist for as long as the full duration of the trace ($\approx 9.5$~weeks). Despite these long-running instances, the overall median instance lifespan in the \IBM trace remains relatively low at $\approx 284$~s.

\begin{figure}[!t]
    \centering
    \subfloat[Instance Concurrency Limit]{
        \includegraphics[width=0.45\columnwidth]{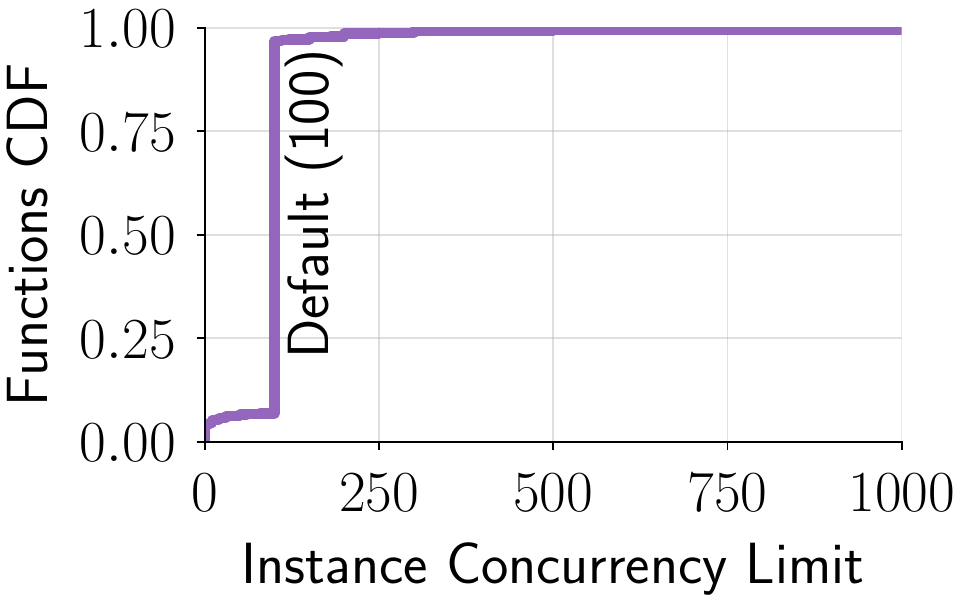}
        \label{fig:ibm26_concurrency_limit_cdf}
    }
    \hfill
    \subfloat[Minimum Instance Scale]{
        \includegraphics[width=0.45\columnwidth]{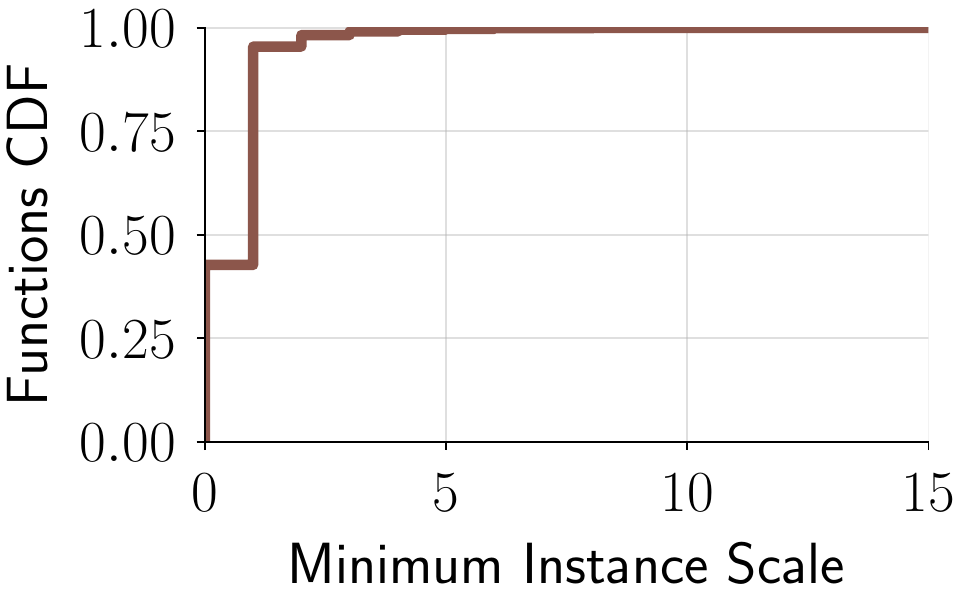}
        \label{fig:ibm26_min_scale_cdf}
    }
    \caption{Instance configuration for \IBM. (a) 90\% of users use the default concurrency limit of 100~requests per instance. (b) Over 50\% of workloads are configured with a minimum scale $\ge 1$, opting out of scale-to-zero.}
    \label{fig:instance_configuration}
\end{figure}

\altparagraph{Concurrency and Scale.}
The capacity to process multiple requests within a single instance alters the relationship between the volume of requests and the required infrastructure. \Cref{fig:max_instances_cdf} compares the peak concurrent instances per function. The \HuaweiFirst trace shows a higher propensity to scale out, with a 99th percentile of $\approx 90$ instances per function. This is characteristic of request-isolated platforms, where high throughput requires a linear increase in pod count. In contrast, the \IBM trace indicates that 90\% of applications never exceed two active instances.

Two configuration factors drive this combined behavior of increased longevity and low peak instances in the \IBM workload. First, the platform relies on an \textit{instance concurrency limit}. As shown in \cref{fig:ibm26_concurrency_limit_cdf}, the median of the configured limit is 100~requests per instance. By multiplexing requests onto a single container, the platform absorbs traffic bursts without scaling out the number of pods, while overlapping requests naturally reset idle timers and extend the instance's lifespan. Second, users actively use the \textit{minimum instance scale} configuration to address their requirements. Our analysis in \cref{fig:ibm26_min_scale_cdf} indicates that the median minimum scale is 1, which means that at least 50\% of workloads effectively opt out of the scale-to-zero model to maintain readiness. This model reduces the aggregate memory footprint and cold start frequency compared to the $1:1$ mapping of request-isolated platforms, although it introduces the potential for resource contention within a single instance if not properly tuned.

\begin{insight}
    Although the low median lifespan of instances confirms that many functions remain ephemeral, the operational reality of modern serverless computing is converging with traditional PaaS: Users are deploying long-lived, concurrent containers to ensure readiness, using the FaaS platform primarily for its simplified deployment and burst-handling capabilities rather than for ephemeral requests and per-request billing.
\end{insight}

\subsection{Resource Usage}
We evaluate resource utilization and provisioning in \AzureFirst, \Alibaba, \HuaweiFirst, \HuaweiSecond, and \IBM, which are the only traces that provide data on allocated limits and/or CPU and memory usage.

\begin{figure}[!t]
    \centering
    \subfloat[\HuaweiFirst]{
        \includegraphics[width=0.45\columnwidth]{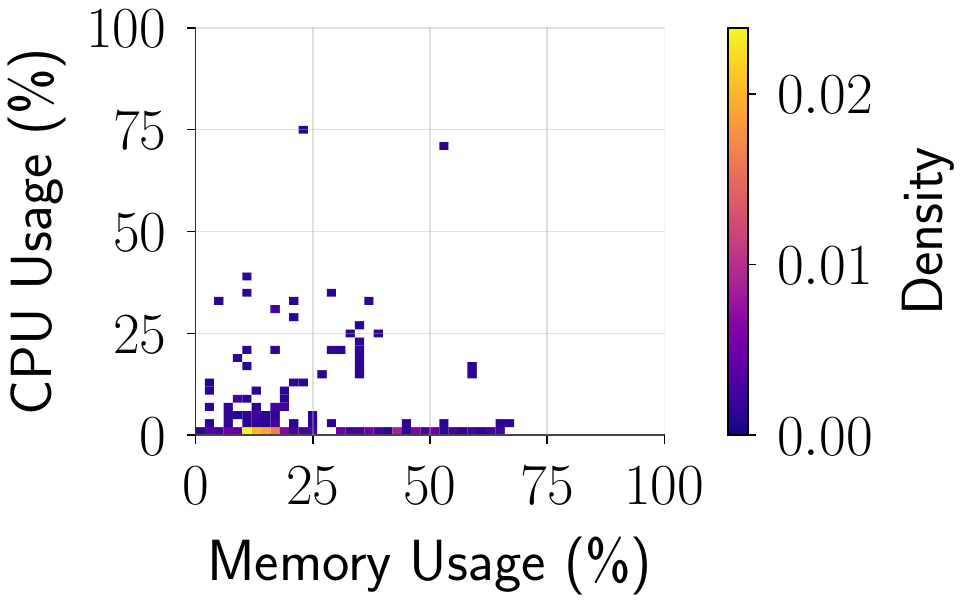}
        \label{fig:huawei23_resource_density}
    }
    \hfill
    \subfloat[\HuaweiSecond]{
        \includegraphics[width=0.45\columnwidth]{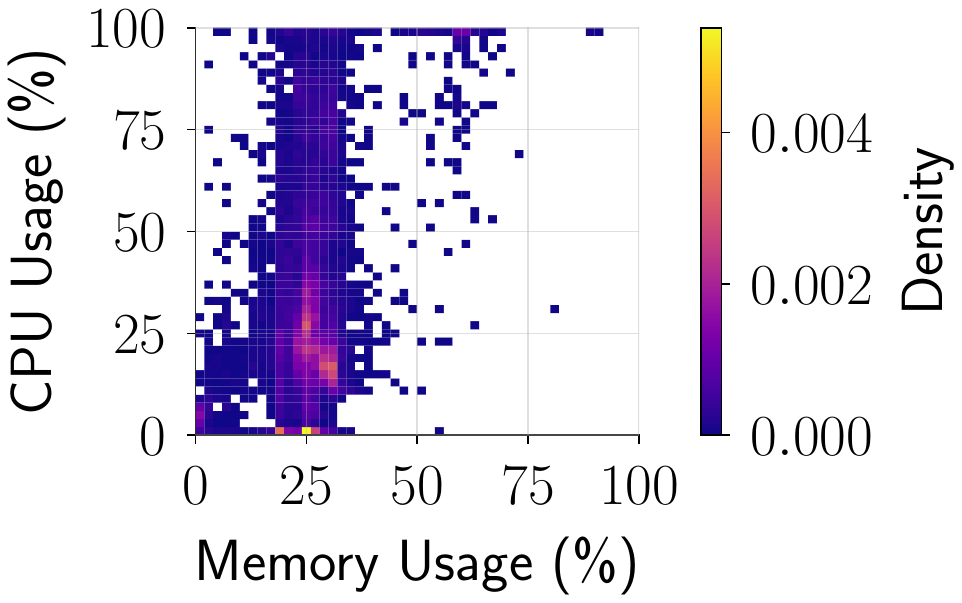}
        \label{fig:huawei25_resource_density}
    }
    \caption{Density of CPU vs. memory utilization. \HuaweiFirst (left) shows extreme over-provisioning with a dense cluster near 0\% for both resources. \HuaweiSecond (right) shows improved relative utilization, with memory usage clustering sharply at $\approx 25\%$ and a wider spread in CPU utilization.}
    \label{fig:resource_usage_density}
\end{figure}

\altparagraph{Memory and CPU Usage.}
The analysis of the \HuaweiFirst and \HuaweiSecond traces shows a pattern of resource over-provisioning, although its severity varies. \Cref{fig:resource_usage_density} presents a density plot of CPU utilization versus memory utilization. For \HuaweiFirst, a dense cluster is visible at low utilization, with functions that use less than 15\% of allocated memory and less than 5\% of allocated CPU. This severe over-provisioning likely stems from the lack of direct cost constraints in this private cloud environment. In contrast, \HuaweiSecond being a public platform shows improved relative utilization, featuring a distinct vertical band at $\approx 25\%$ memory usage and a wider spread of CPU utilization. Nevertheless, the fact that most functions still use significantly less than their allocated bounds points to widespread resource inefficiency.

\Cref{tab:resource_combined_stats} quantifies the memory usage. The memory usage for \HuaweiSecond is low, with a median of only 34~MB and a 99th percentile (P99) of just 423~MB. This supports the assumption that the majority of functions are lightweight and good candidates for snapshotting, which benefits from small memory footprints. However, the distribution has a long tail, with maximum usage reaching over 19~GB, indicating the presence of rare, memory-intensive outliers that require distinct management strategies. In contrast, \HuaweiFirst shows higher absolute usage (median $\approx 320$~MB), whereas \AzureFirst falls between the two with a median usage of 152~MB and a tighter overall distribution.

\begin{figure}[!t]
    \centering
    \subfloat[Memory Limit]{
        \includegraphics[width=0.45\columnwidth]{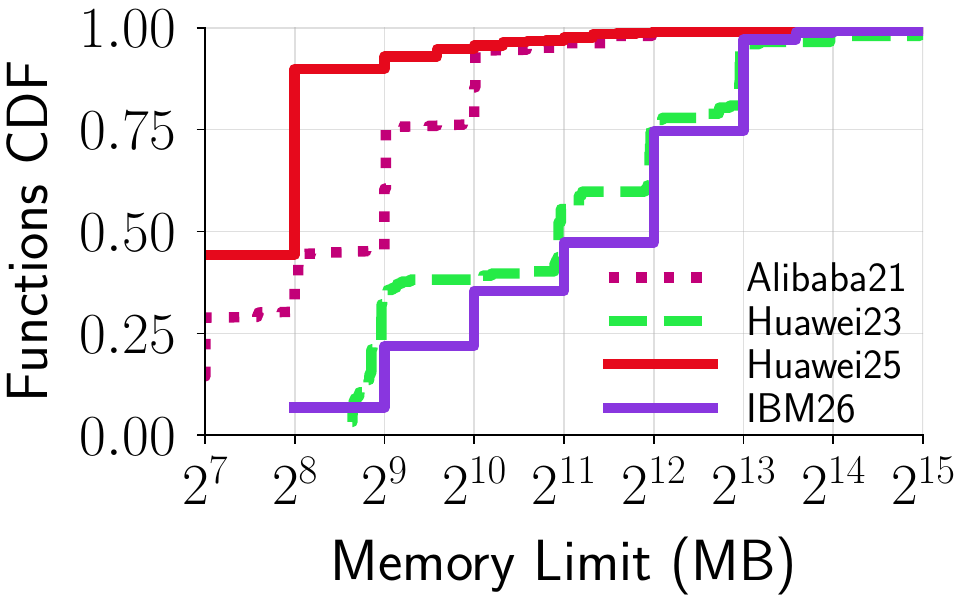}
        \label{fig:memory_limit}
    }
    \hfill
    \subfloat[CPU Limit]{
        \includegraphics[width=0.45\columnwidth]{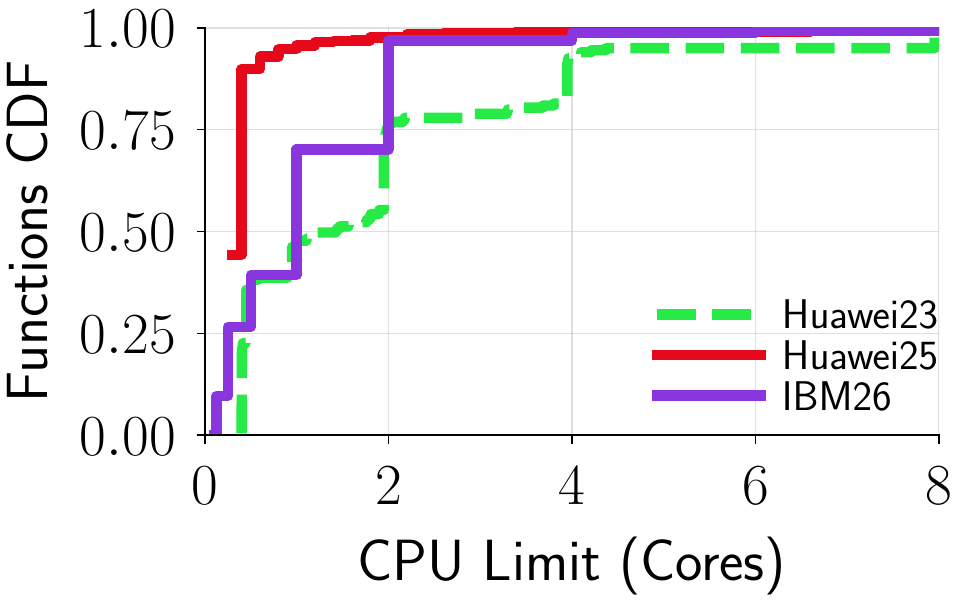}
        \label{fig:cpu_limit}
    }
    \caption{CDFs of configured resource limits. (a) \HuaweiSecond and \Alibaba favor small memory allocations ($<512$~MB), whereas \IBM and \HuaweiFirst use larger containers ($\approx 2$~GB). (b) CPU limits show a similar split, with \IBM defaulting to 1~core.}
    \label{fig:memory_and_cpu_limit}
\end{figure}

\altparagraph{Memory and CPU Limits.}
This inefficiency is further shown by comparing the resource usage with the configured limits. \Cref{fig:memory_and_cpu_limit} shows the CDF of the memory and CPU limits. For \HuaweiSecond, the median memory limit is 128~MB (\cref{tab:resource_combined_stats}), which is nearly $4\times$ higher than the median usage of 34~MB. Similarly, \Alibaba shows a median limit of 512~MB. \IBM and \HuaweiFirst show much higher configurations, with median memory limits of $\approx 2$~GB. This disparity suggests that users often default to generous allocations or lack the tools to right-size their functions.

\begin{figure}[!t]
    \centering
    \subfloat[\HuaweiFirst]{
        \includegraphics[width=0.45\columnwidth]{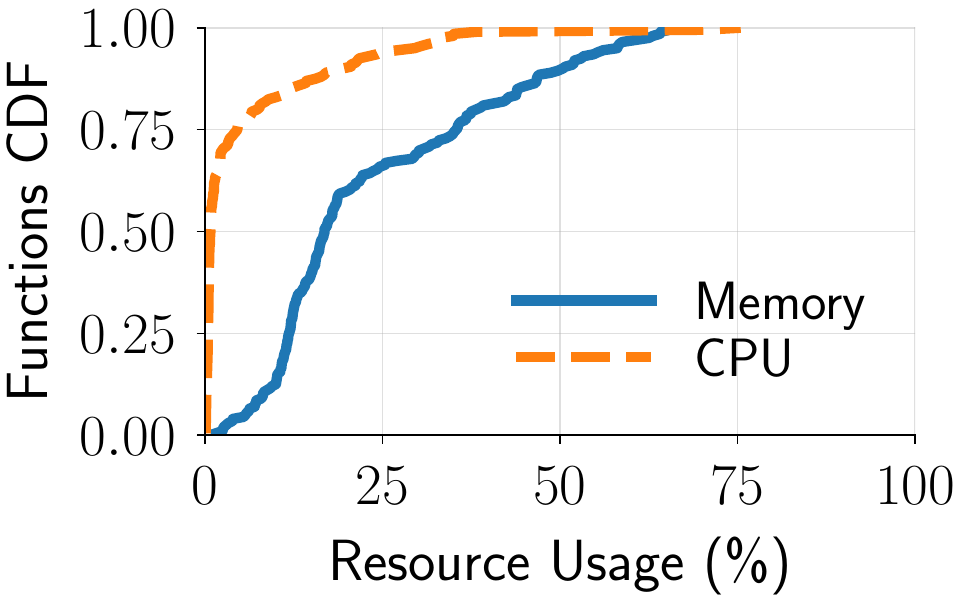}
    }
    \hfill
    \subfloat[\HuaweiSecond]{
        \includegraphics[width=0.45\columnwidth]{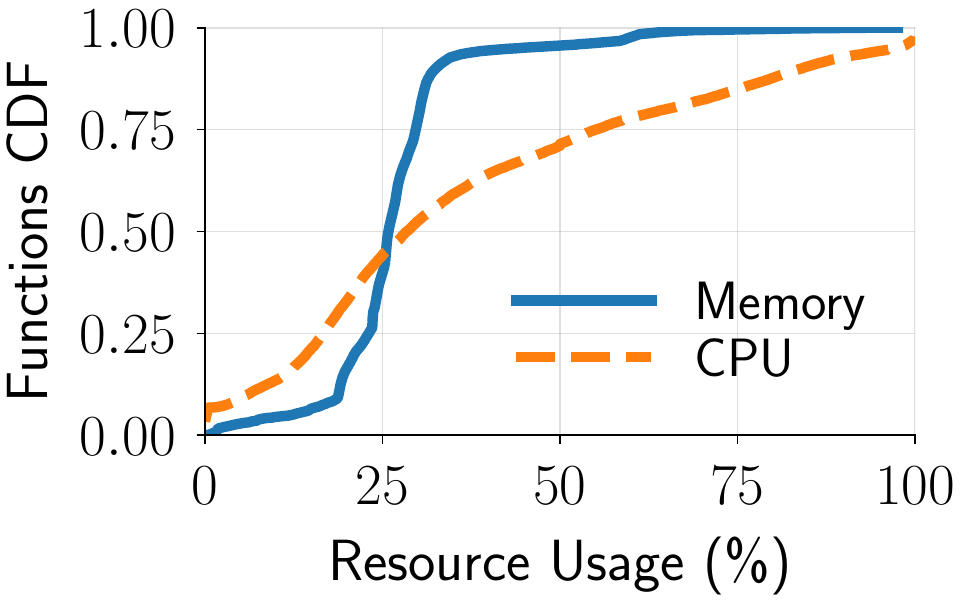}
    }
    \caption{CDFs of resource utilization (as \% of allocated). While \HuaweiFirst functions predominantly use less than 20\% of their allocations, \HuaweiSecond shows higher relative utilization, including a sharp increase at $\approx 25\%$ memory usage.}
    \label{fig:resource_usage_cdf}
\end{figure}

CPU limits follow a similar trend. \HuaweiSecond has a median limit of 0.3~cores, but the median usage is only 0.09~cores (\cref{tab:resource_combined_stats}). \IBM users configure a median of 1~core, whereas \HuaweiFirst users configure a median of $\approx 1.4$~cores. In \cref{fig:resource_usage_cdf}, the gap between these limits and the low utilization (often below 10\% in \HuaweiFirst) confirms substantial room for optimization, although newer platforms such as \HuaweiSecond push limits closer to actual usage.

\begin{table}[!tb]
    \centering
    \caption{Resource usage and limits. Comparing memory and CPU consumption against configured limits shows significant over-provisioning across platforms. Limits are often set generously, and actual usage is typically much lower.}
    \label{tab:resource_combined_stats}
    \resizebox{\columnwidth}{!}{%
    \begin{tabular}{@{}llrrrrr@{}}
        \toprule
        \textbf{Dataset} & \textbf{Metric} & \textbf{Median} & \textbf{Mean} & \textbf{P95} & \textbf{P99} & \textbf{Max} \\
        \midrule
        \multicolumn{7}{@{}l}{\textit{Memory (MB)}} \\
        \cmidrule{1-7}
        \AzureFirst   & Usage & 152 & 163 & 240 & 398 & 1\,368 \\
        \addlinespace
        \Alibaba      & Limit & 512 & 612 & 1\,536 & 3\,072 & 32\,768 \\
        \addlinespace
        \HuaweiFirst  & Usage & 320 & 895 & 4\,248 & 8\,766 & 9\,713 \\
                      & Limit & 1\,963 & 3\,562 & 8\,000 & 31\,963 & 32\,000 \\
        \addlinespace
        \HuaweiSecond & Usage & 34 & 64 & 115 & 423 & 19\,338 \\
                      & Limit & 128 & 343 & 1\,024 & 4\,096 & 32\,768 \\
        \addlinespace
        \IBM          & Limit & 2\,048 & 2\,725 & 4\,096 & 16\,384 & 49\,152 \\
        \midrule
        \multicolumn{7}{@{}l}{\textit{CPU (Cores)}} \\
        \cmidrule{1-7}
        \HuaweiFirst  & Usage & 0.01 & 0.05 & 0.30 & 0.39 & 0.75 \\
                      & Limit & 1.43 & 1.88 & 4.73 & 7.98 & 9.69 \\
        \addlinespace
        \HuaweiSecond & Usage & 0.09 & 0.14 & 0.30 & 0.86 & 6.51 \\
                      & Limit & 0.30 & 0.47 & 1.00 & 3.40 & 25.80 \\
        \addlinespace
        \IBM          & Limit & 1.00 & 0.78 & 1.00 & 4.00 & 8.00 \\
        \bottomrule
    \end{tabular}%
    }
\end{table}

\begin{insight}
    The prevalent underutilization of resources suggests that providers can co-locate more functions on the same physical hardware through resource over-subscription. Furthermore, platforms should provide developers with automated ``right-sizing'' recommendations based on historical usage data to reduce costs and improve cluster efficiency.
\end{insight}

\subsection{Runtimes}
\label{sec:runtimes}
The runtime environment determines both the initialization overhead and the execution performance of serverless functions. We compare function runtimes using \Alibaba and \HuaweiSecond in \Cref{tab:runtimes}, because they are the only traces that publish runtime metadata. \Alibaba only logs cold starts, so its distribution reflects initialized functions rather than all functions and potentially skews results. In both traces, Python is the dominant runtime, accounting for 46.8\% of functions in the \Alibaba trace and 65.6\% in the more recent \HuaweiSecond trace. This aligns with the broader popularity of Python for data-centric tasks and general scripting.

We observe a shift in other runtimes. Although Node.js was a popular runtime in the previous \Alibaba dataset (34.6\%), its share drops to 6.6\% in \HuaweiSecond. Similarly, PHP usage declines from 7.6\% to 1.6\%. In contrast, Java usage has tripled, rising from 6.2\% in \Alibaba to 18.6\% in \HuaweiSecond. This indicates a growing reliance on statically typed languages executed in managed runtimes, despite the higher startup costs associated with the JVM.

Custom containers remain a persistent minority ($\approx 2.4$--$2.5\%$), likely due to the performance penalties associated with unmanaged initialization. For custom runtimes, which accounted for 1.7\% in \Alibaba, there is no information provided in the \HuaweiSecond trace.

\begin{insight}
    Reducing interpreter startup overhead and caching commonly used libraries can provide performance improvements, because Python represents the majority of functions. At the same time, the growing share of Java makes snapshotting techniques important for addressing the relatively expensive initialization of the JVM.
\end{insight}

\begin{figure*}[!t]
\centering
\begin{minipage}{0.49\textwidth}
    \begin{table}[H]
        \centering
        \caption{Distribution of function runtimes. Python remains the primary runtime and Java usage increases significantly in the newer Huawei dataset, whereas usage of Node.js and PHP drops significantly.}
        \resizebox{\columnwidth}{!}{%
            \begin{tabular}{l l@{\hspace{4pt}}r l@{\hspace{4pt}}r}
                \toprule
                & \multicolumn{2}{c}{\Alibaba} & \multicolumn{2}{c}{\HuaweiSecond} \\
                \multicolumn{1}{c}{Runtime} & \multicolumn{2}{c}{(\% Functions)} & \multicolumn{2}{c}{(\% Functions)} \\
                \cmidrule(lr){2-3} \cmidrule(lr){4-5}
                Python           & \db{46.8} & 46.8 & \db{65.6} & 65.6 \\
                Node.js          & \db{34.6} & 34.6 & \db{6.6}  & 6.6  \\
                Java             & \db{6.2}  & 6.2  & \db{18.6} & 18.6 \\
                PHP              & \db{7.6}  & 7.6  & \db{1.6}  & 1.6  \\
                Custom Container & \db{2.5}  & 2.5  & \db{2.4}  & 2.4  \\
                Custom Runtime   & \db{1.7}  & 1.7  & \db{0}    & --  \\
                .NET (C\#)       & \db{0.6}  & 0.6  & \db{0.2}  & 0.2  \\
                Go               & \db{0.03} & 0.03 & \db{0.9}  & 0.9  \\
                Others           & \db{0.03} & 0.03 & \db{3.0}  & 3.0  \\
                Unknown          & \db{0}    & --   & \db{1.3}  & 1.3  \\
                \bottomrule
            \end{tabular}%
        }
        \label{tab:runtimes}
    \end{table}
\end{minipage}
\hfill
\begin{minipage}{0.49\textwidth}
    \begin{table}[H]
        \centering
        \caption{Distribution of function triggers. Timer triggers have surpassed HTTP as the primary invocation source in the modern \HuaweiSecond trace compared to the older \AzureFirst trace.}
        \resizebox{\columnwidth}{!}{%
            \begin{tabular}{l l@{\hspace{4pt}}r l@{\hspace{4pt}}r}
                \toprule
                & \multicolumn{2}{c}{\AzureFirst} & \multicolumn{2}{c}{\HuaweiSecond} \\
                \multicolumn{1}{c}{Trigger} & \multicolumn{2}{c}{(\% Functions)} & \multicolumn{2}{c}{(\% Functions)} \\
                \cmidrule(lr){2-3} \cmidrule(lr){4-5}
                HTTP          & \db{41.9} & 41.9 & \db{36.9} & 36.9 \\
                Timer         & \db{27.6} & 27.6 & \db{42.2} & 42.2 \\
                Queue         & \db{14.6} & 14.6 & \db{0.1}  & 0.1  \\
                Orchestration & \db{7.9}  & 7.9  & \db{2.2}  & 2.2  \\
                Event         & \db{2.6}  & 2.6  & \db{1.8}  & 1.8  \\
                Storage       & \db{2.2}  & 2.2  & \db{2.5}  & 2.5  \\
                Others        & \db{3.3}  & 3.3  & \db{14.2} & 14.2 \\
                \bottomrule
            \end{tabular}%
        }
        \label{tab:triggers}
    \end{table}
\end{minipage}
\end{figure*}

\subsection{Imported Libraries \& Function Code Metrics}
\label{sec:libraries}

\begin{figure}[!t]
    \centering
    \subfloat[Imported Libraries]{
        \includegraphics[width=0.45\columnwidth]{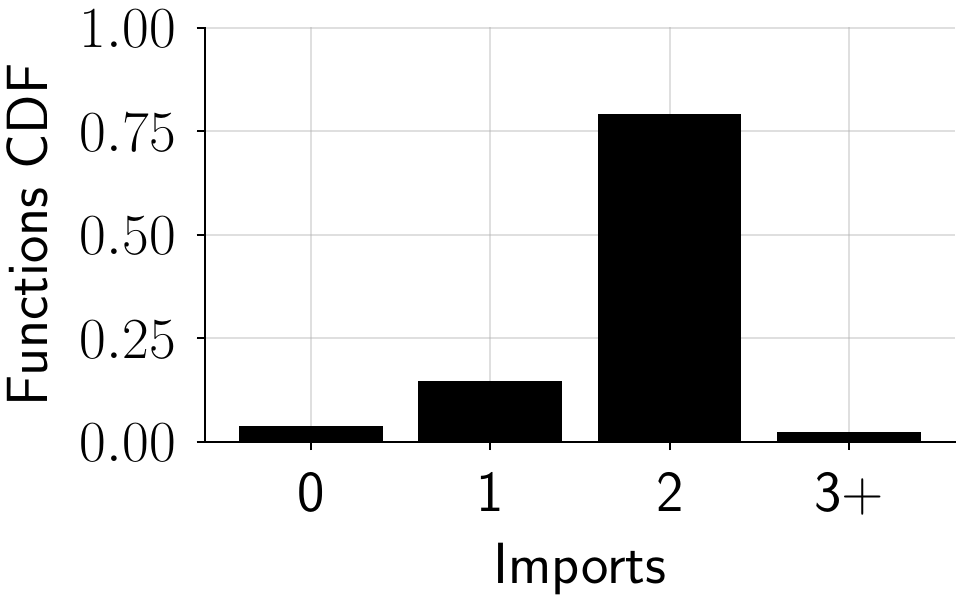}
        \label{fig:package_distribution}
    }
    \hfill
    \subfloat[Response Time Correlation]{
        \includegraphics[width=0.45\columnwidth]{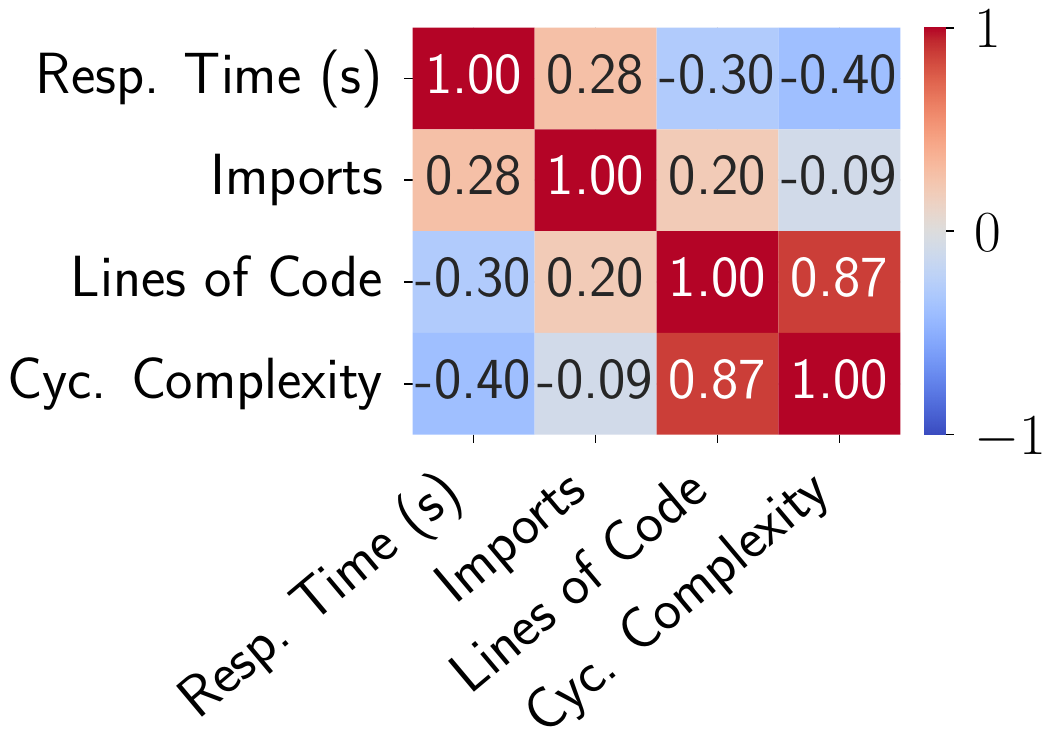}
        \label{fig:globus24_code_correlation}
    }
    \caption{Imported libraries, lines of code, and cyclomatic complexity of functions in the \Globus trace. (a) The majority of functions have one or two dependencies. (b) Spearman correlation matrix for function code metrics and response time. The number of imports is the only code metric positively correlated with latency.}
    \label{fig:globus24_library_analysis}
\end{figure}

The \Globus dataset provides function-level metrics for the number of package imports, lines of code (LoC), and cyclomatic complexity. This platform exclusively supports Python functions, which naturally shapes these specific code and dependency characteristics. \Cref{fig:package_distribution} shows that dependency trees are shallow: Over 90\% of functions import one or two libraries, with two imports being overwhelmingly the most common ($\approx 80\%$). The median function has only 15 lines of code, and the median cyclomatic complexity is 2. This distribution suggests that developers primarily use FaaS for lightweight coordination or simple processing tasks rather than complex, monolithic application logic.

To understand how these metrics impact performance, we computed the Spearman correlation between them and the end-to-end response time. \Cref{fig:globus24_code_correlation} visualizes the results. The number of imports shows a positive correlation (0.28) with the response time, suggesting that functions with more dependencies tend to have higher latency, likely due to increased initialization overhead. Heavy dependency chains increase initialization time, contributing to the long-tail latency distribution discussed in \cref{sec:cold_starts}.

Counterintuitively, both LoC ($-0.30$) and cyclomatic complexity ($-0.40$) show a moderate negative correlation with response time. This suggests that in this scientific workload, longer or more complex code does not necessarily result in longer end-to-end response times. A plausible explanation is that such functions are often I/O-bound or used for orchestration, where compute time is not the dominant factor, whereas short scripts might invoke heavy computation through external library calls. As expected, LoC and cyclomatic complexity are strongly correlated with each other (0.87).

\begin{insight}
    The number of dependencies is a better predictor of response time than the size or complexity of the code itself. This suggests that cold start optimizations should prioritize efficient dependency loading, caching, and layer sharing over simple code size reduction. The combination of shallow dependency trees and popular runtimes (e.g., Python), creates a clear opportunity for package-aware caching. Since most functions rely on a small set of libraries, providers can provision base images pre-loaded with high-frequency packages to eliminate this overhead.
\end{insight}

\subsection{Triggers}
\label{sec:triggers}
Trigger mechanisms define the arrival patterns of function invocations. We analyze trigger mechanisms in \AzureFirst and \HuaweiSecond, which provide categorized distributions of invocation sources. \Cref{tab:triggers} compares the trigger distributions between the older \AzureFirst dataset and the modern \HuaweiSecond dataset to illustrate the evolution of usage patterns. We observe a distinct shift from HTTP-dominated workloads to timer-based invocations. Timer triggers increased from 27.6\% in \AzureFirst to 42.2\% in \HuaweiSecond, surpassing HTTP (36.9\%) as the primary invocation source in the newer dataset. This trend aligns with our findings in \cref{sec:runtimes} regarding the decline in web-centric technologies (Node.js, PHP). Together, these shifts suggest that FaaS is increasingly used for scheduled background tasks and orchestration rather than solely for user-facing web APIs. Queue-based triggers, which were significant in \AzureFirst (14.6\%), appear negligible in the \HuaweiSecond dataset (0.1\%), although this disparity may partly reflect platform categorization differences where internal messaging is grouped under ``Others'' (14.2\%).

\begin{insight}
    The shift from HTTP to timer triggers allows providers to treat resource allocation as a planning problem rather than a prediction problem. Serverless workloads are becoming less driven by user interactions and more by scheduled machine tasks.
\end{insight}

\subsection{Data Interaction \& I/O}
\label{sec:data_io}
Serverless functions interact with external storage services to maintain state, process datasets, or log results. We analyze the \AzureBlob trace, which captures interactions between functions and Azure Blob Storage, to characterize these I/O patterns, focusing on object sizes, access frequencies, and temporal regularity. Furthermore, we use the \Globus and \HuaweiSecond traces to examine payload sizes and I/O latency.

\begin{figure}[!t]
    \centering
    \includegraphics[width=0.8\columnwidth]{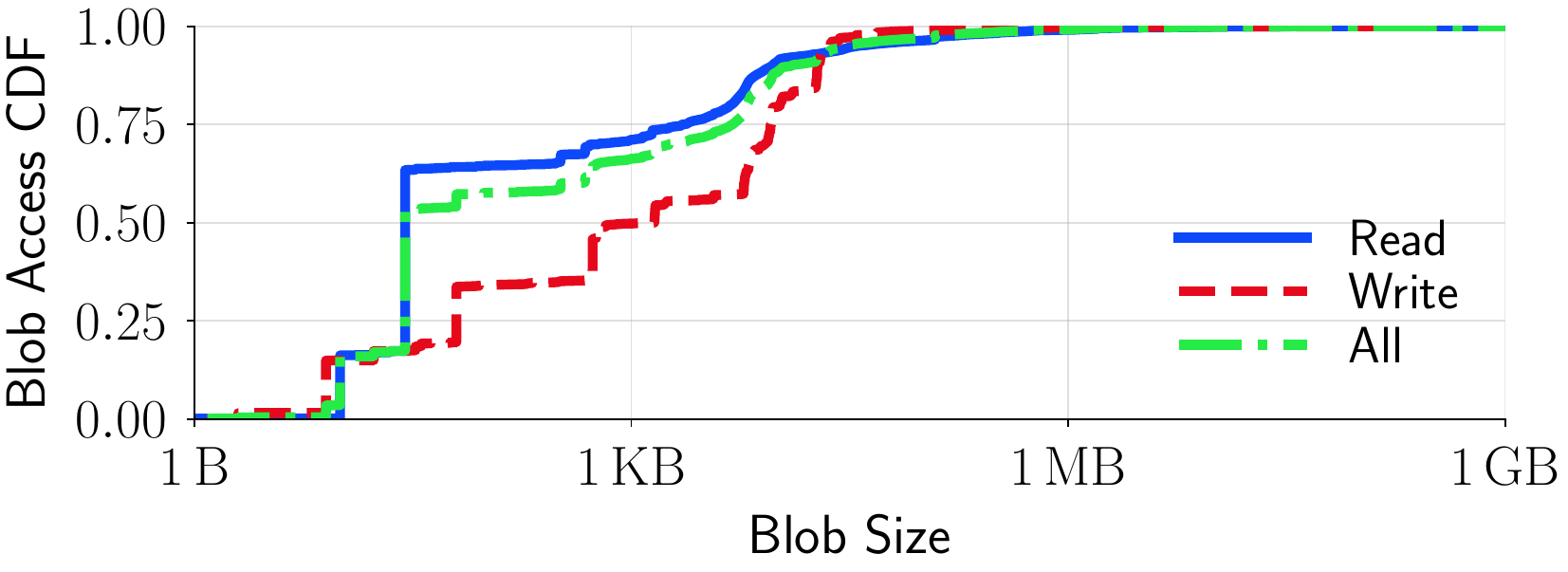}
    \caption{Blob Object Sizes (\AzureBlob). Reads show a heavier tail than writes, whereas writes have a significantly larger median size.}
    \label{fig:blob_size_cdf}
\end{figure}

\altparagraph{External Storage Access Patterns.}
\Cref{fig:blob_size_cdf} illustrates the Cumulative Distribution Function (CDF) of blob sizes for read and write operations. The distribution is heavy-tailed, spanning nine orders of magnitude from bytes to gigabytes.

\Cref{tab:blob_stats} presents the statistical summary of these accesses. An inversion exists between the mean and median values for reads versus writes. The median write size (1\,436~B) is larger than the median read size (28~B), likely due to functions writing structured logs or state checkpoints compared to reading tiny configuration files or metadata. However, the mean read size (68~KB) exceeds the mean write size (46~KB), driven by the long tail of large object reads (e.g., media files or ML models). The 99th percentile for reads is nearly $5\times$ higher than that for writes.

\begin{figure}[!t]
    \centering
    \includegraphics[width=0.8\columnwidth]{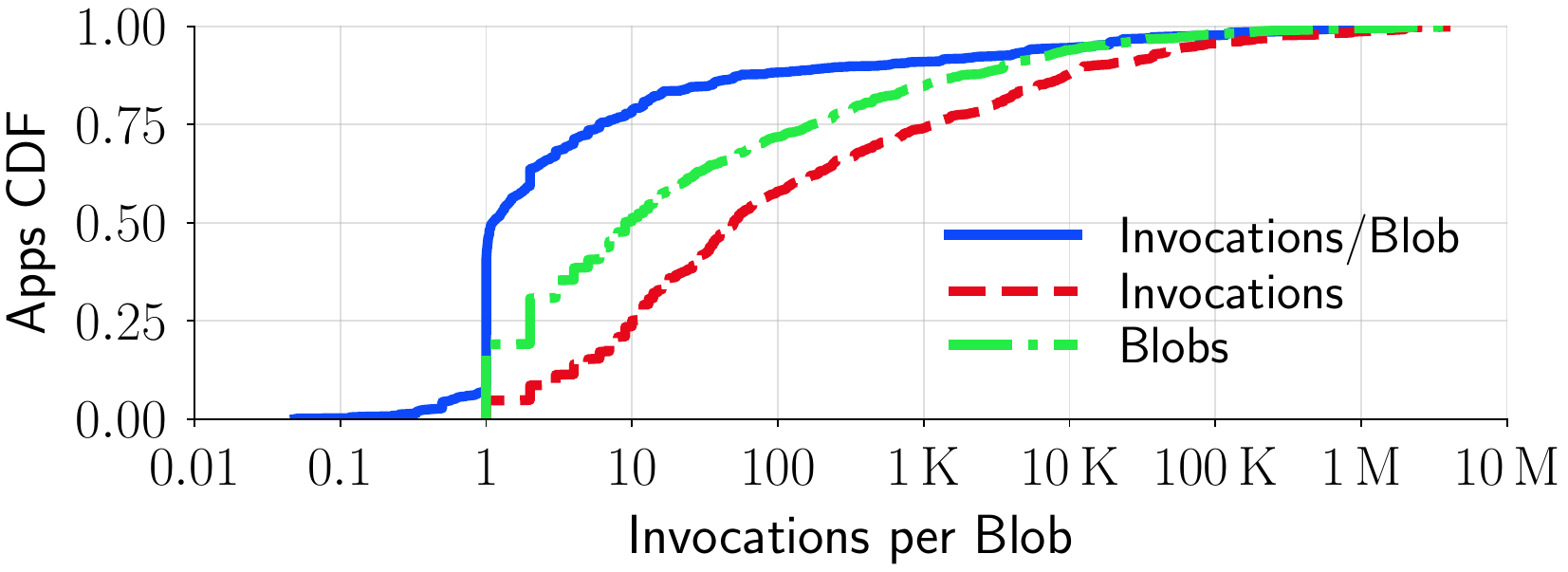}
    \caption{Data ephemerality and contention (\AzureBlob). The median blob is accessed only once, indicating high data ephemerality, whereas the long tail represents highly contended shared objects.}
    \label{fig:blob_access_cdf}
\end{figure}

\begin{table}[!t]
    \centering
    \caption{Blob access statistics (\AzureBlob). Read operations show a heavier tail distribution than writes, whereas the Coefficient of Variation (CV) for inter-arrival times indicates bursty access patterns ($\text{CV} > 1$).}
    \begin{tabular}{l r r r r r}
        \toprule
        \textbf{Metric} & \textbf{Mean} & \textbf{Median} & \textbf{P95} & \textbf{P99} & \textbf{CV} \\
        \midrule
        \multicolumn{6}{l}{\textit{Size per Access (Bytes)}} \\
        \cmidrule{1-6}
        Read Size  & 68\,426 & 28 & 36\,242 & 725\,398 & 25.9 \\
        Write Size & 46\,278 & 1\,436 & 23\,320 & 147\,476 & 42.1 \\
        \midrule
        \multicolumn{6}{l}{\textit{Access Frequency}} \\
        \cmidrule{1-6}
        Inv. per Blob & 11\,049 & 1.1 & 14\,491 & 376\,762 & 7.5 \\
        \bottomrule
    \end{tabular}%
    \label{tab:blob_stats}
\end{table}

\altparagraph{Data Access Frequency.}
\Cref{fig:blob_access_cdf} depicts the distribution of access frequencies (invocations per blob).
The data show that the majority of storage objects are ephemeral. The median number of invocations per blob is $\approx 1.1$, indicating that at least 50\% of objects are accessed only once during the trace period.
Conversely, the distribution tail is long.
The mean is 11\,049 invocations, and the top 1\% of blobs (P99) are accessed more than $376\,000$ times.
This skewed distribution suggests that although most data is transient, a small tail of objects dominates the I/O throughput. Further analysis reveals that data sharing across applications is minimal.
Specifically, 99.5\% of all blobs are bound to a single application.
Even for ``hot'' blobs (those accessed $10$ or more times), this figure
remains high at nearly 89\%. Consequently, contention is heavily
localized within individual application boundaries rather than broadly
distributed across the platform.

\altparagraph{Temporal Regularity.}
We analyzed the regularity of data accesses using the coefficient of variation (CV) of inter-arrival times (IAT). Data access patterns show moderate burstiness rather than strict regularity. The median IAT CV is 1.23, with a mean of 1.20. A $\text{CV} > 1$ indicates that data accesses are generally more clustered/bursty than a Poisson process (random arrivals), contradicting the assumption that storage I/O is purely periodic (i.e., $\text{CV} \approx 0$). This suggests that data caching policies should be resilient to irregular and bursty access intervals rather than relying on predictable schedules.

\begin{figure}[!t]
    \centering
    \includegraphics[width=\columnwidth]{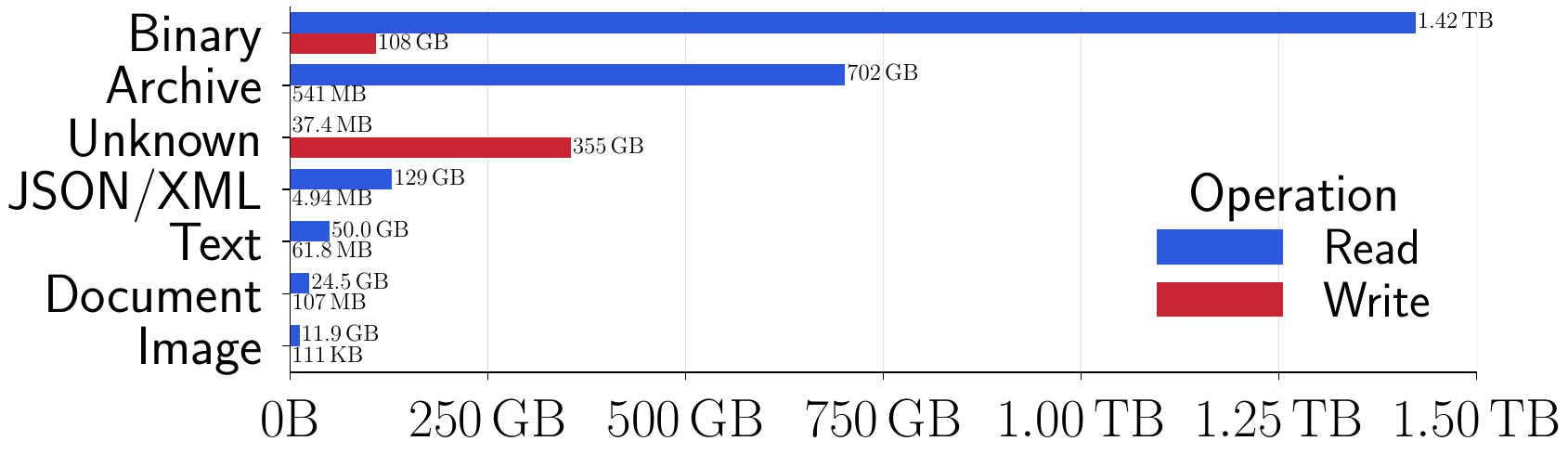}
    \caption{Transfer volume by content type (\AzureBlob). Volume is heavily dominated by Binary, Unknown, and Archive blobs, with significantly smaller contributions from structured formats like JSON/XML and Text.}
    \label{fig:bytes_by_blobtype_grouped}
\end{figure}

\begin{figure}[!t]
    \centering
    \subfloat[Request Body Size]{
        \includegraphics[width=0.45\columnwidth]{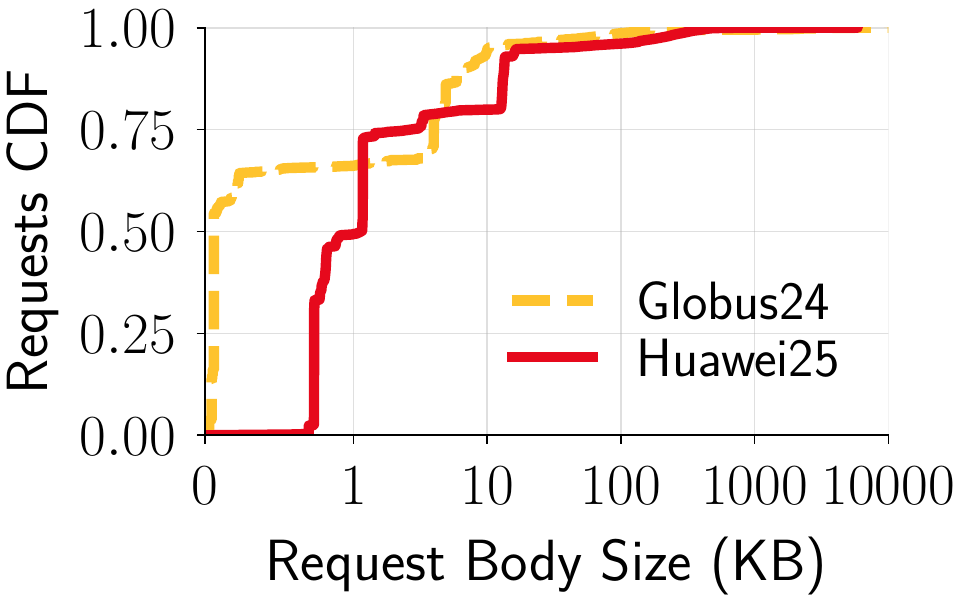}
        \label{fig:request_size_cdf}
    }
    \hfill
    \subfloat[I/O Latency (\HuaweiSecond)]{
        \includegraphics[width=0.45\columnwidth]{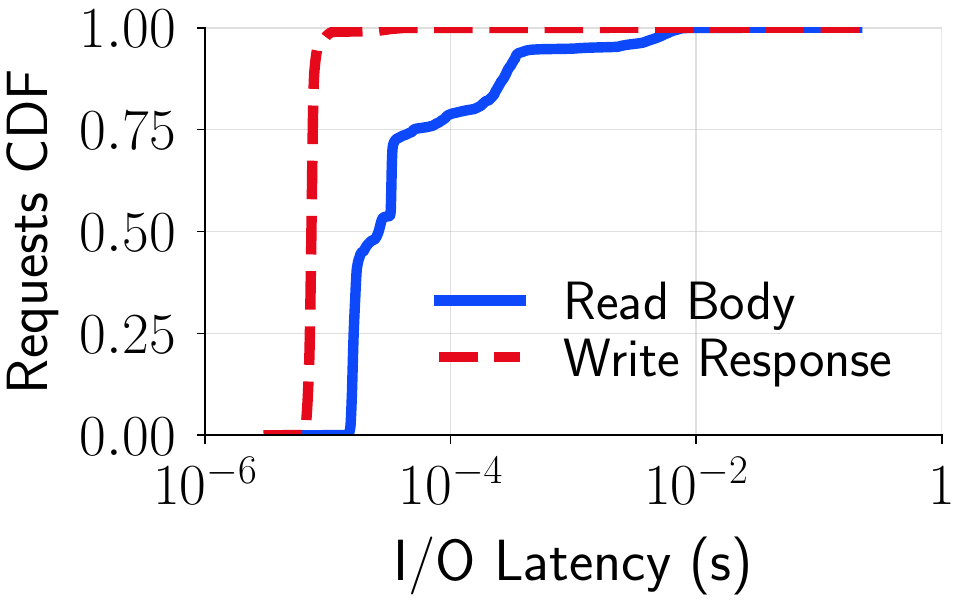}
        \label{fig:io_latency}
    }
    \caption{Request body size and I/O latency. (a) The CDF of request body sizes in the \Globus trace is dominated by small control messages, but features an extreme tail of larger payloads. (b) The I/O latency for \HuaweiSecond shows that reading request bodies dominates the frontend overhead compared to writing rapid responses.}
    \label{fig:io}
\end{figure}

\altparagraph{Transfer Volume by Blob Category.}
To understand what types of data dominate Azure Functions storage traffic, we group each access by high-level \emph{Blob Category} and aggregate transferred bytes from \AzureBlob.
\Cref{fig:bytes_by_blobtype_grouped} shows that the overall volume is heavily skewed toward \emph{Binary} and \emph{Unknown} payloads.
\emph{Binary} objects total $\approx 1.4$~TB for reads and 108~GB for writes. This category is primarily driven by the default content type assigned by the Azure Functions platform when users do not explicitly specify one, indicating that most bandwidth is spent moving opaque binary objects.
\emph{Unknown} payloads (objects that lack a specific content type) account for a massive 355~GB for writes, but only $\approx 37$~MB for reads.
\emph{Archive} formats (e.g., zip/compressed), contribute the next largest share for reads at about 700~GB (despite fewer than 100\,000 accesses), suggesting the packaging of dependencies or datasets.
\emph{JSON/XML} traffic contributes $\approx 130$~GB, and \emph{Text} traffic contributes $\approx 50$~GB.
Notably, although \emph{Text} and \emph{JSON/XML} formats represent the majority of individual accesses (over 25~million combined), their small average sizes limit their total bandwidth footprint compared to the fewer but much larger \emph{Binary}, \emph{Unknown}, and \emph{Archive} blobs.
Spatially, this transfer volume is dense: Among the 28 regions studied, the top 4 regions account for more than 86\% of the total data transfer volume.
Overall, the size distribution indicates that bandwidth demand is driven primarily by binary and compressed objects rather than by strongly typed, human-centric content.

\altparagraph{Request Body Size.}
The \Globus and \HuaweiSecond traces provide insight into the data transfer characteristics of scientific and commercial FaaS workloads, respectively, by capturing the size of function arguments (request bodies).
\Cref{fig:request_size_cdf} illustrates the CDF of these request sizes. For the scientific \Globus workload, the distribution is heavily skewed: Although the median request size is small at 60~B, the mean is significantly larger at $\approx 16.9$~KB.
This disparity is driven by a long tail of data-intensive invocations where the P99 reaches 129~KB and the maximum exceeds 10~MB ($\text{CV} \approx 12.4$).
This aligns with ``bag of tasks'' use cases where functions occasionally receive large data payloads.
In contrast, the commercial \HuaweiSecond trace shows a larger median of 1.13~KB but a smaller coefficient of variation ($\text{CV} \approx 4.0$), suggesting more consistent payload sizes typical of standardized web APIs, although it still shows a heavy tail with a P99 of 319~KB.

\begin{figure}[!t]
    \centering
    \subfloat[Functions per User]{
        \includegraphics[width=0.45\columnwidth]{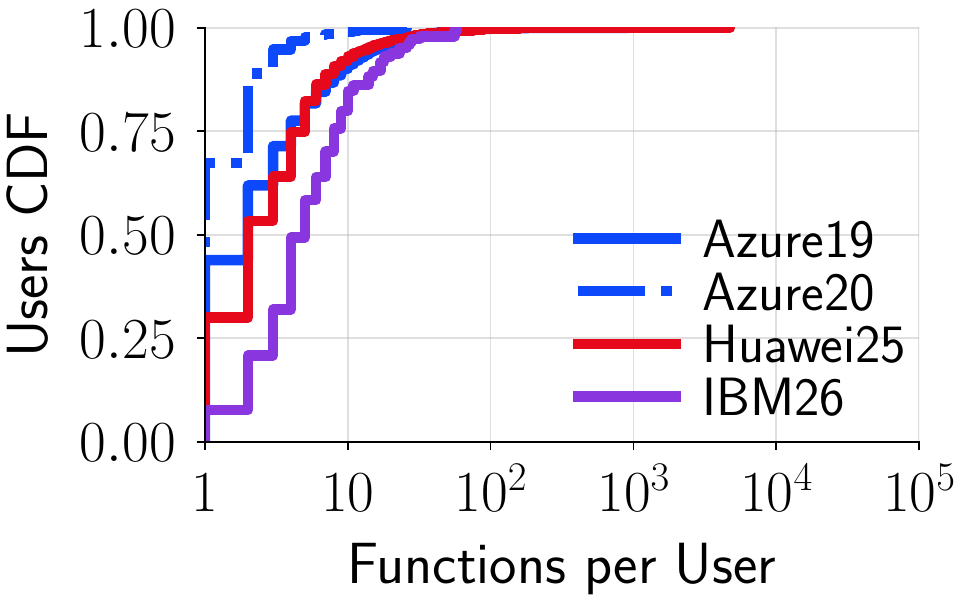}
        \label{fig:functions_per_user}
    }
    \hfill
    \subfloat[Daily Requests per User]{
        \includegraphics[width=0.45\columnwidth]{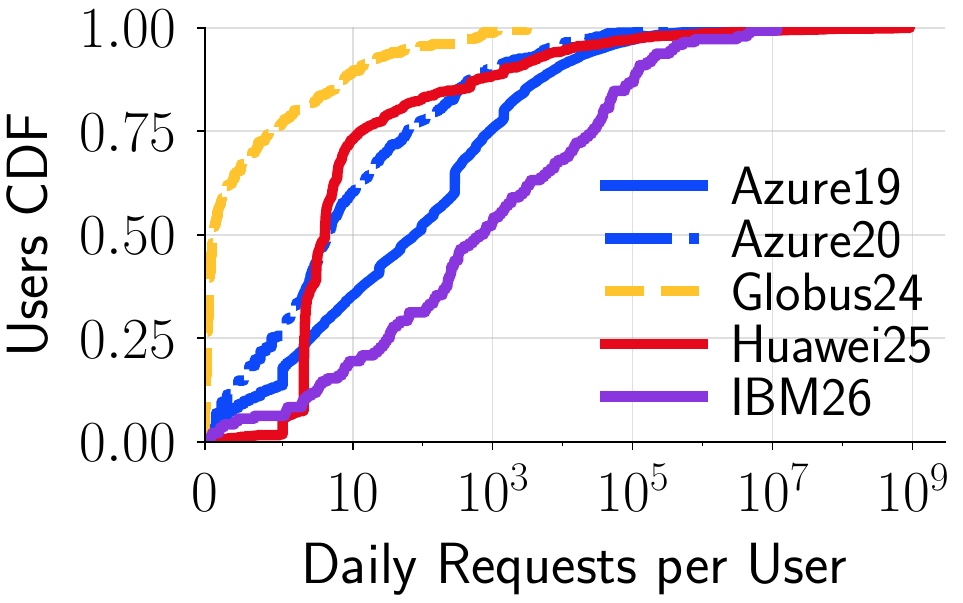}
        \label{fig:requests_per_user}
    }
    \caption{Functions and daily requests per user. Both distributions show heavy tails, demonstrating that a minority of users accounts for the majority of both registered functions and system invocations.}
    \label{fig:users}
\end{figure}

\begin{table}[!tb]
    \centering
    \caption{User tenancy and workload distribution. Commercial platforms generally host many users with few functions (low median), but request volume is heavily skewed toward a small fraction of high-volume users (high CV and Max).}
    \label{tab:user_stats}
    \begin{tabular}{l r r r r}
        \toprule
        \textbf{Dataset} & \textbf{Median} & \textbf{Mean} & \textbf{CV} & \textbf{Max} \\
        \midrule
        \multicolumn{5}{l}{\textit{Functions per User}} \\
        \cmidrule{1-5}
        \AzureFirst    & 2 & 4.9 & 4.8 & 2\,242 \\
        \AzureBlob     & 1 & 1.7 & 1.2 & 26 \\
        \HuaweiSecond  & 2 & 7.6 & 15.1 & 4\,757 \\
        \IBM           & 5 & 7.6 & 1.2 & 58 \\
        \midrule
        \multicolumn{5}{l}{\textit{Daily Requests per User}} \\
        \cmidrule{1-5}
        \AzureFirst    & 77.3 & 6.1e4 & 26.3 & 1.3e8 \\
        \AzureBlob     & 4.6 & 4.7e3 & 10.4 & 1.0e6 \\
        \Globus        & 0.1 & 38.8 & 6.1 & 3.1e3 \\
        \HuaweiSecond  & 4.0 & 1.6e6 & 19.5 & 9.2e8 \\
        \IBM           & 690.0 & 2.0e5 & 5.7 & 1.2e7 \\
        \bottomrule
    \end{tabular}%
\end{table}

\altparagraph{I/O Latency.}
To understand the network overhead of transferring data in and out of functions, we analyze the frontend I/O latencies from the \HuaweiSecond trace in \cref{fig:io_latency}.
Frontend I/O is highly optimized.
Writing responses is nearly instantaneous, showing a median latency of just 7.4~\textmu s and a P99 of 11.4~\textmu s.
This tight distribution indicates that function outputs in this environment are typically small (e.g., status codes or acknowledgments).
In contrast, reading request bodies shows a noticeably heavier tail.
Although the median read latency is still low at 26~\textmu s, the mean is 301~\textmu s, and the P99 latency reaches 6.2~ms.
This disparity directly correlates with the skewed request body sizes observed earlier. Most incoming payloads are tiny, whereas the tail of larger data payloads requires significantly more time to transfer.

\begin{insight}
The dichotomy between the high volume of ephemeral data and the small set of highly frequent objects suggests that traditional LRU caching policies are ineffective for the majority of FaaS I/O, because they will be polluted by single-access objects across tenants. Instead, caching strategies should be size-aware and selective to mitigate the severe tail latencies associated with transferring large payloads. Furthermore, because nearly 90\% of highly accessed objects are bound to a single application, providers can largely rely on application-scoped localized caching tiers rather than cross-tenant distributed caches.
\end{insight}

\subsection{Users}
\label{sec:users}
We characterize the tenancy patterns of FaaS platforms by analyzing the number of functions registered per user and the request volume generated by each user. This analysis helps to understand user behavior and the structural composition of the workload, which is essential for capacity planning. We use \AzureFirst, \AzureBlob, \Globus, \HuaweiSecond, and \IBM, because these datasets include anonymized user identifiers allowing per-user aggregation.

\altparagraph{Functions per User.}
\Cref{fig:functions_per_user} presents the CDF of the number of functions registered by individual users. We exclude \Globus from this analysis because its trace logs the user submitting the task rather than the owner of the function. The distribution is heavily skewed toward users with few functions. As detailed in \cref{tab:user_stats}, platforms such as \AzureFirst, \AzureBlob, and \HuaweiSecond typically show a median of only 1 or 2 functions per user. This suggests that a substantial portion of the user base consists of developers deploying simple applications or microservices with stable configurations. \IBM has a slightly higher median of 5 functions, but shows a tight distribution ($\text{CV} \approx 1.2$) with a maximum of only 58 functions, reflecting a container-centric model where users deploy a constrained set of related services rather than distributed function architectures.

\altparagraph{Daily Requests per User.}
The disparity in workload intensity among users is pronounced. \Cref{fig:requests_per_user} illustrates the CDF of average daily requests per user, showing a heavy-tailed distribution across all datasets. \Cref{tab:user_stats} shows the magnitude of this inequality. For \HuaweiSecond, the median user sends only 4.0~requests per day, but the mean is 1.56~million, driven by high-volume users that generate up to 915~million daily requests ($\text{CV} \approx 19.5$). \AzureFirst follows a similar trend, where the mean daily request count (60\,797) is nearly three orders of magnitude higher than the median (77.3). \IBM stands out with a significantly higher baseline, showing a median of 690~daily requests, although it still exhibits a heavy tail with a maximum of 11.6~million daily requests. \AzureBlob, tracking specific storage interactions, mirrors this high skew ($\text{CV} \approx 10.4$) despite a low median of 4.6~daily requests. \Globus shows a lower overall volume consistent with sporadic scientific processing (median 0.1~daily requests), yet maintains a high coefficient of variation ($\text{CV} \approx 6.1$). This confirms that even in scientific computing, a small fraction of users drives the majority of the system load, whereas most users submit tasks sporadically.

\begin{insight}
The high skew in request generation creates a dichotomy in system management. High-volume ``whale'' users naturally keep their functions warm, whereas the long tail of low-volume users generates the majority of cold starts and resource fragmentation. Providers should therefore balance dedicated capacity and isolation for heavy hitters with aggressive idle-resource reclamation strategies for the long tail.
\end{insight}

\section{Discussion}
\label{sec:discussion}

\subsection{Implications for FaaS Platform Design}
As function execution times have dropped to double-digit milliseconds, the relative cost of platform overhead has grown proportionally. Critically, our breakdown of cold start latency in \HuaweiSecond shows that scheduling (mean: 1.17~s) and dependency deployment (0.95~s) dominate over code download (0.43~s). Combined with the strong positive correlation between dependency count and response time in \Globus, this points clearly toward data-aware scheduling and proactive dependency pre-fetching as higher-value targets than code-size reduction. The growing dominance of timer-based triggers (42.2\% in \HuaweiSecond) and diurnal cycles further amplify this opportunity, since invocation schedules are known to the platform in advance, enabling precise rather than statistical pre-warming.

The resource over-provisioning we observe represents a largely untapped efficiency gain. The \IBM trace reinforces this direction from a different angle: The ``Serverless 2.0'' concurrency model, where a single container multiplexes up to 100 concurrent requests, achieves a P99 instance count of just 6 compared to 90 for the request-isolated \HuaweiFirst, demonstrating that concurrency-aware container management is a more resource-efficient path than linear scale-out.

Finally, the extreme skew in function popularity, per-user request volumes, and data access volume has direct implications for keep-alive and eviction strategies. High-volume ``whale'' users naturally keep their functions warm, whereas the long tail of low-volume users generates the majority of cold starts and resource fragmentation. Platforms should therefore apply asymmetric policies: sustained keep-alive for popular functions and aggressive idle-resource reclamation for the long tail, with per-function keep-alive windows tuned to empirical IAT distributions.

\subsection{Implications for FaaS Benchmark Design}
A recurring theme across all datasets is that summary statistics such as mean or median are poor representatives of production workloads. Execution times, cold start latencies, request patterns, and other characteristics all show heavy tails. Therefore, simple generative models fail to capture the characteristics of serverless workloads. Future benchmarks should adopt trace-driven replay to accurately reflect these structural and temporal dynamics. Furthermore, we emphasize the importance of multi-trace evaluation. An algorithm tuned to an older trace (e.g., \AzureFirst) will likely fail under the high-concurrency workloads of \IBM or the scientific workloads of \Globus. Systems should be evaluated across multiple, diverse traces to ensure generalizability.

\subsection{Limitations}
\label{sec:discussion_limitations}
Although our study provides a general overview, it is subject to several limitations. First, we analyzed ``black box'' systems. The observed performance metrics are a result of both the inherent workload characteristics and the specific policies (e.g., cold start optimization, scheduling, caching) of the underlying platform. Without insight into these internal mechanisms, it is challenging to definitively separate workload effects from platform effects. For example, a low cold start rate could stem from high keep-alive times.

Second, each dataset represents the user base and common use cases of a specific platform. For example, the \Globus trace is dominated by scientific computing tasks and may therefore not be representative of Azure workloads, whereas the \HuaweiFirst trace captures internal private-cloud workloads that may not generalize to public cloud workloads. Therefore, although we identified common patterns, generalizing findings to other FaaS platforms should be done with caution.

Finally, the FaaS ecosystem is evolving rapidly. Some of our datasets, particularly \AzureFirst, are several years old. As our comparison between the \AzureFirst and \AzureSecond traces showed, workload characteristics can change significantly in a relatively short period. Our findings represent a snapshot in time, and continuous characterization with new public datasets will be necessary to track the ongoing evolution of serverless workloads.

\section{Related Work}
\label{sec:related_work}

Workload characterization is fundamental for understanding the behavior of serverless platforms and designing system optimizations. The literature in this field mainly divides into empirical studies analyzing production traces from cloud providers and efforts to generate synthetic workloads when real-world data are unavailable. We review key contributions in both areas to contextualize our cross-platform analysis.

\subsection{Production Workload Traces}
Researchers rely on these traces to design and validate system optimizations, e.g., observed workload patterns have motivated new policies for resource management~\cite{shahrad2020serverless}, caching~\cite{fuerst2021faascache}, predictive pre-warming~\cite{roy2022icebreaker}, autoscaling~\cite{qiu2023aware}, and scheduling~\cite{akbari2025hiku}.
Beyond motivating new algorithms, these same traces often serve as a benchmark to evaluate the performance of algorithms by replaying real-world request patterns from a trace. Our work builds on this foundation by synthesizing insights from the key public traces that have shaped the field.

The foundational work by Shahrad et al.~\cite{shahrad2020serverless} on the Azure Functions trace (\AzureFirst) was one of the first large-scale characterizations of a production FaaS workload. The authors identified key properties such as the bimodal distribution of function lifetimes and strong diurnal patterns, which motivated their design of a predictive keep-alive policy. Many subsequent works have used this trace to evaluate new ideas, from caching techniques~\cite{fuerst2021faascache} to novel pre-warming strategies~\cite{roy2022icebreaker}. Our study re-examines this seminal work in newer traces to understand how these initial findings have evolved.

Subsequent workload studies have broadened the scope to other platforms and use cases. In 2021, Romero et al.~\cite{romero2021faa} released an additional Azure trace (\AzureBlob) that specifically captured interactions with Blob Storage, allowing for the first detailed analysis of data access patterns in FaaS applications. Zhang et al.~\cite{zhang2021faster} released a newer Azure trace (\AzureSecond), highlighting the prevalence of short-lived functions and exploring the potential for using ``harvested'' VMs. Wang et al.~\cite{wang2021faasnet} provided a view into the Alibaba Cloud Function Compute platform (\Alibaba), which was used to inform the design of a container provisioning system. Joosen et al. have released two datasets from Huawei Cloud (\HuaweiFirst and \HuaweiSecond)~\cite{joosen2023does, joosen2025serverless}, with the former characterizing long-term trends and the latter offering deep insights into cold start events, their internal components, and multi-region deployments. Concurrently, Bauer et al.~\cite{bauer2024globus} provided a unique perspective on a federated FaaS model from the scientific computing domain with the Globus Compute trace (\Globus), capturing edge-to-cloud endpoint characteristics. Recently, Nasiri et al.~\cite{nasiri2026serverless} provided a trace from the IBM Cloud Code Engine (\IBM), offering insights into a Knative-based ``Serverless 2.0'' environment that supports both functions and general-purpose containers alongside advanced user configurations like pod concurrency.

Although these individual studies have provided snapshots, they have largely been analyzed in isolation. A few studies have performed limited comparisons of different traces, for example, Li et al.~\cite{li2023alpaserve} referenced the burstiness characteristics of both Azure traces to motivate a system for deep learning serving. However, a comprehensive, side-by-side comparative analysis has been missing across this wide range of modern datasets. Our work directly addresses this gap by synthesizing the findings from these disparate traces into a unified characterization. By analyzing eight datasets spanning four major public cloud providers and a specialized scientific platform over a seven-year period, we provide a more robust and generalizable understanding of serverless workloads.

\subsection{Synthetic Workload Traces}
In the absence of sufficient public production data, some research has turned to simulated or synthetic workloads. Benchmarking suites such as the Serverless Benchmark Suite (SeBS)~\cite{sebs} and FunctionBench~\cite{kim2019functionbench} provide a standardized set of functions for performance evaluation, whereas SeBS also provides simulated workloads. Others have developed entire simulation platforms, such as SimFaaS~\cite{mahmoudi2021simfaas}, to predict performance metrics. Although simulated data are useful for controlled experiments, they often simplify real-world complexities and may not capture the full spectrum of behaviors such as burstiness or heavy-tailed distributions. Synthetic data can offer more realism by statistically mirroring production traces, but its quality is fundamentally dependent on the availability of representative real-world data. Therefore, these approaches cannot fully replace the analysis of large-scale production traces, which remains the ``gold standard'' for understanding real-world system behavior.

\section{Conclusions \& Future Work}
\label{sec:conclusion}

In this paper, we conducted a large-scale, cross-platform characterization of serverless workloads using eight publicly available FaaS workload traces. Our analysis provides a comprehensive and up-to-date view of the state of FaaS, moving beyond the reliance on single, often outdated, data sources. We found that despite the diversity of platforms, several key patterns hold true: FaaS workloads are driven by strong, predictable diurnal cycles and extreme skewness in function popularity and user request volumes. Performance is characterized by fast execution times, which are becoming increasingly short, coupled with heavy-tailed latency distributions. This makes system overheads such as cold starts even more critical to mitigate. Our analysis further showed widespread resource over-provisioning, a shift toward persistent ``Serverless 2.0'' concurrency models, and a dependency landscape where most functions import few libraries, pointing to clear avenues for optimization in resource allocation and data locality. To support reproducibility and further research, we have open-sourced our analysis: \textcolor{darkblue}{\url{https://github.com/saman-akbari/characterizing-serverless-workloads}}.

Future work should build on these findings in two main directions: First, we encourage the development of algorithms grounded in our empirical data, such as predictive autoscalers and dependency-aware schedulers. Second, establishing a common data format for new public traces would greatly accelerate research and help the community track the ongoing evolution of serverless computing.

\section*{Acknowledgments}

We thank Jonathan W\"orner who supported this work
in the scope of a bachelor's thesis.

\balance


\begin{thebibliography}{10}
\expandafter\ifx\csname url\endcsname\relax
  \def\url#1{\texttt{#1}}\fi
\expandafter\ifx\csname urlprefix\endcsname\relax\def\urlprefix{URL }\fi
\expandafter\ifx\csname href\endcsname\relax
  \def\href#1#2{#2} \def\path#1{#1}\fi

\bibitem{shahrad2020serverless}
M.~Shahrad, R.~Fonseca, {\'I}.~Goiri, G.~Chaudhry, P.~Batum, J.~Cooke, E.~Laureano, C.~Tresness, M.~Russinovich, R.~Bianchini, Serverless in the wild: Characterizing and optimizing the serverless workload at a large cloud provider, in: Proceedings of the USENIX Annual Technical Conference (ATC), 2020, pp. 205--218, trace released in 2019.

\bibitem{awslambda}
{Amazon Web Services}, \href{https://aws.amazon.com/lambda/}{{AWS Lambda}} (2014).
\newline\urlprefix\url{https://aws.amazon.com/lambda/}

\bibitem{joosen2025serverless}
A.~Joosen, A.~Hassan, M.~Asenov, R.~Singh, L.~Darlow, J.~Wang, Q.~Deng, A.~Barker, Serverless cold starts and where to find them, in: Proceedings of the ACM European Conference on Computer Systems (EuroSys), 2025, pp. 938--953.

\bibitem{romero2021faa}
F.~Romero, G.~I. Chaudhry, {\'I}.~Goiri, P.~Gopa, P.~Batum, N.~J. Yadwadkar, R.~Fonseca, C.~Kozyrakis, R.~Bianchini, {Faa\$t}: A transparent auto-scaling cache for serverless applications, in: Proceedings of the ACM Symposium on Cloud Computing (SoCC), 2021, pp. 122--137.

\bibitem{zhang2021faster}
Y.~Zhang, {\'I}.~Goiri, G.~I. Chaudhry, R.~Fonseca, S.~Elnikety, C.~Delimitrou, R.~Bianchini, Faster and cheaper serverless computing on harvested resources, in: Proceedings of the ACM Symposium on Operating Systems Principles (SOSP), 2021, pp. 724--739.

\bibitem{wang2021faasnet}
A.~Wang, S.~Chang, H.~Tian, H.~Wang, H.~Yang, H.~Li, R.~Du, Y.~Cheng, {FaaSNet}: Scalable and fast provisioning of custom serverless container runtimes at {Alibaba Cloud Function Compute}, in: Proceedings of the USENIX Annual Technical Conference (ATC), 2021, pp. 443--457.

\bibitem{joosen2023does}
A.~Joosen, A.~Hassan, M.~Asenov, R.~Singh, L.~Darlow, J.~Wang, A.~Barker, How does it function? {Characterizing} long-term trends in production serverless workloads, in: Proceedings of the ACM Symposium on Cloud Computing (SoCC), 2023, pp. 443--458.

\bibitem{bauer2024globus}
A.~Bauer, H.~Pan, R.~Chard, Y.~Babuji, J.~Bryan, D.~Tiwari, I.~Foster, K.~Chard, The {Globus Compute} dataset: An open function-as-a-service dataset from the edge to the cloud, Future Generation Computer Systems 153 (2024) 558--574.

\bibitem{nasiri2026serverless}
N.~Nasiri, N.~Munshi, S.~D. Moser, M.~Pirvu, V.~Sundaresan, D.~Maier, T.~Premnath, N.~B{\"o}wing, S.~Gopalakrishnan, M.~Shahrad, In-production characterization of an open source serverless platform and new scaling strategies, in: Proceedings of the ACM European Conference on Computer Systems (EuroSys), 2026.

\bibitem{dask}
{Dask Development Team}, \href{https://www.dask.org/}{Dask: Library for dynamic task scheduling} (2016).
\newline\urlprefix\url{https://www.dask.org/}

\bibitem{fuerst2021faascache}
A.~Fuerst, P.~Sharma, {FaasCache}: keeping serverless computing alive with greedy-dual caching, in: Proceedings of the ACM International Conference on Architectural Support for Programming Languages and Operating Systems (ASPLOS), 2021, pp. 386--400.

\bibitem{roy2022icebreaker}
R.~B. Roy, T.~Patel, D.~Tiwari, {IceBreaker}: Warming serverless functions better with heterogeneity, in: Proceedings of the ACM International Conference on Architectural Support for Programming Languages and Operating Systems (ASPLOS), 2022, pp. 753--767.

\bibitem{qiu2023aware}
H.~Qiu, W.~Mao, C.~Wang, H.~Franke, A.~Youssef, Z.~T. Kalbarczyk, T.~Ba{\c{s}}ar, R.~K. Iyer, {AWARE}: Automate workload autoscaling with reinforcement learning in production cloud systems, in: Proceedings of the USENIX Annual Technical Conference (ATC), 2023, pp. 387--402.

\bibitem{akbari2025hiku}
S.~Akbari, M.~Hauswirth, {Hiku}: Pull-based scheduling for serverless computing, in: Proceedings of the IEEE International Symposium on Cluster, Cloud and Internet Computing (CCGrid), 2025, pp. 450--461.

\bibitem{li2023alpaserve}
Z.~Li, L.~Zheng, Y.~Zhong, V.~Liu, Y.~Sheng, X.~Jin, Y.~Huang, Z.~Chen, H.~Zhang, J.~E. Gonzalez, I.~Stoica, {AlpaServe}: Statistical multiplexing with model parallelism for deep learning serving, in: Proceedings of the USENIX Symposium on Operating Systems Design and Implementation (OSDI), 2023, pp. 663--679.

\bibitem{sebs}
M.~Copik, G.~Kwasniewski, M.~Besta, M.~Podstawski, T.~Hoefler, {SeBS}: a serverless benchmark suite for function-as-a-service computing, in: Proceedings of the ACM/IFIP Middleware Conference, 2021, pp. 64--78.

\bibitem{kim2019functionbench}
J.~Kim, K.~Lee, {FunctionBench}: A suite of workloads for serverless cloud function service, in: Proceedings of the IEEE International Conference on Cloud Computing (CLOUD), 2019, pp. 502--504.

\bibitem{mahmoudi2021simfaas}
N.~Mahmoudi, H.~Khazaei, {SimFaaS}: A performance simulator for serverless computing platforms, in: International Conference on Cloud Computing and Services Science (CLOSER), 2021, pp. 23--33.

\end{thebibliography}
\end{document}